\documentclass[a4paper,11pt]{article}
\pdfoutput=1

\usepackage{jheppub}
\usepackage{hyperref}
\usepackage[T1]{fontenc}

\usepackage{amsmath}
\usepackage{amsthm}

\usepackage{bm}
\usepackage{mathtools}
\usepackage{caption}
\usepackage{slashed}
\usepackage{tikz-cd}
\usepackage{eufrak}
\usepackage{upgreek}
\usepackage{bm}

\DeclareRobustCommand{\varlambda}{\text{\usefont{OML}{txmi}{m}{it}\symbol{"15}}}

\def\A{\mathcal A}

\newcommand{\bea}{\begin{eqnarray}}
\newcommand{\eea}{\end{eqnarray}}

\newcommand{\be}{\begin{equation}}
\newcommand{\ee}{\end{equation}}
\renewcommand{\exp}[1]{\operatorname{exp}\left\{#1\right\}}

\newcommand\AffiliationTextUnimib{Dipartimento di Fisica, Università di Milano-Bicocca, Piazza della Scienza 3, I-20126 Milan, Italy}
\newcommand\AffiliationTextInfnMib{INFN, Sezione di Milano-Bicocca, Piazza della Scienza 3, I-20126 Milan, Italy}
\newcommand\AffiliationTextCern{Theoretical Physics Department, CERN, 1211 Geneva 23, Switzerland}
\newcommand\AffiliationTextColorado{Department of Physics, Colorado State University, Fort Collins, CO 80523, USA}

\title{Spectral densities from Euclidean correlators via integral transforms: theoretical framework}
\author[a,b,c]{\!\! L.~Giusti}
\author[a,d]{\!\!, M.~Saccardi}
\author[a,b]{\!\!, and D.~Toniolo}

\affiliation[a]{\AffiliationTextUnimib}
\affiliation[b]{\AffiliationTextInfnMib}
\affiliation[c]{\AffiliationTextCern}
\affiliation[d]{\AffiliationTextColorado}

\emailAdd{leonardo.giusti@unimib.it}
\emailAdd{matteo.saccardi@colostate.edu}
\emailAdd{d.toniolo4@campus.unimib.it}

\abstract{Spectral densities link experimental measurements to
  dynamical properties of a quantum field theory which, in turn, can
  be resolved non-perturbatively from the Euclidean time-dependence of
  correlation functions. By making extensive use of integral 
  transforms, we present analytic formulae to carry out the inverse
  Laplace transform so as to extract spectral densities from either
  the continuum or the discrete sampling of correlation functions in
  the Euclidean time.
  Formulae extend to regulated and/or smeared spectral densities as well.
  We explicitly show that the proposed lattice solution tends to its
  continuum counterpart up to $O(a^2)$ effects in the lattice
  spacing $a$ if the lattice correlator is $O(a)$-improved. In practical
  computations, lattices have necessarily a finite Euclidean temporal extent,
  a lack of knowledge which suggests to introduce incomplete integral transforms
  and the corresponding incomplete smeared spectral densities. The contribution
  from the unknowns to a smeared spectral density can then be rigorously bound
  and kept under control if the integral transform of the smearing function decays fast
  enough with the conjugate variable. Conversely, the bound can be used to plan lattices so
  as to achieve a given target precision on the reconstructed spectral density of
  interest. The formulae presented here in the context of lattice field theory
  can be easily applied or extended to other areas of research.
}  
\preprint{CERN-TH-2026-150}

\begin{document}

\maketitle

\section{Introduction}\label{sec:intro}
Spectral densities play a central r\^ole in field theories. They link dynamical properties of a theory to observables that can be
measured in experiments. In the presence of strongly interacting fields, such as in Quantum Chromodynamics (QCD), non-perturbative
computations of spectral densities from first principles are essential for a detailed comprehension of the dynamics of the theory,
e.g. its particle and resonance content and their properties, and for the comparison against experimental results. Smeared spectral
densities from physics-motivated kernels are interesting as well. These classes of observables give access to inclusive hadronic
cross sections, semi-leptonic decay rates, or non-static properties
of the quark-gluon plasma, to name a few (see for instance
Refs.~\cite{Poggio:1975af,Weinberg:1995mt,Hansen_2017,Gambino:2020crt,Jeon:1995zm,Meyer:2007ic}).

Correlation functions of (composite) fields in Euclidean space-time are the primary quantities that are computed
non-perturbatively on the lattice by Monte Carlo evaluations of path-integrals. Thanks to the K\"all\'en–Lehmann decomposition,
two-point correlators are related to the corresponding spectral densities by a Laplace transform. The direct extraction of (smeared)
spectral densities from lattice data thus entails the Inverse Laplace Transform (ILT) of correlators, a notorious difficult inverse problem.
The conceptual, algorithmic and technological progress in lattice QCD, however, unlocked unprecedented levels of precision in modern
numerical calculations~\cite{FlavourLatticeAveragingGroupFLAG:2024oxs,Aliberti:2025beg}.
Correlation functions at larger and larger time-distances are becoming accessible thanks
to more efficient statistical estimators and to the development of multi-level integration techniques in the presence of
fermions~\cite{Ce:2016idq,Ce:2016ajy,DallaBrida:2020cik}. Larger and larger lattices with a denser and denser
spectrum of states
are being simulated in master-field simulations~\cite{Luscher:2017cjh,Giusti:2018cmp,Francis:2019muy}. Therefore, it has become
interesting and timely to reconsider the direct extraction of (smeared) spectral densities from lattice
data~\cite{Hansen_2017,HLT,Bailas:2020qmv}.

Recently, we have proposed explicit analytic formulae to extract smeared spectral densities
from the Euclidean time-dependence of correlation functions~\cite{Bruno:2024fqc}.
By making extensive use of integral transforms, the aim of this paper is to
generalize and extend these results so as to be able to
extract regulated and/or smeared spectral densities from either the continuous
or discrete sampling of
correlation functions in the Euclidean time. We pay particular
attention to discretization effects and to rigorously bind the systematic uncertainties
due to the finite extension of the time direction.
The practical implementation of the strategy proposed here, and its
first application to real lattice QCD data generated by a multi-level Monte Carlo will be
presented in full details in a companion paper~\cite{Giusti:2026lg2}.\\[0.25cm]

\section{Spectral densities from integral transforms}\label{sec:transforms}
In this paper we focus on two-point connected correlation functions in
Euclidean space-time\footnote{In the main sections of this paper we
 assume the time-momentum representation of correlators, omitting the
 momentum dependence from the notation,
 and consider the zero momentum case. Our results, however, are fully
 generic and applicable to non-zero momentum correlators defined
 in Appendix~\ref{app:KL}. Note also that our derivation applies
 straightforwardly to higher $n$-point functions,
 involving multiple Euclidean-time separations~\cite{Bulava:2019kbi,Patella:2024cto}.} 
\be\label{eq:corrt}
    C(t) = \int\, \langle {\mathcal O}_1(t, \mathbf x) {\mathcal O}_2(0, \mathbf 0)
    \rangle_{\rm c}\; d^3 \mathbf x \,,  
\ee
where ${\mathcal O}_1$ and ${\mathcal O}_2$ are two renormalized scalar  fields\footnote{Generalization 
to non-zero spin can be found in
many textbooks, see for instance Ref.~\cite{Weinberg:1995mt}.}. Thanks
to the K\"all\'en–Lehmann spectral decomposition, see
Appendix~\ref{app:KL} for the notation, two-point correlators are related to the
corresponding spectral densities $\rho(\omega) $ by
the Laplace transform  
\begin{equation}
    C(t) = \int_0^{\infty} \rho(\omega) \, e^{-\omega t}\, d \omega \,.
    \label{eq:ct}
\end{equation}
The spectral density is a function of the energy
 $\omega$, with support over $[\omega_0\! > \! 0,\infty)$ in channels
 with a mass gap $\omega_0$, which encodes the dynamical information of the
 theory in the channel identified by the quantum numbers of the fields
in Eq.~(\ref{eq:corrt}). When combined with the short-distance (operator product
expansion) analysis, these properties fix the asymptotic behaviour
of correlation functions at the origin and at infinity to be 
\be\label{eq:asymptc}
\lim_{t\rightarrow 0^+} C(t) = O(t^{-p})\,, \qquad \lim_{t\rightarrow +
  \infty} C(t) = O(e^{-\omega_0 t}) \, ,
\ee
where $p$ is close to an integer in QCD. This,  
together with Eq.~(\ref{eq:ct}), in turn implies that 
\be\label{eq:asymptrho}
\rho(\omega)\Big|_{\omega\in [0,\omega_0)} = 0\,, \qquad
\lim_{\omega\rightarrow +
  \infty} \rho(\omega) = O(\omega^{p-1}) \, .
\ee
To determine the spectral density from the correlator, 
the Eq.~(\ref{eq:ct}) has to be inverted, or equivalently
the ILT of the correlator has to be computed. For the clarity of the 
presentation, in the first part of this section we present a rather
general scheme of how integral transforms can lead to the extraction
of the spectral density. The underlying assumptions and the details of
the particular transforms (e.g. basis functions, domain of integration, etc.)
are fully spelled out in the following subsections.

For the time being,
given the asymptotic behaviours in Eqs.~(\ref{eq:asymptc}) and (\ref{eq:asymptrho}),
let us assume that the correlator and the spectral density in Eq.~(\ref{eq:ct})
admit integral transforms, e.g. one of those in Appendix~\ref{app:Trsf}, defined as
\be\label{eq:transforms}
{\overline {\rm C}}_s  =  \int C(t)\, \bar {\rm u}_s(t)\, dt\, ,
\qquad
\hat \uprho_s = \int \rho(\omega) \, \hat {\rm u}_s(\omega)\, d\omega\, ,
\ee
and their inverses, and that the basis functions $\bar {\rm u}_s(t)$ and $\hat {\rm u}_s(\omega)$ are related by
\be\label{eq:ubar2uhat}
\varlambda_s \, \hat {\rm u}_s(\omega) = \int e^{-\omega t}\, \bar {\rm u}_s(t)\, dt \, ,
\ee
where $\varlambda_s$ is a numerical coefficient and $s$ is the real conjugate variable,
see below. By inserting Eq.~(\ref{eq:ubar2uhat}) into the second of Eqs.~(\ref{eq:transforms}), and 
then by using Eq.~(\ref{eq:ct}), it immediately follows that 
\be\label{eq:ILTalg}
{\overline {\rm C}}_s = \varlambda_s\, \hat \uprho_s\, . 
\ee
As a result, {\it the basic integral equation} (\ref{eq:ct}) {\it becomes an algebraic relation
between the integral transforms of the correlator and of the spectral density} provided
$\varlambda_s\neq 0$ for all $s$. This is
the main advantage for considering integral transforms in this context. If we require
the orthogonality condition
\be\label{eq:ortuhatg}
\int \hat {\rm u}^{*}_s(\omega)\,\hat {\rm u}_{s'}(\omega) \, d\omega = \delta( s - s') \,,
\ee
then Eq.~(\ref{eq:ubar2uhat}) leads to
\be\label{eq:uAug}
\int \int \bar {\rm u}^{*}_s(t)\, A(t+t')\, \bar {\rm u}_{s'}(t')\, dt\, dt' =
\vert \varlambda_s \vert^2 \delta( s - s') \,,
\ee
where $A(t+t')$ is the kernel of a (quasi) Carleman operator\footnote{The (quasi) Carleman operators
are real and they always admit a basis of real eigenvectors, see Appendix~\ref{app:carleman} for details.
We prefer, however, to keep the discussion more general, and leave the possibility that the basis functions are complex.}
defined in
Eq.~(\ref{eq:quasi-carl}) of Appendix~\ref{app:carleman}. If the
$\bar {\rm u}_s(t)$ are chosen to be eigenfunctions of $A$,  
\be
    \int A(t+t')\, \bar {\rm u}_{s}(t')\, dt' = \vert\varlambda_{s}\vert^2\, \bar {\rm u}_{s}(t) \,,
\ee
then also the vectors $\bar {\rm u}_s(t)$ form an orthonormal basis and satisfy
\be \label{eq:uhat2ubar}
\varlambda^{*}_s\, \bar {\rm u}_s(t) = \int e^{-t \omega}\, \hat {\rm u}_s(\omega)\, d\omega \, .
\ee
Therefore, once the integral transform of the correlator is computed,
by using the completeness relation
\be\label{eq:compluhatg}
\int \hat {\rm u}^{*}_s(\omega)\,\hat {\rm u}_{s}(\omega') \, ds = \delta(\omega - \omega') \,,
\ee
the spectral density can be readily computed as 
\be\label{eq:rhogen}
\rho(\omega) = \int \frac{{\overline {\rm C}}_s}{\varlambda_s} \,
\hat {\rm u}^{*}_s(\omega)\, ds\, .
\ee
Notice that typically $\vert \varlambda_s \vert$ decreases
exponentially with $s$ as well as $\vert {\overline {\rm C}}_s\vert$,
and a delicate cancellation between numerator and denominator is at work
on the r.h.s. of Eq.~(\ref{eq:rhogen}).

\subsection{Mellin transform}
Maybe the simplest choice to implement the strategy outlined above is the
Mellin transform, see Ref.~\cite{Bruno:2024fqc} and references therein.
In this subsection we first examine the case 
of correlators which are $L^2(\mathbb R^+)$. The corresponding 
spectral densities are $L^2(\mathbb R^+)$ as well, and do not need
to be regulated, see Appendix~\ref{app:KL}. The more
general case of regulated spectral densities follows. 

\subsubsection{Spectral density}
If $C(t) \in L^2(\mathbb R^+)$, i.e. $p<1/2$ in Eq.~(\ref{eq:asymptc}),
the basis functions and eigenvalues
for the Mellin transform either for the correlator or for the
spectral density are\footnote{Notice the change in the sign of
the label $s$ when moving from the basis $u_s(t)$ for the
correlator to $u_s^\ast(\omega) = u_{-s}(\omega)$ for the
spectral density.}
\be
\bar {\rm u}_s(t) = u_s(t) = \frac{t^{is}}{\sqrt{2\pi t}}\, , \quad 
\hat {\rm u}_s(\omega) = u^*_s(\omega)\, , \quad
\varlambda_s = \lambda_s = \Gamma\Big(\frac{1}{2}+is\Big)\,,\quad s \in {\mathbb R}\, , 
\ee
since the Carleman operator is the one given in Eq.~(\ref{eq:quasi-carl1}), 
see Appendix~\ref{app:carleman} for more details. The Mellin transforms of the correlator and
the spectral density read
\be
\overline C_s = \int_{0}^{\infty} C(t)\, u_s(t)\, dt\, , \qquad
\hat \rho_s = \int_{0}^{\infty} \rho(\omega)\, u^*_s(\omega)\,  d\omega\,  ,
\ee
and thanks to the analogous of Eq.~(\ref{eq:ubar2uhat}), see
Appendix~\ref{app:carleman}, 
\be\label{eq:bellam0}
\lambda_s u_s^\ast(\omega) = \int_0^{\infty} e^{-\omega t}
u_s(t) \, d t\, ,
\ee
we have
\be
\overline C_s  = \lambda_s\, \hat \rho_s\, . 
\ee
Notice that $\overline C_s$ and $\hat \rho_s$ are analytic in the strip of the
complex plane where they exist, see for instance Refs.~\cite{Wong:1989ll,Flajolet:1995ll}.
The spectral density is then obtained as~\cite{Bruno:2024fqc}
\be\label{eq:rhosub1}
\rho(\omega) = \int_{-\infty}^{+\infty} \frac{\overline C_s}{\lambda_s}\,
u_s(\omega)\, d s\, ,
\ee
and, as anticipated, a delicate cancellation between numerator and denominator
is at work.

\subsubsection{Regulated spectral density\label{sec:mellinregu}}
When in Eq.~(\ref{eq:asymptc}) $p\geq 1/2$, the Mellin transform of the correlator can be
defined as
\be\label{eq:cbarM}
\overline C_{m,s} = \frac{1}{\sqrt{2\pi}} \int_{0}^{\infty} C(t)\, t^{m-1/2+is} dt =
 \int_{0}^{\infty} C(t)\, t^{m}\, u_s(t)\, dt\, ,\quad
m\in(p-1/2,+\infty)\, ,
\ee
where $s \in {\mathbb R}$, see Appendix \ref{app:Trsf} for more
details. It amounts to multiplying $C(t)$ by $t^m$, with $m>p-1/2$, so that
the resulting function is in $L^2(\mathbb R^+)$ and can again be expanded on the
basis of the $u_s(t)$. Analogously, the Mellin transform of the spectral density
can be reformulated as 
\be\label{eq:rhohatM}
\hat \rho_{m,s} = \frac{1}{\sqrt{2\pi}} \int_{0}^{\infty} \rho(\omega)\, \omega^{-m-1/2-is} d\omega =
\int_{0}^{\infty} \rho_m(\omega)\, u^*_s(\omega)  d\omega
\,, \quad m\in(p-1/2,+\infty)\, ,
\ee
and it corresponds to the Mellin transform of
the regulated spectral density $\rho_m(\omega) = \rho(\omega)/\omega^m\in L^2(\mathbb R^+)$,
see Appendix~\ref{app:KL}, which again
can be expanded on the basis of the $u^*_s(\omega)$.
From Eq.~(\ref{eq:bellam0}) it follows
\be\label{eq:bella}
\lambda_{m,s} \, \omega^{-m}\, u^*_s(\omega)\! =\!  
\lambda_s \Big(\!\!-\! \frac{\partial}{\partial \omega}\Big)^m\!\! u^*_s(\omega)
 =  \Big(\!\!-\! \frac{\partial}{\partial \omega}\Big)^m\!\!\!\! \int_{0}^{\infty}\!\! e^{-\omega t}\,
u_s(t)\, dt
=\!\!  \int_{0}^{\infty}\!\!\!\! e^{-\omega t}\, t^m u_s(t)\,  dt\, , 
\ee
where $\lambda_{m,s} = \Gamma\big(m+\frac{1}{2}+is\big)$,
which implies again that the basic integral equation (\ref{eq:ct}) becomes an
algebraic relation between the Mellin transforms of the correlator and of the
spectral density
\be\label{eq:fundamental}
\overline C_{m,s}  = \lambda_{m,s} \, \hat \rho_{m,s}\, .
\ee
By using the inversion formula in Eq.~(\ref{eq:invmellin}) of
Appendix~\ref{app:Trsf}, the regulated spectral density can finally be written
as~\cite{Bruno:2024fqc}
\be\label{eq:rhosubf}
\rho_m(\omega) = \frac{\rho(\omega)}{\omega^{m}} = \int_{-\infty}^{+\infty}
\frac{\overline C_{m,s}}{\lambda_{m,s}}
u_s(\omega)\, d s\; , \qquad m\geq 0\; .
\ee

\subsection{Kontorovich–Lebedev and Mehler-Fock transforms\label{sec:KLMF}}
Lattice correlators are often affected by very large discretization effects
at short time-distances. It is therefore appropriate to consider integral
transforms defined by integrating over $t\geq t_0>0$. In this subsection we
first examine the case of correlators with spectral densities which
are $L^1(\omega_0,\infty)$, i.e. $p<0$ in Eq.~(\ref{eq:asymptc}),
which do not need to be regulated to define the Kontorovich–Lebedev
transform and its inverse. The more general case of regulated
spectral densities follows. 

\subsubsection{Spectral density}
When in Eq.~(\ref{eq:asymptc}) $p<0$, the spectral density is
$L^1(\omega_0,\infty)$ and the Kontorovich–Lebedev transform and its inverse
can be defined. The basis functions and eigenvalues
for the related Mehler-Fock and Kontorovich–Lebedev transforms for 
the correlator and the spectral density are
\bea
\bar {\rm u}_s(t) & = & \bar u_s(t,t_0) = \sqrt{\frac{s \tanh(\pi s)}{t_0}}\,
{}_2F_1\Big(\frac{1}{2}+is,\frac{1}{2}-is;1;\frac{t_0-t}{2 t_0}\Big)\, , \quad \varlambda_s = \vert \lambda_s \vert \,,
\\[0.25cm] 
\hat {\rm u}_s(\omega) & = & \hat u_s(\omega,t_0)= \frac{\sqrt{2 s \sinh(\pi s)} }{\pi} \
\frac{K_{is}(\omega t_0)}{\sqrt{\omega}} \, , \quad
t \geq t_0>0\, \;\; s\in {\mathbb R}^+\, , \label{eq:uhat_s}
\eea
where ${}_2F_1$ and $K_{\nu}$ are the Gauss’s hypergeometric function
and the modified Bessel function of the second
kind respectively~\cite{Gradshteyn:1702455},
see Appendix~\ref{app:carleman} for more details. The corresponding
quasi-Carleman operators are defined in Eqs.~(\ref{eq:quasi-carl2}) and
(\ref{eq:quasi-carl3}).

The Mehler-Fock and the Kontorovich–Lebedev transforms of the correlator
and of the spectral density then read\footnote{We use the same notation for
these transforms and the corresponding Mellin ones since any ambiguity is resolved from
the context. The same applies to the regulated ones below. For the clarity of the notation,
we leave the dependence on $t_0$ implicit.}
\be\label{eq:MFKL}
\overline C_s = \int_{t_0}^{\infty} C(t)\, \bar u_s(t,t_0)\, dt\, , \qquad
\hat \rho_s = \int_{0}^{\infty} \rho(\omega)\, \hat u_s(\omega,t_0)\,  d\omega\,  ,
\quad s \in {\mathbb R^+}\, , 
\ee
while the analogous of Eq.~(\ref{eq:ubar2uhat}), see
Eq.~(\ref{eq:ubartouhat}) in Appendix~\ref{app:carleman}, becomes
the ``incomplete'' Laplace transform  
\be\label{eq:bellam1}
\vert \lambda_s \vert\, \hat u_s(\omega,t_0) =
\int_{t_0}^{\infty} e^{-\omega t}\, \bar u_s(t,t_0)\, dt\, .
\ee
This implies that 
\be
\overline C_s  = \vert \lambda_s\vert\, \hat \rho_s\, ,
\ee
and the spectral density is then obtained as
\be\label{eq:rhosub2}
\rho(\omega) = \int_{0}^{\infty} \frac{\overline C_s}{\vert\lambda_s\vert}\,
\hat u_s(\omega,t_0)\, d s\, 
\ee
where a delicate cancellation between numerator and denominator
is again at work.

\subsubsection{Regulated spectral density}
When in Eq.~(\ref{eq:asymptc}) $p\geq 0$, we can apply the procedure
of the previous subsection but on a suitably regulated correlator and
spectral density. Following the standard line of argumentation summarized
in Appendix~\ref{app:KL}, we define the  
regulated correlator and spectral density as 
\be\label{eq:fndsub}
C_m(t) = \int_0^\infty \rho_m(\omega)\,  e^{-\omega t}\, d\omega\, , \qquad
\rho_m(\omega) =\frac{\rho(\omega)}{\omega^m}\, , \qquad m>p\, .  
\ee
By noticing that
\be\label{eq:Cmdiff}
\Big(- \frac{\partial}{\partial t}\Big)^m C_m(t) = C(t)\, ,
\ee
and that, in the presence of a mass gap, the regulated correlator and its derivatives
are null at infinity, the $C_m(t)$ can be written as
\be\label{eq:Cmint}
C_m(t)  = \frac{1}{\Gamma(m)} \int_t^{\infty} C(t')\, (t'-t)^{m-1}\, d t'\,, \quad m>0\, , 
\ee
while $C_0(t)=C(t)$ for $m=0$. The Mehler-Fock transform of $C_m(t)$ reads
\be\label{eq:Cbarms}
\overline C_{m,s} = \int_{t_0}^{\infty} C_m(t)\, \bar u_s(t,t_0)\, dt =
\int_{t_0}^{\infty} C(t)\, \bar u_{m,s}(t,t_0)\, dt\, , \quad m\geq 0\, ,
\ee
where $\bar u_{0,s}(t,t_0)=\bar u_s(t,t_0)$ while, by using
Eqs.~(7.137.9) and (8.702) of Ref.~\cite{Gradshteyn:1702455}, 
\bea
\bar u_{m,s}(t,t_0) & = & \frac{1}{\Gamma(m)}\, \int_{t_0}^{t} \bar u_s(t',t_0) \, (t-t')^{m-1}\,
d t'\\[0.25cm]
& = & \sqrt{\frac{s \tanh(\pi s)}{t_0}}\, \frac{\,\,(t-t_0)^m}{\Gamma(m+1)}\,\,
{}_2F_1\Big(\frac{1}{2}+is,\frac{1}{2}-is;1+m;\frac{t_0-t}{2 t_0}\Big)\, ,\quad m \geq 0\, . \nonumber
\eea
The Kontorovich–Lebedev transform of $\rho_m(\omega)$ is
\be\label{eq:rhatms}
\hat \rho_{m,s} = \int_{0}^{\infty} \rho_m(\omega)\, \hat u_s(\omega,t_0)\,  d\omega =
\int_{0}^{\infty} \rho(\omega)\, \hat u_{m,s}(\omega,t_0)\,  d\omega \, ,
\ee
where 
\be
\hat u_{m,s}(\omega,t_0) = \frac{1}{\omega^m}\, \hat u_s(\omega,t_0)\,  ,
\ee
and 
\be\label{eq:bellam2}
\vert\lambda_s\vert \hat u_{m,s}(\omega,t_0) =
\int_{t_0}^{\infty} e^{-\omega t}\, \bar u_{m,s}(t,t_0)\, dt\, ,
\ee
with the last two equations which can also be explicitly derived
by using Eqs.~(8.702) and (7.141.5) of Ref.~\cite{Gradshteyn:1702455}.
By inserting Eq.~(\ref{eq:fndsub})
into Eq.~(\ref{eq:Cbarms}) and by using Eq.~(\ref{eq:bellam2}), it follows that 
\be\label{eq:CmsMFKL}
\overline C_{m,s}  = \vert \lambda_s\vert\, \hat \rho_{m,s}\, ,
\ee
and the regulated spectral density reads
\be\label{eq:rhosub3}
\rho_m(\omega) = 
\int_{0}^{\infty} \frac{\overline C_{m,s}}{\vert\lambda_s\vert}\,
\hat u_s(\omega,t_0)\, d s\,, \quad m \geq 0\, .
\ee
Notice that, if this very same procedure is adopted for the case of the Mellin transform,
it leads to the results given in Section~\ref{sec:mellinregu} where in particular
$\bar u_{m,s}$ is replaced by $t^m u_s(t)$ up to a ratio of Gamma functions.

\section{Smeared spectral densities}\label{sec:smeared}
A smeared spectral density is defined by integrating a 
(regulated) density with a known kernel function $\kappa(\omega)$,
\begin{equation}
 \rho_{\kappa;m} =  \int_0^{\infty}  \rho_m(\omega)  \, \kappa(\omega)\, d \omega \,.
    \label{eq:rhokappa}
  \end{equation}
Smeared spectral densities are interesting for several reasons. 
For physics-motivated kernels, they give access
to phenomenologically relevant observables, such as inclusive hadronic
cross sections or semi-leptonic decay rates to name a few (see for
instance Refs.~\cite{Hansen_2017,Gambino:2020crt,ETMCtau23,ETMCtau24}).
From a more theoretical point
of view, we have that often correlation functions are computed in a
finite spatial volume. In those cases, the associated spectral
densities are weighted sum of Dirac $\delta$-functions with possibly
large spacings between successive levels. Integrating them with
kernels which are representations of a Dirac $\delta$-function peaked
around a certain energy $\omega_\ast$ with a smearing width $\sigma$
allows one to define the infinite volume limit smoothly~\cite{Hansen_2017}.
For the clarity of the presentation, we first extend the general picture
provided  at the beginning of Section~\ref{sec:transforms} by introducing a
rather general scheme for computing smeared spectral densities via integral transforms.
The discussion is later made  precise for the Mellin and Kontorovich-Lebedev cases.

Let us assume that the smearing function
$\kappa(\omega)$ admits an integral transform defined as
\be\label{eq:transformk}
{\hat {\rm k}}_s = \int \kappa(\omega) \, \hat {\rm u}^*_s(\omega)\, d\omega\, .
\ee
By inserting the Eq.~(\ref{eq:rhogen}) into the Eq.~(\ref{eq:rhokappa})
we arrive to the Parseval-like formula
\be\label{eq:parseval}
\rho_{\kappa} = \int \hat \uprho_s \, {\hat {\rm k}}_s\, ds
= \int \frac{{\overline {\rm C}}_s}{\varlambda_s} \, {\hat {\rm k}}_s\, ds\, ,
\ee
which can be directly used for the calculation of $\rho_{\kappa}$. The Eq.~(\ref{eq:parseval})
may alternatively be derived by assuming that $\kappa(\omega)$ admits not only the integral transform
but also its inverse, i.e. satisfies the conditions described in Section~\ref{sec:transforms}.
Then, by inserting the analogous of Eq.~(\ref{eq:rhogen}) for the smearing
function into the Eq.~(\ref{eq:rhokappa}), Eq.~(\ref{eq:parseval}) is readily obtained. This
way $\rho_m(\omega)$ is only required to admit the integral
transform, a condition that can be more easily met by spectral densities which
become distributions at some finite value of $\omega$.

When a definite kernel is needed for illustration, in this paper we consider the
Breit-Wigner smearing defined as  
\be\label{eq:BWsm}
 \kappa(\omega) =   \frac1\pi
 \frac{\sigma}{(\omega-\omega_*)^2+\sigma^2}\, ,
\ee
while a more general discussion will be reported in
Ref.~\cite{Giusti:2026lg2}.

\subsection{Mellin case}
Again the Mellin transform is maybe the simplest option
for computing a smeared spectral density in the continuum. By following the line
of argumentation presented above, we define
\be\label{eq:khatm}
\hat \kappa_s = \int_{0}^{\infty} \kappa(\omega) \, u_s(\omega)\,  d\omega\, .
\ee
By inserting the Eq.~(\ref{eq:rhosubf}) into the Eq.~(\ref{eq:rhokappa})
we then obtain
\be\label{eq:rhosubks}
\rho_{\kappa;m}  = \int_{-\infty}^{+\infty} \frac{\overline C_{m,s}}{\lambda_{m,s}}\,
\hat \kappa_s\,  d s\, , \qquad m\geq 0\, ,
\ee
where $\overline C_{m,s}$ is defined in Eq.~(\ref{eq:cbarM}). For the Breit-Wigner, by using
Eq.~(3.197.1) of Ref.~\cite{Gradshteyn:1702455}, the transform reads
\bea
\hat \kappa_s  & = & \frac{\vert \lambda_s \vert^2}{2\pi i}
\Big[u_s(-\omega_* - i\sigma) - u_s(-\omega_* + i\sigma)\Big]\nonumber\\[0.25cm]
& = & \frac{1}{\cosh({\pi s})}
\Bigg[\sqrt{\frac{b+\omega_*}{2 b}}
\cosh(s\vartheta) -i \sqrt{\frac{b-\omega_*}{2 b}} \sinh(s\vartheta)\Bigg]\, u_s(b)\, , 
\label{eq:rhosubf_smea_bw}
\eea
where $b=\sqrt{\omega_*^2+\sigma^2}$ and $\cos \vartheta = -\frac{\omega_*}{b}$. 

\subsection{Kontorovich–Lebedev case\label{sec:KLsmea}}
For the case of the Kontorovich–Lebedev transform, we proceed analogously to the
previous subsection. The transform of the smearing
function is defined as 
\be
\hat \kappa_s = \int_{0}^{\infty} \kappa(\omega) \, \hat u_s(\omega,t_0) \,  d\omega\,.
\ee
By inserting the Eq.~(\ref{eq:rhosub3}) into the Eq.~(\ref{eq:rhokappa}) we obtain
\be\label{eq:rhosubkss}
\rho_{\kappa;m}  = \int_{0}^{\infty} \frac{\overline C_{m,s}}{\vert \lambda_s\vert }\,
\hat \kappa_s\,  d s\, , \quad m\geq 0\, ,
\ee
where $\overline C_{m,s}$ is defined in Eq.~(\ref{eq:Cbarms}).
For the Breit-Wigner, by using Eqs.~(6.562.3) and (8.432.8) of Ref.~\cite{Gradshteyn:1702455},
the Kontorovich–Lebedev transform reads
\bea\label{eq:BW_KL_transform}
\hat\kappa_s & = & \displaystyle
    \frac{\sqrt{s \sinh(\pi s)\, t_0}}{8\pi^2}\,
    \mathrm{Im}\Biggr\{2\left|\Gamma\left(-\frac{1}{4}+ i\, \frac{s}{2}\right)\right|^2
    \!\!_1F_2\left(1;\frac{5}{4}-i\, \frac{s}{2}, \frac{5}{4}+i\, \frac{s}{2};\frac{(\omega_\ast +i\sigma)^2t_0^2}{4}\right)
    +
    \nonumber\\[0.25cm]
&+& \displaystyle t_0\, (\omega_\ast+i\sigma) \left |\Gamma\left(-\frac{3}{4} + i\, \frac{s}{2}\right)\right|^2
\!\!_1F_2\left(1;\frac{7}{4}-i\, \frac{s}{2}, \frac{7}{4}+i\, \frac{s}{2};\frac{(\omega_\ast +i\sigma)^2t_0^2}{4}\right)
+
\\[0.25cm]
&+& 8 \sqrt{2}\left|\Gamma\left(\frac{1}{2}+is\right)\right|^2
\frac{K_{is}(-(\omega_\ast+i\sigma)t_0)}{\sqrt{-(\omega_\ast+i\sigma)\, t_0}} \Biggr\}\,,\nonumber
\eea
where $_1F_2$ is a generalized hypergeometric series~\cite{Gradshteyn:1702455}.

\section{Smeared spectral densities from incomplete ones}\label{sec:incomplete}
Correlation functions are often known for $0 \leq t \leq t_{\rm max}$
only\footnote{Throughout this paper we assume
the interval  $0 \leq t \leq t_{\rm max}$ to be chosen so that the effects
due to the boundary conditions in the time-direction are negligible.
Otherwise, the additional systematics can be easily quantified
by applying standard arguments.}. It is therefore necessary
to study the effect of this limitation on the extraction of
(smeared) spectral densities. To this aim, from this section to the end of
the paper, we focus on the Kontorovich–Lebedev and Mehler-Fock transforms
since they are the continuum limit counterparts of those used
in Section~\ref{sec:discr} for the ILT of correlation functions sampled
on a discrete lattice.
In particular, we introduce the incomplete Mehler-Fock transform and the
incomplete (smeared) spectral densities, and scrutinize their behaviour.
We then bind the difference between the target smeared spectral
density and its incomplete counterpart so as to have a rigorous
estimate of the systematic error associated to the ILT.
An analogous (simpler) discussion can be applied to the Mellin transform as well.
\begin{figure}
    \centering
    \includegraphics[width=0.90\linewidth]{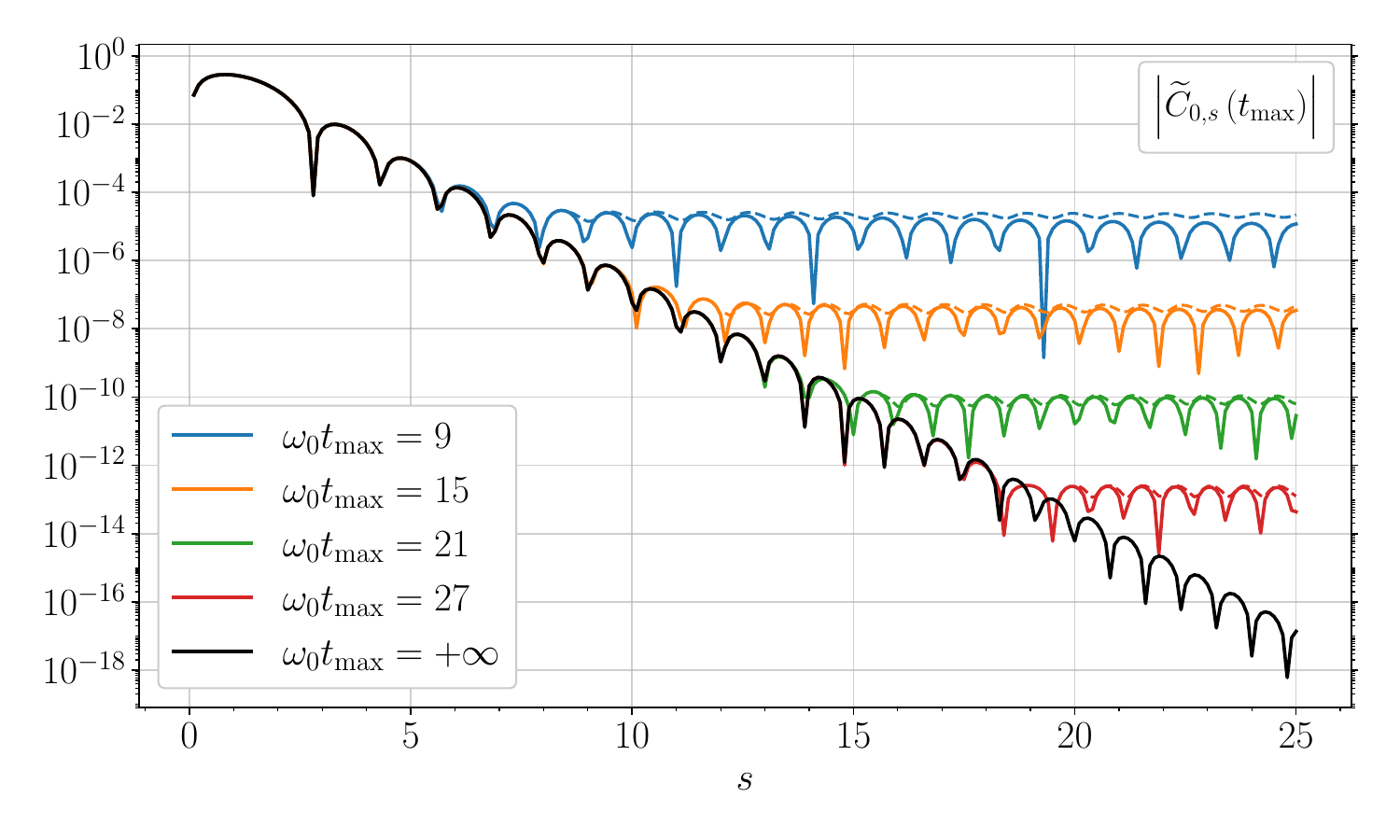}
    \caption{Absolute value of the complete (black) and incomplete Mehler-Fock transforms for $m=0$ of the function $e^{-\omega_0
        t}$ with $\omega_0=0.9$ and $t_0=1$ for various values of $\omega_0 t_{\rm
        max}$. The dashed lines represent the bounds obtained in Eq.~(\ref{eq:systCtilde}) which, when
        $\vert \overline C_{m,s}\vert\ll\vert \widetilde C_{m,s}(t_{\rm max})\vert$, de-facto binds
        $\vert \widetilde C_{m,s}(t_{\rm max})\vert$ too. The norm of the
        (incomplete) Mehler-Fock transform vanishes in the region of downward oscillations represented in
        the plot. However, for clarity, the $s$-grid has been chosen to avoid representing these zeros.
 \label{fig:MT_theo}}
\end{figure}

\subsection{Incomplete Mehler-Fock transform\label{sec:inc_MF}}
When a correlation function is sampled on a finite temporal
domain $t_0 \leq t \leq t_{\rm max}$, we have access to the
incomplete Mehler-Fock  transform
\be\label{eq:Cbarmss}
\widetilde C_{m,s}(t_{\rm max}) = \int_{t_0}^{t_{\rm max}} C(t)\, \bar u_{m,s}(t,t_0)\, dt\, ,
\quad m\geq 0\, ,
\ee
which differs from the complete one by
\be
\overline C_{m,s} - \widetilde C_{m,s}(t_{\rm max}) = \int_{t_{\rm max}}^{\infty}
C(t)\, \bar u_{m,s}(t,t_0)\, dt\, . 
\ee
Thanks to Eq.~(\ref{eq:ct}) we can assume that, for large enough values of
$t_{\rm max}$, it holds~\cite{Bruno:2024fqc}
\be\label{eq:Cbound}
\vert C(t) \vert \leq \vert C(t_{\rm max})\vert\, e^{-\omega_0
  (t-t_{\rm max})}\, , \quad t\geq  t_{\rm max}\, .  
\ee
This implies that 
\be\label{eq:systCtilde}
\Big\vert \overline C_{m,s} - \widetilde C_{m,s}(t_{\rm max}) \Big\vert
\leq \Big\vert C(t_{\rm max})\Big\vert\, b_{m,s}(t_{\rm max}) \, ,
\ee
where\footnote{The dependence of $b_{m,s}(t_{\rm max})$ on $\omega_0$ and $t_0$ is omitted
for the clarity of the notation.} 
\be
b_{m,s}(t_{\rm max}) =
\int_{t_{\rm max}}^{\infty} e^{-\omega_0 (t-t_{\rm max})}\, \vert \bar u_{m,s}(t,t_0) \vert \, dt\, .
\ee
For asymptotically large values  of $t$, it holds
\be
\bar u_{m,s}(t,t_0) = 2 \frac{\vert \lambda_s \vert }{\vert \lambda_{m,s} \vert}\, t^m\,\Big\{
{\rm Re}\Big[e^{i \phi_{m,s}}\,u_s(t)\Big]\, + O\Big(\frac{t_0}{t}\Big)\Big\}\,,
\quad
e^{i \phi_{m,s}} = \frac{\Gamma(is)}{\vert \Gamma(is) \vert}\,
\frac{\vert \lambda_{m,s} \vert}{\lambda_{m,s}} \left(\frac{t_0}{2}\right)^{-is}\, ,
\ee
leading to
\be\label{eq:bndbm}
b_{m,s}(t_{\rm max}) \leq \sqrt{\frac{2}{\pi}}\, \frac{\vert \lambda_s \vert}{\vert \lambda_{m,s} \vert}\,
\frac{t^{m}_{\rm max}}{\omega_0 \sqrt{t_{\rm max}}}\, + \dots\, , 
\ee
which in turn implies that the r.h.s of Eq.~(\ref{eq:systCtilde})
is exponentially suppressed in $t_{\rm max}$. As a
representative example, in Figure~\ref{fig:MT_theo} we show 
the absolute value of the complete (black) and the incomplete Mehler-Fock transforms
for $m=0$ of a single exponential $C(t)=e^{-\omega_0 t}$
with $\omega_0=0.9$ in units of $t_0=1$ for various values of $\omega_0 t_{\rm max}$  together with
the bound in Eq.~(\ref{eq:systCtilde}). Analogous results are obtained
for higher values of $m$ where, as expected from Eq.~(\ref{eq:bndbm}), a weak dependence
on $s$ is observed for $b_{m,s}(t_{\rm max})$. On the one
hand the norm
of the Mehler-Fock transform of a correlation function of the
type considered in this paper, see Eq.~(\ref{eq:ct}), is expected to decrease
exponentially with $s$ up to harmful oscillations. On the other hand, the r.h.s of
Eq.~(\ref{eq:systCtilde}) is exponentially suppressed in $t_{\rm max}$
but very weakly dependent on $s$. Therefore, $\widetilde C_{m,s}$ is an excellent
approximation of $\overline C_{m,s}$ when $t_{\rm max}$ is large and $s$ is small,
see below. But the larger $s$ gets, the worse the approximation becomes, until the delicate
cancellation between the numerator and the denominator in
the r.h.s. of Eq.~(\ref{eq:rhosub3}) is not at work anymore if
$\overline C_{m,s}$ is approximated by $\widetilde C_{m,s}$ in that formula.

\subsection{Incomplete spectral density}
The previous considerations suggest to define an incomplete spectral density as 
\be\label{eq:rhosubf_smax}
\rho_m(\omega, s_{\rm max}) = 
\int_{0}^{s_{\rm max}} \frac{\overline C_{m,s}}{\vert\lambda_s\vert}\,
\hat u_s(\omega,t_0)\, d s\,, \quad m \geq 0\, ,
\ee
which, for $\omega=\omega_*$, can also be interpreted as a smeared spectral density
with smearing function
\be
\kappa(\omega)= \int_0^{s_{\rm max}} \hat u_s(\omega_*,t_0)\,\hat u_{s}(\omega,t_0) \, ds\,.
\ee
Thanks to the upper cut-off $s_{\rm max}$, the
replacement of ${\overline C}_{m,s}$ with ${\widetilde C}_{m,s}$ is well justified,
and therefore we introduce
\be\label{eq:hatrhosubf_smax}
\tilde \rho_m(\omega, s_{\rm max}) = \int_{0}^{s_{\rm max}}
\frac{\widetilde C_{m,s}(t_{\rm max})}{\vert\lambda_s\vert}\,
\hat u_s(\omega,t_0)\, d s\,, \quad m \geq 0\, .
\ee
Its difference from $\rho_m(\omega, s_{\rm max})$ satisfies
\be
\big\vert \rho_m(\omega, s_{\rm max})  - \tilde \rho_m(\omega, s_{\rm max})
\big\vert \leq \Delta  \tilde \rho_m(\omega, s_{\rm max})\,  ,
\ee
where
\be
\Delta  \tilde \rho_m(\omega, s_{\rm max}) =
\Big\vert C(t_{\rm max})\Big\vert\, \int_{0}^{s_{\rm max}}
\frac{b_{m,s}(t_{\rm max})\, \vert \hat u_s(\omega,t_0) \vert}{\vert\lambda_s\vert}\,
 ds
\ee
with the integral on the r.h.s that can be computed
numerically. At asymptotically large values of $s_{\rm max}$,
that integral scales approximatively proportionally to
$e^{\pi s_{\rm max}/2}/s^{m}_{\rm   max}$. Therefore, to meet a given absolute
precision on the incomplete spectral density, $s_{\rm max}$ {\it has to scale
approximately proportionally to} $\omega_0 t_{\rm max}$.
\begin{figure}[t!]
    \centering
    \includegraphics[width=0.48\linewidth]{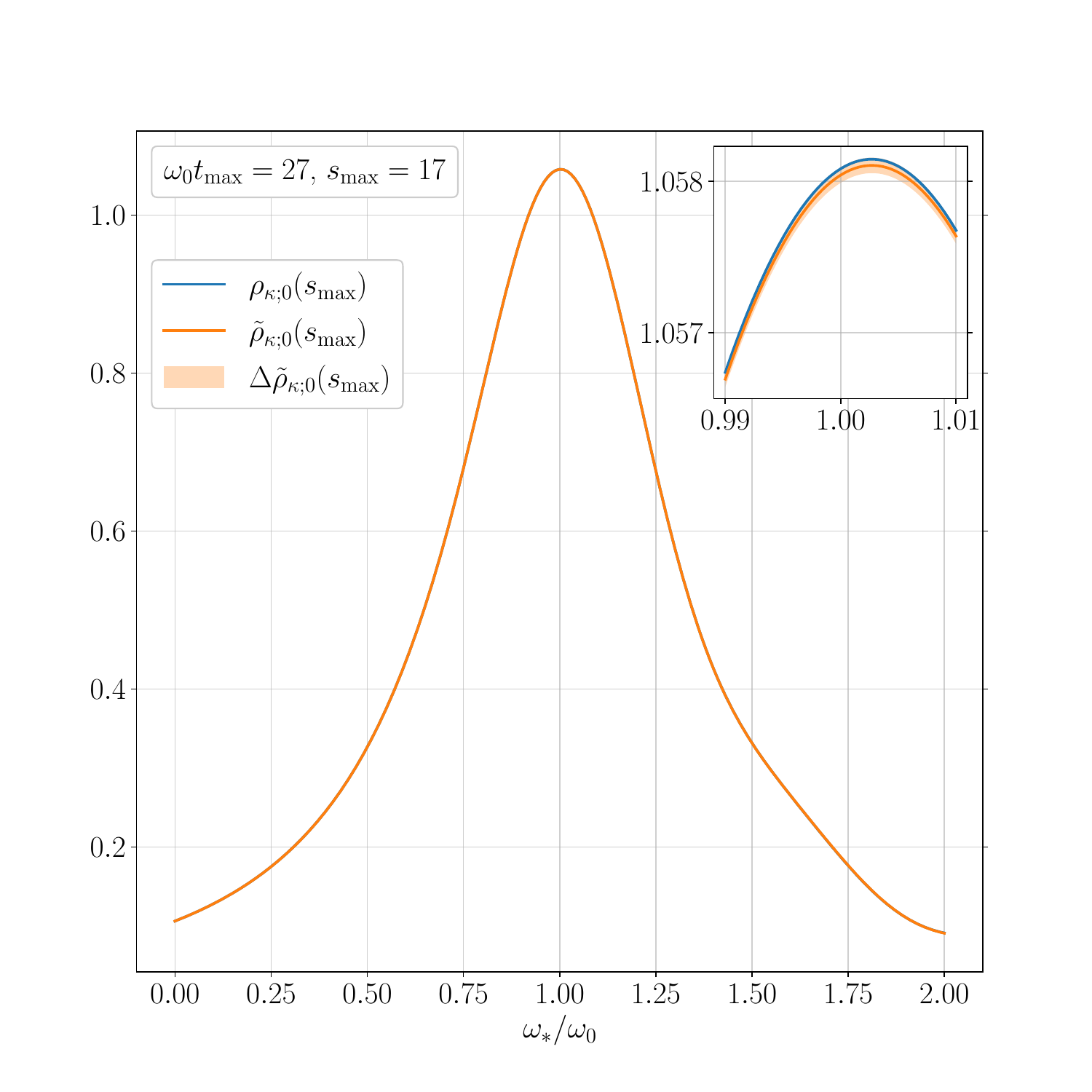}%
     \includegraphics[width=0.48\linewidth]{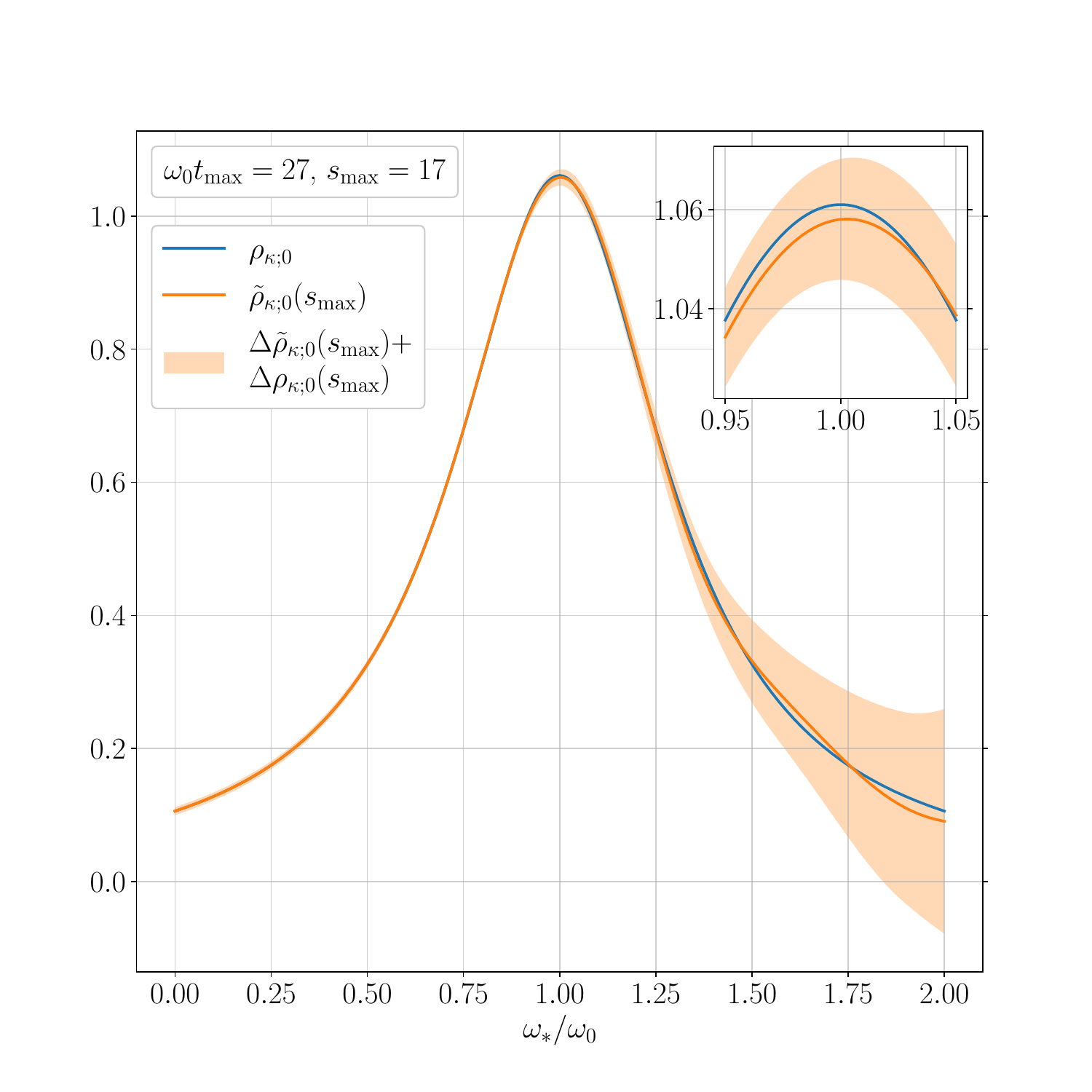}
    \caption{Left: The approximated incomplete smeared spectral density $\tilde \rho_{\kappa;0}(s_{\rm max})$
with the error estimated as in Eq.~(\ref{eq:deltahatrhomk}), against
the incomplete smeared spectral density $\rho_{\kappa;0}(s_{\rm max}) $ for
the single exponential $e^{-\omega_0 t}$ with $\omega_0=0.9$, $t_0=1$.
Right: The approximated incomplete smeared spectral
density $\tilde \rho_{\kappa;0}(s_{\rm max})$
with the total error estimated as in Eq.~(\ref{eq:syst}), against
the exact smeared spectral density $\rho_{\kappa;0}$ for the same exponential.}
    \label{fig:rho_theo}
\end{figure}

\subsection{Incomplete smeared spectral densities}
Analogously to the previous subsection, we can define the 
incomplete smeared spectral densities as 
\be\label{eq:rhosubf_smea_inc}
\rho_{\kappa;m}(s_{\rm max})  = \int_{0}^{s_{\rm max}} \frac{\overline C_{m,s}}{\vert \lambda_s\vert }\,
\hat \kappa_s\,  d s\, , \quad m\geq 0\, .
\ee
Also in this case, if we define
\be\label{eq:rhosubf_smea_inchat}
\tilde \rho_{\kappa;m}(s_{\rm max})  = \int_{0}^{s_{\rm max}}
\frac{\widetilde C_{m,s}(t_{\rm max})}{\vert \lambda_s\vert }\,
\hat \kappa_s\,  d s\, , \quad m\geq 0\, ,
\ee
its difference from $\rho_{\kappa;m}(s_{\rm max}) $ satisfies
\be
\big\vert \rho_{\kappa;m} (s_{\rm max})  - \tilde \rho_{\kappa;m} (s_{\rm max})
\big\vert \leq \Delta  \tilde \rho_{\kappa;m} (s_{\rm max})\,  ,
\ee
where
\be\label{eq:deltahatrhomk}
\Delta  \tilde \rho_{\kappa;m} (s_{\rm max}) =
\Big\vert C(t_{\rm max})\Big\vert\, \int_{0}^{s_{\rm max}}
b_{m,s}(t_{\rm max})\,
\frac{\vert \hat \kappa_s \vert}{\vert\lambda_s\vert}\, ds \, ,
\ee
and again the integral on the r.h.s can be computed
numerically. As a representative example, in the left panel of
Figure~\ref{fig:rho_theo} we show $\tilde \rho_{\kappa;0}(s_{\rm max})$
with its error, estimated as in Eq.~(\ref{eq:deltahatrhomk}), 
against $\rho_{\kappa;0}(s_{\rm max}) $ both extracted from the
exponential considered in Section~\ref{sec:inc_MF}. The
smearing function is a Breit-Wigner with $\sigma=\omega_0/3$.

\subsection{Smeared spectral densities\label{sec:systsmea}}
An incomplete smeared spectral density becomes a satisfactory approximation of the
complete one if $s_{\rm max}\sim \omega_0 t_{\rm max}$  is chosen large
enough so as to guarantee a reconstruction error  below the
target precision. To quantify this systematics, 
we decompose a smeared density as  
\be
\rho_{\kappa;m} =  \tilde \rho_{\kappa;m} (s_{\rm max}) +
\Big[\rho_{\kappa;m} (s_{\rm max})  - \tilde \rho_{\kappa;m} (s_{\rm max})\Big] +
\Big[\rho_{\kappa;m} - \rho_{\kappa;m}(s_{\rm max})\Big] \, .
\ee
The norm of the second term on the r.h.s of this equation is bounded
by $\Delta  \tilde \rho_{\kappa;m} (s_{\rm max}) $ in
Eq.~(\ref{eq:deltahatrhomk}). We are therefore left to bind
the difference between the target smeared spectral density and
its incomplete counterpart. To do so, we notice that
from Eqs.~(\ref{eq:CmsMFKL}), (\ref{eq:rhatms}) and
(\ref{eq:uhat_s}) it follows 
\bea
\Bigg\vert \frac{\overline C_{m,s}}{\lambda_s} \Bigg\vert & \leq & 
\frac{\sqrt{2 s \sinh(\pi s)} }{\pi}
\int_{0}^{\infty} \Big\vert \frac{\rho_m(\omega)}{\sqrt{\omega}}\Big\vert\,
\Big\vert K_{is}(\omega t_0) \Big\vert \, d\omega\, \nonumber\\[0.25cm]
& \leq &
\frac{\sqrt{2 s \sinh(\pi s)} }{\pi} \max_{\;\;\omega\,\in\, {\mathbb R^+}}
\Big\vert K_{is}(\omega t_0) \Big\vert\, r_m
\, , 
\eea
where
\be
r_m = \int_{0}^{\infty} \Big\vert \frac{\rho_m(\omega)}{\sqrt{\omega}}\,\Big\vert \, d\omega\, .
\ee
The modified Bessel function $K_{is}(x)$ oscillates for $x<s$ while
it decays exponentially for $x>s$ due to the change in nature of the
differential equation it satisfies. As a consequence, its absolute
maximum is the last peak of the oscillatory phase which occurs
for $x$ just below $s$. By assuming that
$\rho_m(\omega)$ and therefore $C_m(t)$ have a definite sign,
and by using Eq.~(\ref{eq:fundamental}) for $s=0$, $r_m$ can be re-written as 
\be\label{eq:rm}
r_m = \frac{1}{\Gamma(m+\frac{1}{2})}\, \int_{0}^{\infty} \vert C(t)\vert \, t^{m-1/2}\, dt\, .
\ee
The integral in Eq.~(\ref{eq:rm}) can be decomposed in the short-, intermediate- and
long-distance contributions $r_m^{\rm s}$, $r_m^{\rm i}$ and $r_m^{\rm l}$ corresponding
to integrating in the ranges $[0,t_0]$, $[t_0,t_{\rm max}]$, and $[t_{\rm max},\infty]$
respectively. The short- and the intermediate-distance contributions can be
bound or determined from the knowledge\footnote{For small enough values of $t_0$,
perturbation theory is an interesting
viable option to bind $C(t)$ in $[0,t_0]$, and therefore $r_m^{\rm s}$.}
of the correlator up to $t_{\rm max}$, respectively.
By using Eq.~(\ref{eq:Cbound}), the long distance contribution satisfies
\be
r_m^{\rm l} \leq r_m^{\rm l, \max} = \frac{e^{\omega_0 t_{\rm max}}}{\omega_0^{m+1/2}}\,
\frac{\Gamma(m+\frac{1}{2},\omega_0 t_{\rm max})}{\Gamma(m+\frac{1}{2})}\, \vert C(t_{\rm max})\vert \,,
\ee
with $\Gamma(a,x)$ being the upper incomplete
$\Gamma$-function which can easily be computed
numerically~\cite{Abramowitz:lg}. For asymptotically large
values  of $x$, it holds
\be
\Gamma\big(a,x\big) =
x^{a-1} e^{-x}\, + \dots
\ee
leading to
\be\label{eq:bound2}
r_m^{\rm l, \max} = \frac{1}{\Gamma(m+\frac{1}{2})}\, \frac{t_{\rm max}^{m-1/2}}{\omega_0}\, 
\vert C(t_{\rm max})\vert \, + \dots
\ee
which is exponentially suppressed in $t_{\rm max}$. The difference between
the smeared spectral density and the incomplete one can then be bounded as 
\be
\Big\vert \rho_{\kappa;m} - \rho_{\kappa;m}(s_{\rm max}) \Big\vert \leq
\Delta \rho_{\kappa;m} (s_{\rm max}) \, ,
\ee
with
\be\label{eq:deltarhomk}
\Delta  \rho_{\kappa;m} (s_{\rm max}) = r^{\rm \max}_m
\int_{s_{\rm max}}^{\infty}
\frac{\sqrt{2 s \sinh(\pi s)} }{\pi} \max_{\;\;\omega\,\in\, {\mathbb R^+}}
\Big\vert K_{is}(\omega t_0) \Big\vert\,
\Big\vert \hat \kappa_s \Big\vert\, d s\, , 
\ee
where $r^{\rm \max}_m$ is the sum of the bounds on $r_m^{\rm s}$, $r_m^{\rm i}$ and $r_m^{\rm l}$.
By using Eqs.~(\ref{eq:deltahatrhomk}) and
(\ref{eq:deltarhomk}), it follows that 
\be\label{eq:syst}
\Big\vert\rho_{\kappa;m} -  \tilde \rho_{\kappa;m} (s_{\rm
  max})\Big\vert \leq
\Delta  \tilde \rho_{\kappa;m} (s_{\rm max}) + 
\Delta  \rho_{\kappa;m} (s_{\rm max})\, . 
\ee
Therefore, the approximated incomplete smeared spectral density $\tilde
\rho_{\kappa;m} (s_{\rm max})$, when supplemented by the systematic
error $[\Delta  \tilde \rho_{\kappa;m} (s_{\rm max}) + \Delta  \rho_{\kappa;m} (s_{\rm max})]$,
becomes a rigorous estimate of the smeared spectral density.
As a representative example, in the right panel of
Figure~\ref{fig:rho_theo} we show $\tilde \rho_{\kappa;0}(s_{\rm max})$
together with the total error, estimated as in Eq.~(\ref{eq:syst}), 
against $\rho_{\kappa;0}$ both extracted from the
exponential considered in subsection~\ref{sec:inc_MF}. The smearing
function is the same Breit-Wigner used in the left panel of the same figure.
If instead a Breit-Wigner with a varying width but with $\sigma/\omega_*$
kept constant as a function of $\omega_*$ was chosen, see Section~\ref{sec:discuss},
the error bound would have been more uniform as a function of $\omega_*$.

The $s$-dependence of the integrand in Eq.~(\ref{eq:deltarhomk}) is de-facto
the one of the function $\vert \hat \kappa_s \vert$, since the remaining
factor depends very weakly on $s$. Therefore, the dependence on $s$ of
the norm of the Kontorovich-Lebedev transform of the smearing function
determines the rate of convergence to zero of $\Delta  \rho_{\kappa;m} (s_{\rm max})$ 
as a function of $s_{\rm max}$.
As discussed in Section~\ref{sec:inc_MF},
when a correlation function is only known on a finite temporal
domain $t \leq t_{\rm max}$, we have access to $\overline C_{m,s}$
for $s\leq s_{\rm max} \sim \omega_0 t_{\rm max}$ only, a lack of knowledge which
prevents us to reconstruct exactly the (regulated) spectral density. If, however,
we restrict ourselves to smeared spectral densities with $\vert \hat \kappa_s \vert$
decaying fast enough with $s$, the effect of the unknowns can be
bounded and kept under control. It is therefore wise to choose smearing functions
with a fast decaying integral transform if the problem under investigation allows
to do so. In this respect, the Breit-Wigner
is a very conservative example since its transform decays
approximatively as\footnote{A similar bound can be derived for the Mellin transform. In that
case, the transform of the smearing function decays exponentially with $s$,
see Eq.~(\ref{eq:rhosubf_smea_bw}), and the
bound on the error is tighter. However,
in order to extract the spectral density, the Mellin transform requires the knowledge of $C(t)$ in
the interval $[0,t_0]$  and not only to bind its contribution to
the associated systematic error.}
$1/s^2$, but it has the advantage of having a simple expression which can be
managed easily in the numerical examples considered in this paper.

\section{Spectral densities from discrete sampling}\label{sec:discr}
When QCD is regularized on a lattice, the correlators are defined at discrete
Euclidean times only,
\be\label{eq:ctd}
    {\cal C}(t_0+an) = \int_{0}^{\infty} \varrho(\omega) \, e^{-\omega(t_0+an)}\, d \omega \,, \quad
    n=0,1,2,\dots, \infty\, ,
\ee
where $a$ is the lattice spacing. Similarly to the continuum, the Eq.~(\ref{eq:ctd})
must be inverted to extract the spectral density. We proceed by mimicking the
derivation in Section~\ref{sec:KLMF} with $t_0$ as a freely selectable
parameter to leave open the possibility of avoiding potentially large discretization
effects at short time-distances.

\subsection{Spectral density}
When a correlator is sampled on regularly-spaced discrete points,
the Eqs.~(\ref{eq:MFKL}) are replaced by
\be\label{eq:transformsD}
{\overline {\cal C}}_s  = a \sum_{n=0}^\infty {\cal C}(t_0+an)\, \overline v_s(n,t_0,a)\, , 
\qquad
\hat \varrho_s = \int_{0}^{\infty}
\varrho(\omega) \, \hat v_s(\omega,t_0,a)\, d\omega\, ,
\quad s \in {\mathbb R^+}\, ,\\
\ee
where $t_0 = a n_0$, and the basis functions
are related by the discretized version
of Eq.~(\ref{eq:bellam1}), namely 
\be\label{eq:hatv2barv}
\vert\lambda_s\vert\, \hat v_s(\omega,t_0,a) = 
 a \sum_{n=0}^{\infty} e^{- \omega (t_0+an)}\, \bar v_s(n,t_0,a)\, . 
\ee
From this equation it immediately follows
\be
{\overline {\cal C}}_s = \vert\lambda_s\vert\, \hat \varrho_s\, ,
\ee
which is exact at finite lattice spacing. Similarly to the continuum, if we
require the orthogonality condition
\be\label{eq:ortvhatg}
\int_{0}^{\infty} \hat v_s(\omega,t_0,a)\, \hat v_{s'}(\omega,t_0,a)\, d\omega\, =
\delta(s-s')
\ee
to hold, then Eq.~(\ref{eq:hatv2barv}) implies 
\be\label{eq:discr-sonc-tg}
a^2 \sum_{n,n'=0}^\infty \overline v_s(n,t_0,a) \A(n + n') \, \overline v_{s'}(n',t_0,a) =
\vert\lambda_s\vert^2\, \delta(s-s')\,,
\ee
a straightforward discretization of Eq.~(\ref{eq:uAug}) with $\A(n+n')$ being
the kernel of the discretized quasi-Carleman operator defined in
Eq.~(\ref{eq:AaC}) of Appendix~\ref{app:hilbert}. We then choose 
the $\overline v_s(n,t_0,a)$ to be the eigenvectors of $\A$
\be\label{eq:discr-sonc-tgg}
a \sum_{n'=0}^\infty \A(n + n') \, \overline v_s(n',t_0,a) =
\vert\lambda_s\vert^2\, \overline v_s(n,t_0,a) \,,
\ee
which form a complete and orthonormal basis for the discretized
functions in $\ell^2(\mathbb Z^+)$,
see Appendix~\ref{app:hilbert} for their definition and properties.
Notice that, as expected, the eigenvalues $\vert\lambda_s\vert^2$
of $\A$ are the same as those of the continuum
operator~\cite{Yafaev:2017dry}.
The spectral density can then be extracted as
\be\label{eq:rhodisc}
\varrho(\omega) = \int_0^{\infty} \frac{{\overline {\cal C}}_s}{\vert \lambda_s \vert} \,
\hat v_s(\omega,t_0,a)\, ds\, .
\ee

\noindent {\it Continuum limit at fixed $t_0$}\\[0.125cm]
To study the approach to the continuum limit of $\varrho(\omega)$ at fixed $t_0$, we start by
analyzing the leading discretization effects in ${\overline {\cal C}}_s$
defined in Eq.~(\ref{eq:transformsD}). By assuming that discretization effects in
${\cal C}(t_0+an)$ start at $O(a^2)$, by using Eq.~(\ref{eq:fiuuu1}) in Appendix~\ref{app:hilbert},
and by noticing that the trapezoidal rule is adopted in the first of Eqs.~(\ref{eq:transformsD})
up to a factor $1/2$ in the first contribution to the sum, we obtain
\be
{\overline {\cal C}}_s = \overline C_s + \frac{a}{2}\, C(t_0)\, \bar u_s(t_0,t_0) +
\frac{a}{4 t_0}  \int_{t_0}^{\infty} C(t)\, \Big[1+ 2 t \frac{\partial}{\partial t} \Big] \, \bar u_s(t,t_0)\, dt
+ O(a^2)\, ,
\ee
where $\overline C_s$ is defined in Eq.~(\ref{eq:MFKL}). By integrating by parts
the last integral, it follows that
\be
{\overline {\cal C}}_s = \Big[1- \frac{a}{4 t_0} \Big]\, \overline C_s
- \frac{a}{2 t_0} \int_{t_0}^{\infty} t\, \Big[\frac{\partial}{\partial t} C(t)\Big]\, \bar u_s(t,t_0)\, dt 
+ O(a^2)\, , 
\ee
and by using Eq.~(\ref{eq:ct}) and integrating again by parts we finally arrive at
\be\label{eq:blsma}
{\overline {\cal C}}_s = \vert \lambda_s \vert \Big\{\Big[1+ \frac{a}{4 t_0} \Big]\, \hat \rho_s +
\frac{a}{2 t_0}\, \int_{0}^{\infty} \Big[\omega' \frac{\partial}{\partial\omega'}\rho(\omega')\Big]\,
\hat u_s(\omega',t_0)\,  d\omega' + O(a^2)\Big\}\, .
\ee
Notice that, since both $\overline C_s$ and ${\overline {\cal C}}_s$ are suppressed exponentially
in $s$, discretization effects are suppressed exponentially in $s$ as well. By inserting
Eqs.~(\ref{eq:blsma}) and (\ref{eq:fiuuu2}) into
Eq.~(\ref{eq:rhodisc}) we obtain
\bea
\varrho(\omega) & = & \Big[1 - \frac{a}{2 t_0} \omega \frac{\partial}{\partial\omega}\Big]\, \rho(\omega)\, +
\nonumber\\[0.25cm]
& & + \frac{a}{2 t_0} \int_{0}^{\infty} \Big\{
\int_{0}^{\infty}
\Big[\omega' \frac{\partial}{\partial \omega'}\, \rho(\omega')  \Big]\, \hat u_s(\omega',t_0)\,  d\omega'
\Big\}\, \hat u_s(\omega,t_0)\, ds  + O(a^2)\, .
\eea
By assuming that the Kontorovich–Lebedev transform and its inverse exist not only for
$\rho(\omega)$ but also for $\omega \frac{\partial}{\partial \omega}\, \rho(\omega)$,
we finally arrive to
\be
\varrho(\omega) = \rho(\omega) +  O(a^2)\, ,
\ee
which explicitly shows that the inversion procedure does not introduce $O(a)$-effects. 
If ${\cal C}(t_0+an)$ had no discretization effects and if $\rho(\omega)$
admitted the transform and its inverse on the basis of the $\hat v_s(\omega,t_0,a)$,
then the extracted spectral density would not be affected by discretization effects.

\subsection{Regulated spectral density}
The Eqs.~(\ref{eq:Cmint}) and (\ref{eq:Cbarms}) can be discretized as 
\be\label{eq:trapcm}
{\cal C}_m(t_0+an) = \frac{a}{\Gamma(m)}
\sum_{n'=n}^\infty w_{n'-n}\, {\cal C}(t_0+an') (an'-an)^{m-1}\, , \quad m>0\, ,  
\ee
where $w_0=\frac{1}{2}$ and $w_i=1$ for $i>0$ to implement the trapezoidal rule\footnote{
By changing the weights $w_i$, different Newton-Cotes rules could be implemented as well.}, 
and
\be\label{eq:transformCs}
{\overline {\cal C}}_{m,s}  = a \sum_{n=0}^\infty {\cal C}_m(t_0+an)\, \overline v_s(n,t_0,a)
= a \sum_{n=0}^\infty {\cal C}(t_0+an)\, \overline v_{m,s}(n,t_0,a)
\, , \quad m\geq 0\,  , 
\ee
where ${\cal C}_0(t_0+an)={\cal C}(t_0+an)$,
\be
\overline v_{m,s}(n,t_0,a) = \frac{a}{\Gamma(m)} \sum_{n'=0}^n w_{n-n'}\, (an-an')^{m-1}\, \overline v_s(n',t_0,a)\, ,
\ee
and $\overline v_{0,s}(n,t_0,a)=\overline v_{s}(n,t_0,a)$. By defining
\be\label{eq:hatv2barvm}
\vert\lambda_s\vert \hat v_{m,s}(\omega,t_0,a) = 
 a \sum_{n=0}^{\infty} e^{- \omega (t_0+an)}\, \bar v_{m,s}(n,t_0,a)\, , 
\ee
after some algebra it is possible to show that
\be
\hat v_{m,s}(\omega,t_0,a) = \frac{1}{\upomega^m}\, \hat v_s(\omega,t_0,a)\,,\qquad 
\frac{1}{\upomega^m} = \frac{a}{\Gamma(m)} \Big(-\frac{\partial}{\partial \omega}\Big)^{m-1}
\Big[ \frac{1}{1-e^{-a\omega}} - \frac{1}{2}\Big]\, ,
\ee
and therefore we define
\be
\hat \varrho_{m,s} =
\int_{0}^{\infty} \varrho(\omega) \, \hat v_{m,s}(\omega,t_0,a)\, d\omega = 
\int_{0}^{\infty}
\frac{\varrho(\omega)}{\upomega^m} \, \hat v_s(\omega,t_0,a)\, d\omega\, .
\ee
Notice that, as in the continuum, the dependence on $m$ and $s$ in $\hat v_{m,s}$
is factorized. By inserting Eq~(\ref{eq:ctd}) into
Eq.~(\ref{eq:transformCs}), it immediately follows that
\be
{\overline {\cal C}}_{m,s} = \vert\lambda_s\vert\, \hat \varrho_{m,s}\, ,
\ee
and therefore
\be\label{eq:rhomlat}
\varrho_m(\omega) = \frac{\varrho(\omega)}{\upomega^m} =  
\int_0^{\infty} \frac{{\overline {\cal C}}_{m,s}}{\vert \lambda_s \vert} \,
\hat v_s(\omega,t_0,a)\, ds\, .
\ee
Since $\upomega^m/\omega^m=1+O(a^2)$, in the continuum we recover the expected
subtraction power and, by following the same line of argumentation
of the previous section, it follows that
\be
\varrho_m(\omega) = \rho_m(\omega) +  O(a^2)\, .
\ee
An alternative procedure to extract regulated spectral densities
is discussed in Appendix~\ref{app:appsub}, where the starting
point is to consider the correlator $C(t)\, t^m$ and extract
the corresponding spectral density with the formula analogous
to Eq.~(\ref{eq:rhodisc}). The latter is then linked
to the target regulated spectral density, see Eqs.~(\ref{eq:rho_m})
and (\ref{eq:rho_m_smea}). When applied to discrete sampling, discretization
effects are different with respect to the procedure described
in this section.

\subsection{Smeared spectral densities}
For smeared spectral densities, we proceed analogously to
Section~\ref{sec:KLsmea}. We define the transform of the smearing
function as
\be\label{eq:kslat}
\hat {\it k}_s = \int_{0}^{\infty} \kappa(\omega)\,
\hat v_s(\omega,t_0,a)\,  d\omega\,.
\ee
By inserting Eq.~(\ref{eq:rhomlat}) into the analogous of Eq.~(\ref{eq:rhokappa}) for
$\varrho_m$ we obtain\footnote{For a single
exponential, an explicit check of discretization effects induced by the ILT reveals that
for $m=0$ there are none, while for $m>0$ they are very moderate as expected from
the expression of $\upomega^m/\omega^m$. The more involved procedure in Appendix~\ref{app:appsub}
removes them by construction.}
\be\label{eq:rhosubkssd}
\varrho_{\kappa;m} = \int_{0}^{\infty} \frac{{\overline {\cal C}}_{m,s}}{\vert\lambda_s\vert}\,
\hat {\it k}_s\,  d s\, , \qquad m\geq 0\,  ,
\ee
where ${\overline {\cal C}}_{m,s}$ is defined in Eq.~(\ref{eq:transformCs}).

\section{Smeared spectral densities from discrete incomplete ones}\label{sec:discrsme}
To compute a smeared spectral density from a discrete
sampling of a correlator known for $t_0\leq t \leq t_{\rm max}$ only,
we adapt the strategy outlined
in Section~\ref{sec:incomplete} to
the discrete case. In particular, we introduce the incomplete discrete
Mehler-Fock transform and the corresponding incomplete (smeared) spectral density.
We then bind the difference between the target smeared spectral
density and its incomplete counterpart so as to have a rigorous
estimate of the systematic uncertainty associated to the ILT.
\begin{figure}
    \centering
    \includegraphics[width=0.90\linewidth]{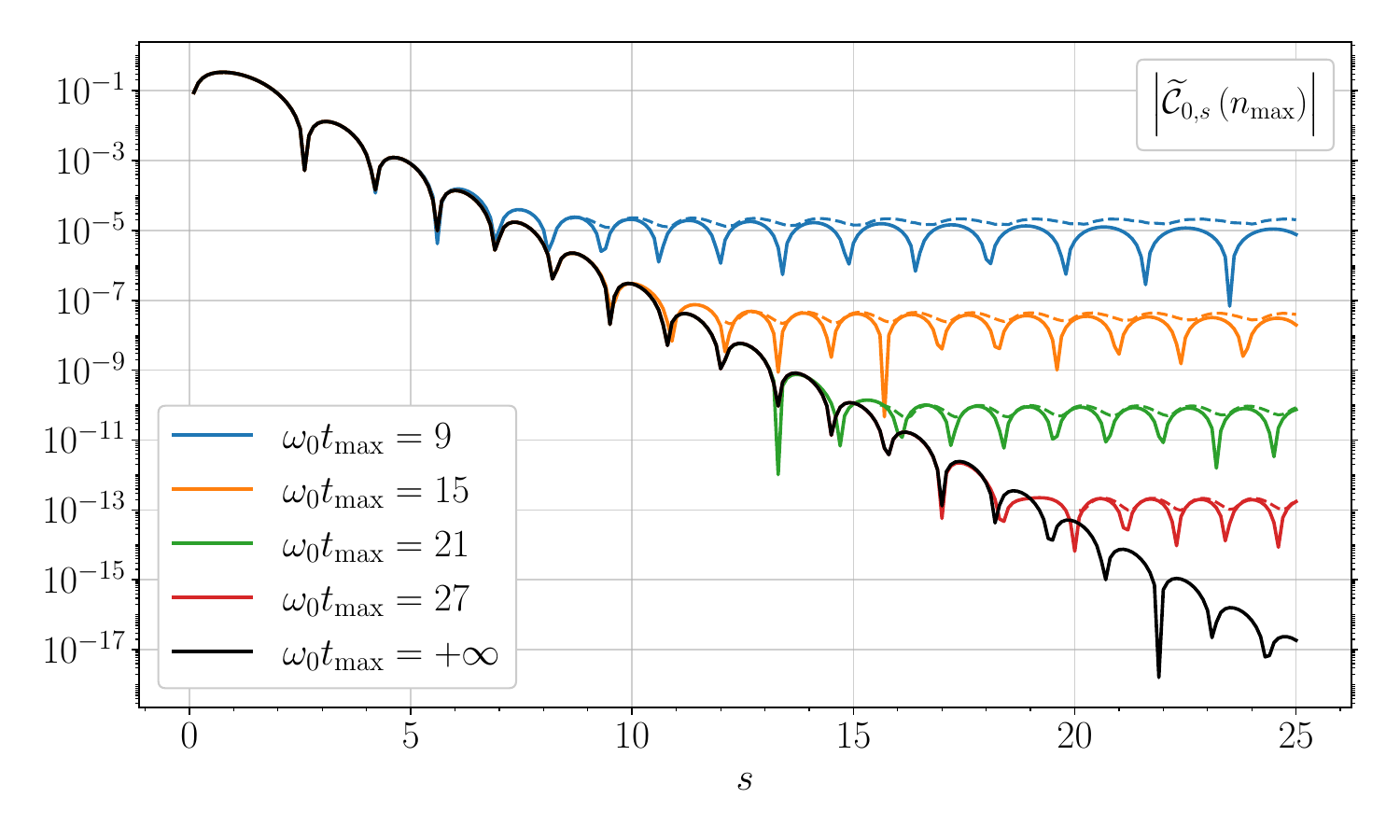}
    \caption{Absolute value of the complete (black) and incomplete discrete Mehler-Fock transforms for $m=0$ of the
             function $e^{-\omega_0(t_0 + an)}$ with $a \omega_0=0.3$, $t_0=3a$, and
             $t_{\rm max} = t_0 + an_{\rm max}$ for various values of
             $\omega_0\, t_{\rm max}$. The dashed lines represent the bounds obtained from
             Eq.~(\ref{eq:systCtilded}). As in the continuum, the norm of the (incomplete) discrete
             Mehler-Fock transform vanishes in the region of downward oscillations represented in
             the plot. However, for clarity, the $s$-grid has been chosen to avoid representing
             these zeros.
 \label{fig:MT_theo_disc}}
\end{figure}

\subsection{Incomplete discrete Mehler-Fock transform\label{sec:inc_MF_disc}}
When a correlation function is sampled on regularly-spaced discrete points  of
a finite temporal domain $t_0 \leq t \leq t_{\rm max}$, we have access to the
incomplete discrete Mehler-Fock transform
\be\label{eq:Cbarmss_disc}
{\widetilde {\cal C}}_{m,s}(n_{\rm max})  =
a \sum_{n=0}^{n_{\rm max}} {\cal C}(t_0+an)\, \overline v_{m,s}(n,t_0,a)
\, , \quad m\geq 0\, ,
\ee
with $an_{\rm max} = t_{\rm max} - t_0$, 
which differs from the complete one by
\be
{\overline {\cal C}}_{m,s} - {\widetilde {\cal C}}_{m,s}(n_{\rm max})  =
a \sum_{n=n_{\rm max}+1}^{\infty} {\cal C}(t_0+an)\, \overline v_{m,s}(n,t_0,a)
\, .
\ee
Thanks to Eq.~(\ref{eq:ctd}) we can assume that, for large enough values of
$n_{\rm max}$, it holds
\be\label{eq:Cboundd}
\vert {\cal C}(t_0+an) \vert \leq
\vert C(t_0+a n_{\rm max})\vert\, e^{-a \omega_0
  (n-n_{\rm max})}\, , \quad n\geq  n_{\rm max}\, .
\ee
This implies that 
\be\label{eq:systCtilded}
\Big\vert
{\overline {\cal C}}_{m,s} - {\widetilde {\cal C}}_{m,s}(n_{\rm max})
\Big\vert
\leq \Big\vert {\cal C}(t_0+a n_{\rm max})\Big\vert\, {\mathfrak b}_{m,s}(n_{\rm max}) \, ,
\ee
where
\be
{\mathfrak b}_{m,s}(n_{\rm max}) =
a \sum_{n=n_{\rm max}+1}^{\infty}
e^{-a \omega_0 (n-n_{\rm max})}\, \vert \overline v_{m,s}(n,t_0,a) \vert \, .
\ee
As a representative example, in Figure~\ref{fig:MT_theo_disc} we show 
the complete (black) and the incomplete discrete Mehler-Fock transforms
for $m=0$ of a single exponential $e^{-\omega_0 (t_0 + a n)}$
with $a \omega_0=0.3$ and $t_0=3a$ for various values of $\omega_0 t_{\rm max}$
together with the bound obtained from Eq.~(\ref{eq:systCtilded}).
Analogous results are obtained for higher values of $m$. Comments
similar to those reported at the end of Section~\ref{sec:inc_MF} apply also
in this case.
In particular the larger $s$ gets, the worse $\widetilde C_{m,s}$ approximates
$\overline C_{m,s}$, until the delicate cancellation between the numerator and
the denominator in the r.h.s. of Eq.~(\ref{eq:rhomlat}) is not at work anymore.

\subsection{Incomplete spectral density from discrete sampling}
The previous considerations suggest to define an incomplete spectral density as 
\be\label{eq:rhosubf_smax_disc}
\varrho_m(\omega,s_{\rm max})  =
\int_0^{s_{\rm max}} \frac{{\overline {\cal C}}_{m,s}}{\vert \lambda_s \vert} \,
\hat v_s(\omega,t_0,a)\, ds\, , \quad m \geq 0\, .
\ee
If we define
\be\label{eq:hatrhosubf_smax_disc}
\tilde \varrho_m(\omega, s_{\rm max}) =  \int_{0}^{s_{\rm max}}
\frac{{\widetilde {\cal C}}_{m,s}(n_{\rm max})}{\vert\lambda_s\vert}\,
\hat v_s(\omega,t_0,a)\, d s\,, \quad m \geq 0\, ,
\ee
its difference from $\varrho_m(\omega,s_{\rm max})$ satisfies
\be
\big\vert \varrho_m(\omega,s_{\rm max})  - \tilde \varrho_m(\omega, s_{\rm max})
\big\vert \leq \Delta  \tilde \varrho_m(\omega, s_{\rm max})\,  ,
\ee
where
\be
\Delta  \tilde \varrho_m(\omega, s_{\rm max}) =
\Big\vert {\cal C}(t_0+a n_{\rm max})\Big\vert\,
\int_{0}^{s_{\rm max}}
\frac{{\mathfrak b}_{m,s}(n_{\rm max})\,
\vert \hat v_s(\omega,t_0,a)\vert}{\vert\lambda_s\vert}\, ds
\ee
with the integral on the r.h.s that can be computed
numerically.
\begin{figure}[t!]
    \centering
    \includegraphics[width=0.48\linewidth]{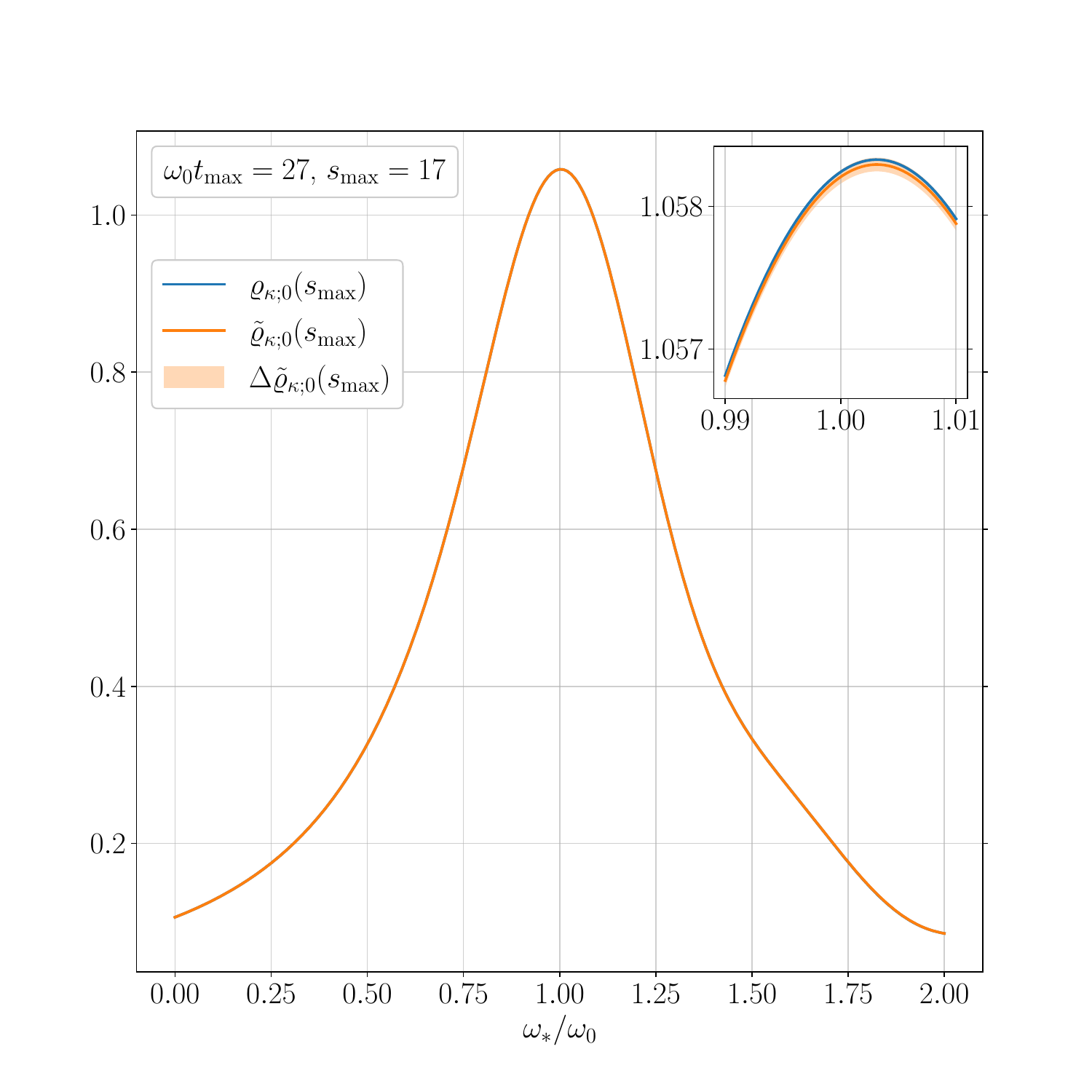}%
     \includegraphics[width=0.48\linewidth]{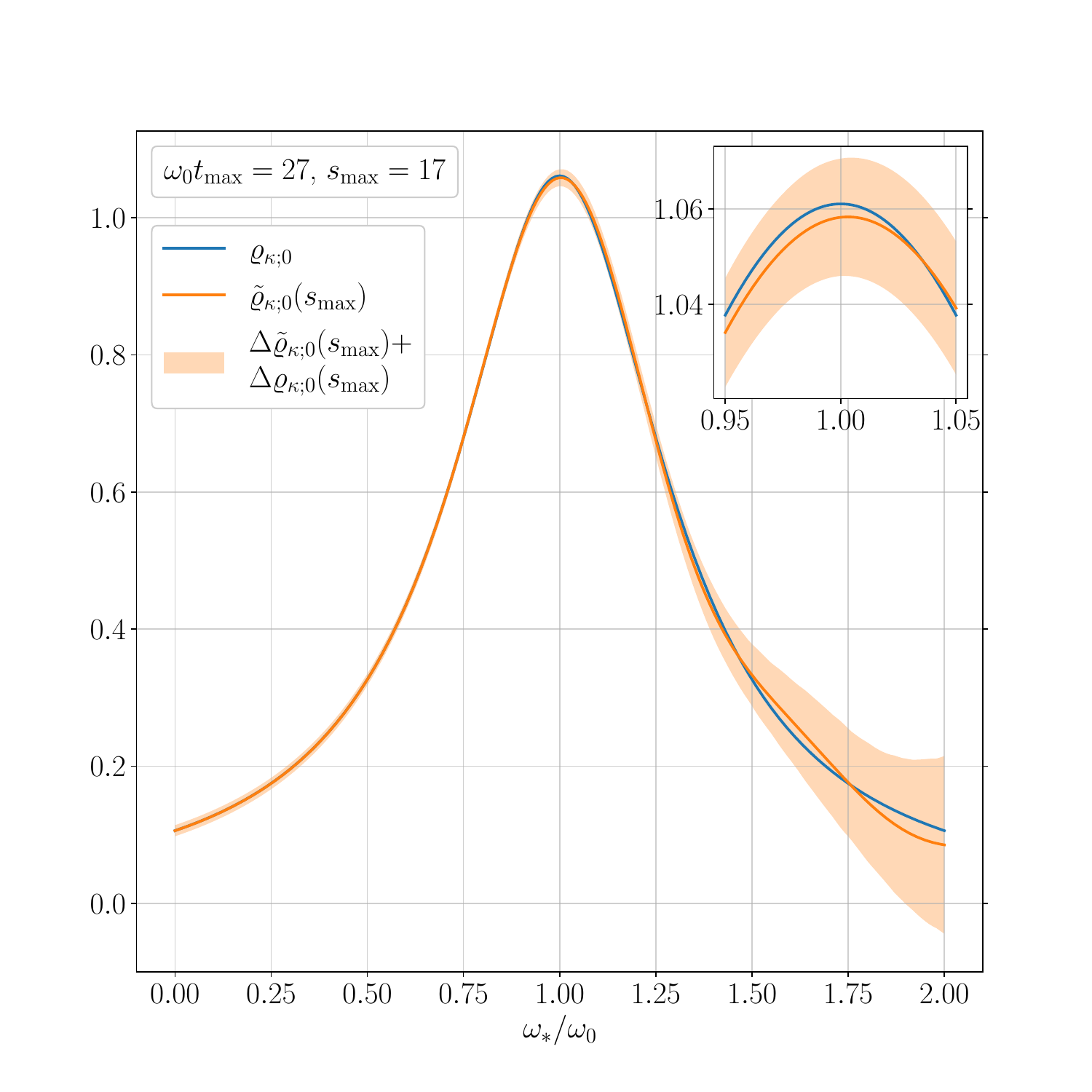}
    \caption{Left: The approximated incomplete smeared spectral density $\tilde
    \varrho_{{\it k};0}(s_{\rm max})$
with the error estimated as in Eq.~(\ref{eq:deltahatrhomk_disc}), against
the incomplete smeared spectral density $\varrho_{\kappa;0}(s_{\rm max}) $ for
the single exponential $e^{-\omega_0 (t_0 + a n)}$ with $a \omega_0=0.3$
and $t_0=3a$. Right: The approximated incomplete smeared spectral
density $\tilde \varrho_{\kappa;0}(s_{\rm max})$
with the total error estimated as in Eq.~(\ref{eq:syst_disc}), against
the exact spectral density $\varrho_{\kappa;0}$ for the same exponential.}
    \label{fig:rho_theo_disc}
\end{figure}

\subsection{Incomplete smeared spectral densities from discrete sampling}
Analogously to the continuum, we can define the 
incomplete smeared spectral densities as 
\be\label{eq:rhosubf_smea_inc_disc}
\varrho_{\kappa;m}(s_{\rm max})
= \int_{0}^{s_{\rm max}} \frac{{\overline {\cal C}}_{m,s}}
{\vert \lambda_s\vert }\,
{\hat {\it k}}_s\,  d s\, , \quad m\geq 0\, ,
\ee
where ${\hat {\it k}}_s$ is defined in Eq.~(\ref{eq:kslat}).
Again, the truncation given by $s_{\rm max}$ allows us to define
\be\label{eq:rhosubf_smea_inchat_disc}
\tilde \varrho_{\kappa;m}(s_{\rm max})  = \int_{0}^{s_{\rm max}}
\frac{\widetilde {\cal C}_{m,s}(n_{\rm max})}{\vert \lambda_s\vert }\,
{\hat {\it k}}_s\,  d s\, , \quad m\geq 0\, .
\ee
Its difference from $\varrho_{\kappa;m}(s_{\rm max})$ satisfies
\be
\big\vert \varrho_{\kappa;m} (s_{\rm max})  - \tilde \varrho_{\kappa,m} (s_{\rm max})
\big\vert \leq \Delta  \tilde \varrho_{\kappa;m} (s_{\rm max})\,  ,
\ee
where
\be\label{eq:deltahatrhomk_disc}
\Delta  \tilde \varrho_{\kappa;m} (s_{\rm max}) =
\Big\vert {\cal C}(t_0+a n_{\rm max})\Big\vert\,
\int_{0}^{s_{\rm max}}
{\mathfrak b}_{m,s}(n_{\rm max})\,
\frac{\vert {\hat {\it k}}_{s} \vert}{\vert\lambda_s\vert}\, ds \, , 
\ee
and again the integral on the r.h.s can be computed
numerically. As a representative example, in the left panel of
Figure~\ref{fig:rho_theo_disc} we show $\tilde \varrho_{\kappa;m}(s_{\rm max})$
with its error, estimated as in Eq.~(\ref{eq:deltahatrhomk_disc}), 
against $\varrho_{\kappa;m}(s_{\rm max}) $ both extracted from the
exponential considered in Section~\ref{sec:inc_MF_disc}. The
smearing function is a Breit-Wigner with $\sigma=\omega_0/3$.

\subsection{Smeared spectral densities from discrete sampling\label{sec:systsmea_disc}}
Analogously to the continuum, an incomplete smeared spectral density becomes a satisfactory
approximation of the complete one if $\omega_0\, t_{\rm max}$  is chosen large
enough so as to guarantee a reconstruction error below the target precision.
To quantify this systematics, we decompose a smeared density as\footnote{A decomposition
in which the three contributions are $O(a)$-improved independently is beyond
the scope of this paper.}
\be
\varrho_{\kappa;m} =  \tilde \varrho_{\kappa;m} (s_{\rm max}) +
\Big[\varrho_{\kappa;m} (s_{\rm max})  - \tilde \varrho_{\kappa;m} (s_{\rm max})\Big] +
\Big[\varrho_{\kappa;m} - \varrho_{\kappa;m}(s_{\rm max})\Big] \, .
\ee
The norm of the second term on the r.h.s of this equation is bounded
by $\Delta  \tilde \varrho_{\kappa;m} (s_{\rm max}) $ in
Eq.~(\ref{eq:deltahatrhomk_disc}). To bind the difference between
the target smeared spectral density and its incomplete counterpart,
we notice that from Eqs.~(\ref{eq:vhat}) and (\ref{eq:2F1toP})
it follows that
\bea
\Bigg\vert \frac{{\overline {\cal C}}_{m,s}}{\lambda_s} \Bigg\vert & \leq & 
\frac{\sqrt{2 s \sinh(\pi s)} }{\pi}
\int_{0}^{\infty} \vert\varrho_m(\omega)\vert
\frac{a}{\sqrt{1-e^{-a\omega}}}\,
\Big\vert {\cal K}_{is}(\omega,t_0) \Big\vert \, d\omega\, \nonumber\\[0.25cm]
& \leq &
\frac{\sqrt{2 s \sinh(\pi s)} }{\pi} \max_{\;\;\omega\,\in\, {\mathbb R^+}}
\Big\vert {\cal K}_{is}(\omega,t_0) \Big\vert\, {\mathfrak r}_m\, , 
\eea
where 
\be
{\cal K}_{is}(\omega,t_0) = \sqrt{\frac{\pi}{a}}\, 
\Big\vert\Gamma\Big(2t_0/a\!-\!1/2\!+\!is\Big)\Big\vert \,
\sqrt{\frac{e^{-a\omega}}{1-e^{-a\omega} }}\,
P^{1-2t_0/a}_{-\frac{1}{2} + is}\Big(\coth\Big(\frac{a \omega}{2}\Big)\Big)\, ,
\ee
and
\be
{\mathfrak r}_m = a
\int_{0}^{\infty} \vert\varrho_m(\omega)\vert
\sqrt{\frac{1}{1-e^{-a\omega}}}\, d\omega\, . 
\ee
Analogously to the continuum the function ${\cal K}_{is}(\omega,t_0)$
first oscillate and then it decades exponentially, and  its absolute
maximum can be easily computed numerically.
By assuming that $\varrho_m(\omega)$ and therefore ${\cal C}(t_0+an)$ has a definite
sign, and by remembering that
\be
\frac{1}{\sqrt{1-e^{-a\omega}}} = \sum_{n=0}^{\infty} (-1)^{n} \binom{-\frac{1}{2}}{n}\, e^{-a\omega n} \, ,
\ee
we obtain
\be\label{eq:rmd}
{\mathfrak r}_m =  a \sum_{n=0}^{\infty} (-1)^{n} \binom{-\frac{1}{2}}{n}\, {\cal C}_m(an) = 
a \sum_{n=0}^{\infty} b^n_m\, {\cal C}(an) \, , 
\ee
where from Eq.~(\ref{eq:trapcm})
\be
b^n_m = \frac{a}{\Gamma(m)}
\sum_{n'=0}^{n} w_{n-n'}\, (-1)^{n'}
\binom{-\frac{1}{2}}{n'}\, (an-an')^{m-1}\, , \quad m\geq 1\, ,  
\ee
while for $m=0$ we use the expression in the middle of Eq.~(\ref{eq:rmd}). 
Analogously to the continuum, the sum in Eq.~(\ref{eq:rmd}) can be decomposed in
the short-, intermediate- and
long-distance contributions ${\mathfrak r}_m^{\rm s}$, ${\mathfrak r}_m^{\rm i}$
and ${\mathfrak r}_m^{\rm l}$ corresponding
to sum in the ranges $[0,t_0]$, $[t_0,t_{\rm max}]$, and $[t_{\rm max},\infty]$
respectively. For the short- and the intermediate-distance contributions considerations
analogous to those in the continuum apply.
By using Eq.~(\ref{eq:Cboundd}), the long distance contribution satisfies
\be
{\mathfrak r}_m^{\rm l} \leq {\mathfrak r}_m^{\rm l, \max}
=
\vert {\cal C}(a n_{\rm max})\vert\,\, a\!\!\!\!\!\!\!\! \sum_{n=n_{\rm max}+1}^{\infty}
b^n_m\, e^{-a \omega_0 (n-n_{\rm max})}\, .
\ee
which is exponentially suppressed in $n_{\rm max}=t_{\rm max}/a$.
The difference between
the smeared spectral density and the incomplete one can then be bounded
as\footnote{With a similar spirit, while we were completing this paper
an approximation of these bounds was proposed in Ref.~\cite{Tsuji:2026zku}. 
Additionally, explicitly enforcing the positivity of spectral functions
recently allowed one to compute alternative bounds on their smeared 
counterpart~\cite{Lawrence:2024hjm,Abbott:2026wdw,Jay:2026qoh}.} 
\be
\Big\vert \varrho_{\kappa;m} - \varrho_{\kappa;m}(s_{\rm max}) \Big\vert \leq
\Delta \varrho_{\kappa;m} (s_{\rm max}) \, ,
\ee
with
\be\label{eq:deltarhomk_disc}
\Delta  \varrho_{\kappa;m} (s_{\rm max}) =
{\mathfrak r}_m^{\max}
\int_{s_{\rm max}}^{\infty} \frac{\sqrt{2 s \sinh(\pi s)} }{\pi}
\max_{\;\;\omega\,\in\, {\mathbb R^+}} \Big\vert {\cal K}_{is}(\omega,t_0) \Big\vert\,
\Big\vert {\hat {\it k}}_s \Big\vert \, d s\, , 
\ee
where ${\mathfrak r}_m^{\max}$ is the sum of the bounds on
${\mathfrak r}_m^{\rm s}$, ${\mathfrak r}_m^{\rm i}$ and ${\mathfrak r}_m^{\rm l}$.
By using Eqs.~(\ref{eq:deltahatrhomk_disc}) and (\ref{eq:deltarhomk_disc}) it
follows that 
\be\label{eq:syst_disc}
\Big\vert\varrho_{\kappa;m} -  \tilde \varrho_{\kappa;m} (s_{\rm
  max})\Big\vert \leq
\Delta  \tilde \varrho_{\kappa;m} (s_{\rm max}) + 
\Delta  \varrho_{\kappa;m} (s_{\rm max})\, . 
\ee
Analogously to the continuum case, the approximated incomplete smeared spectral density $\tilde
\varrho_{\kappa;m} (s_{\rm max})$, when supplemented by the systematic
error $[\Delta  \tilde \varrho_{\kappa;m} (s_{\rm max}) + \Delta  \varrho_{\kappa;m} (s_{\rm max})]$,
becomes a rigorous estimate of the smeared spectral density at finite lattice spacing.
As a representative example, in the right panel of
Figure~\ref{fig:rho_theo_disc} we show $\tilde \varrho_{\kappa;m}(s_{\rm max})$
together with the total error, estimated as in Eq.~(\ref{eq:syst_disc}), 
against $\varrho_{\kappa;m}$ both extracted from the
exponential considered in Subsection~\ref{sec:inc_MF_disc}. Also in this
case, the smearing function is a Breit-Wigner with $\sigma=\omega_0/3$.

\section{Discussion}\label{sec:discuss}
When a correlator and the associated spectral density admit
integral transforms and their inverses with basis functions
related by an equation of the type (\ref{eq:ubar2uhat}), the basic integral
equation in Eq.~(\ref{eq:ct}) becomes a simple algebraic
relation among the integral
transforms of the correlator and of the spectral density,
see Eq.~(\ref{eq:ILTalg}).
The (smeared) spectral density is then readily obtained
as an inverse transform. The key
issue is the delicate cancellation between the numerator and the denominator
in the integral on the r.h.s. of Eq.~(\ref{eq:rhogen}), i.e.
between the exponential decay of the integral transform of the
correlator as a function of the conjugate variable $s$ and the exponentially
decaying numerical coefficient in the denominator.

If the correlator is known for all $t\geq t_0$, the Laplace
transform can be inverted exactly. In practical applications, however,
the correlator is often known only up to $t\leq t_{\rm max}$. This implies that
we only have access to an incomplete transform, see Eq.~(\ref{eq:Cbarmss}),
due to the finiteness of the simulated lattice or possibly to the
further limitation due to the worsening of the signal-to-noise ratio,
see Ref.~\cite{Giusti:2026lg2} for more details.
At small values of $s$, the latter typically approximates the complete
transform exponentially well in $t_{\rm max}$, see Figure~\ref{fig:MT_theo}. But
when $s\geq s_{\rm max}\sim \omega_0 t_{\rm max}$, with $\omega_0$ being the
mass gap in that channel, the norm of the incomplete transform of the correlator
does not decay exponentially with $s$ anymore, a fact which prevents us from
reconstructing exactly the (smeared) spectral density. This is clearly shown in
Figure~\ref{fig:MT_theo} for the Mehler-Fock transform, and an analogous plot
can be obtained for the Mellin one.

This lack of knowledge implies that we have de-facto access to
the incomplete smeared spectral density as defined in Eq.~(\ref{eq:rhosubf_smea_inc}) only.
If we restrict ourselves to smearing functions which have transforms
decaying fast enough with $s$, however, the contributions of the unknowns to the smeared
spectral density can be bounded and kept under control\footnote{An analogous line of argumentation can
be applied to the study of finite volume effects. A complementary approach has been recently
presented in Ref.~\cite{Bresciani:2026kjv}.}, see Figure~\ref{fig:rho_theo}.
Very similar comments apply when the correlation function is sampled on discrete
points only, as discussed in
detail in Section~\ref{sec:discrsme}.

It is instructive to scrutinize these findings for a simple example of 
a correlator made of a sum of 3 exponentials
\be\label{eq:3exp}
C(t) = c_0\, e^{-\omega_0 t} + c_1\, e^{-\omega_1 t} + c_2\, e^{-\omega_2 t}
\ee
with $\omega_0=0.5$, $\omega_1=1.0$, $\omega_2=1.5$ and $c_0=1.0$,
$c_1=0.5$, $c_2=1.25$ in units of $t_0=1$. From this correlator we compute the spectral
density smeared with the Breit-Wigner in Eq.~(\ref{eq:BWsm}) centered in
$\omega_*$ and with width $\sigma$. The Breit-Wigner is a
very conservative example since its Kontorovich–Lebedev transform
decays quite slowly as $1/s^2$, but it has the advantage that its
transform has a simple expression which can be
managed easily.
\begin{figure}[t!]
    \centering
    \includegraphics[width=0.34\linewidth]{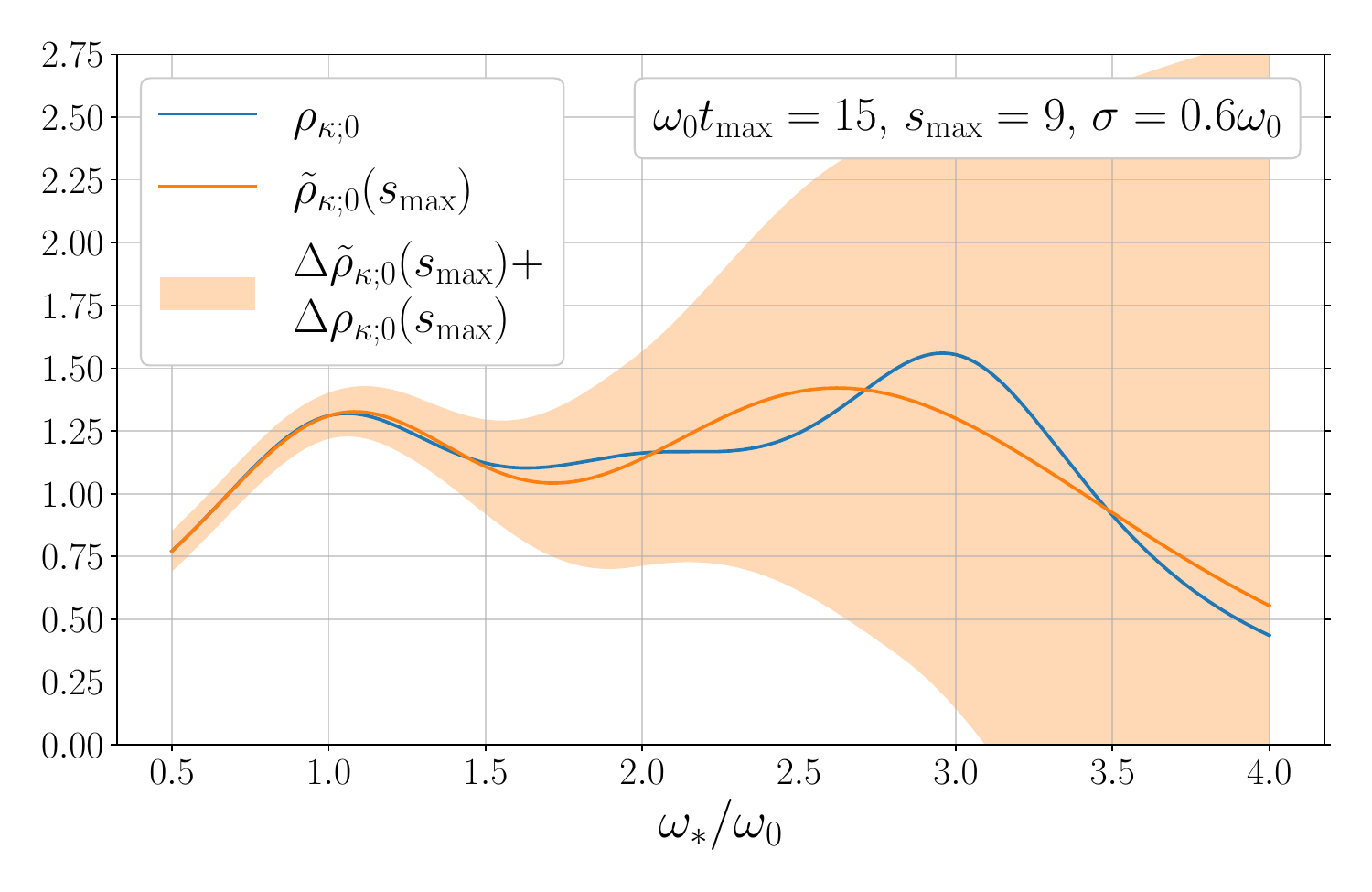}%
    \includegraphics[width=0.34\linewidth]{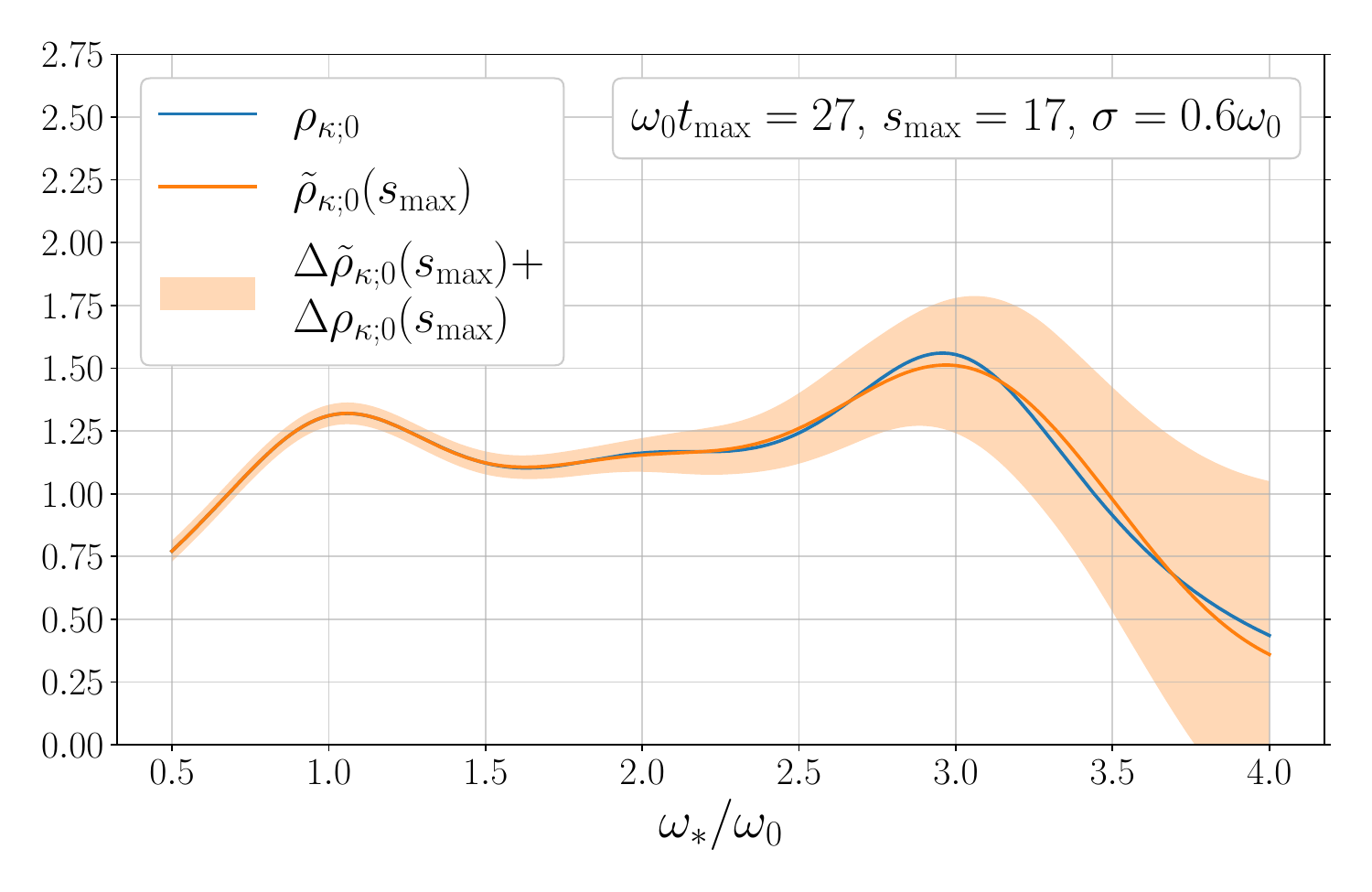}%
    \includegraphics[width=0.34\linewidth]{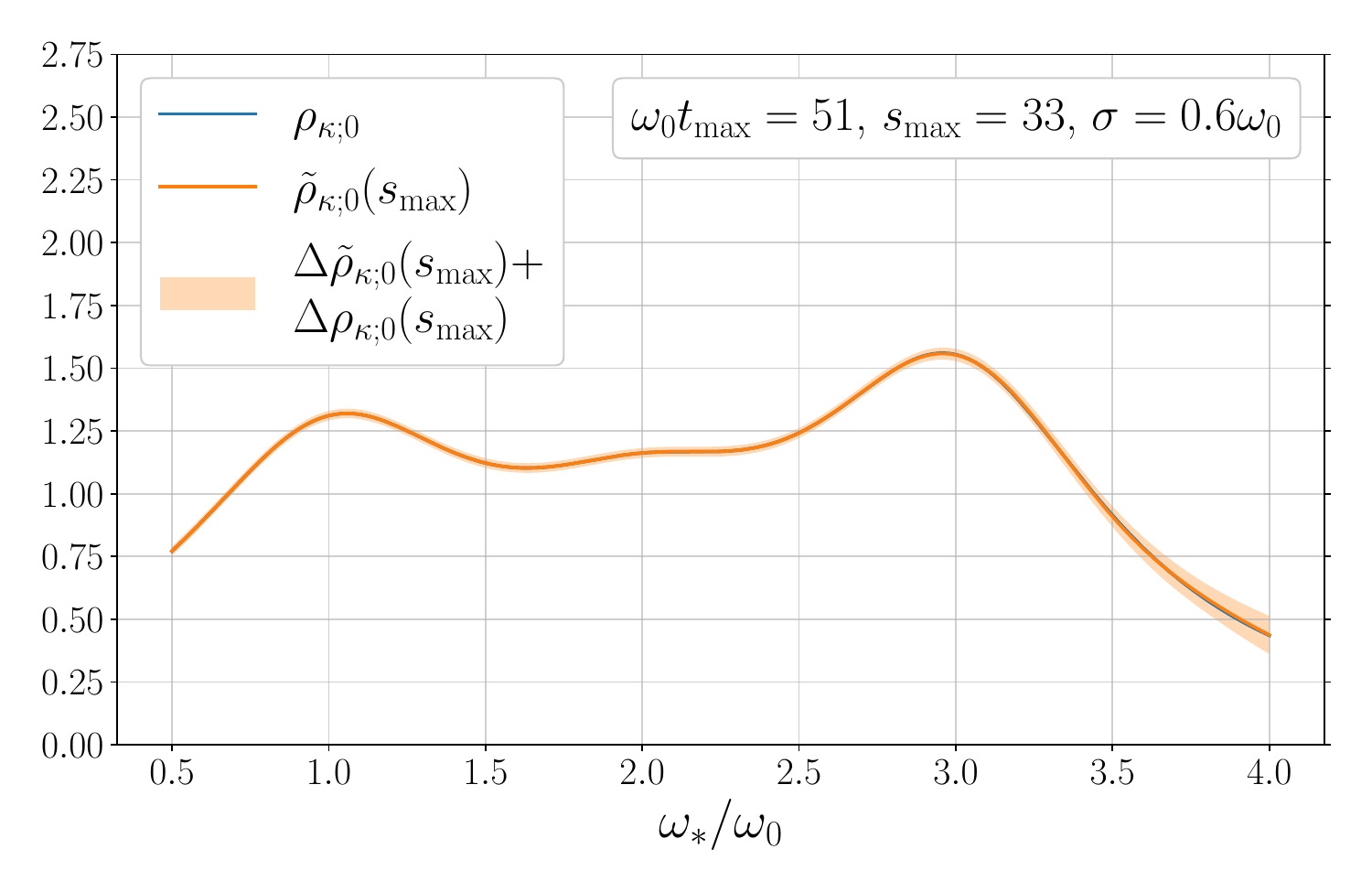}
    \caption{The approximated incomplete smeared spectral density
    $\tilde \rho_{\kappa;0}(s_{\rm max})$
with the total error estimated as in Eq.~(\ref{eq:syst}), against
the exact one $\rho_{\kappa;0}$ for the correlator in
Eq.~(\ref{eq:3exp}) and for a Breit-Wigner smearing centered in
$\omega_*$ and with width $\sigma/\omega_0=0.6$. The three panels
correspond to  $\omega_0 t_{\rm max}=15$ (left), $27$ (center)
and $51$ (right).}
\label{fig:rho_theo_discuss1}
\end{figure}
\begin{figure}[h!]
    \centering
    \includegraphics[width=0.34\linewidth]{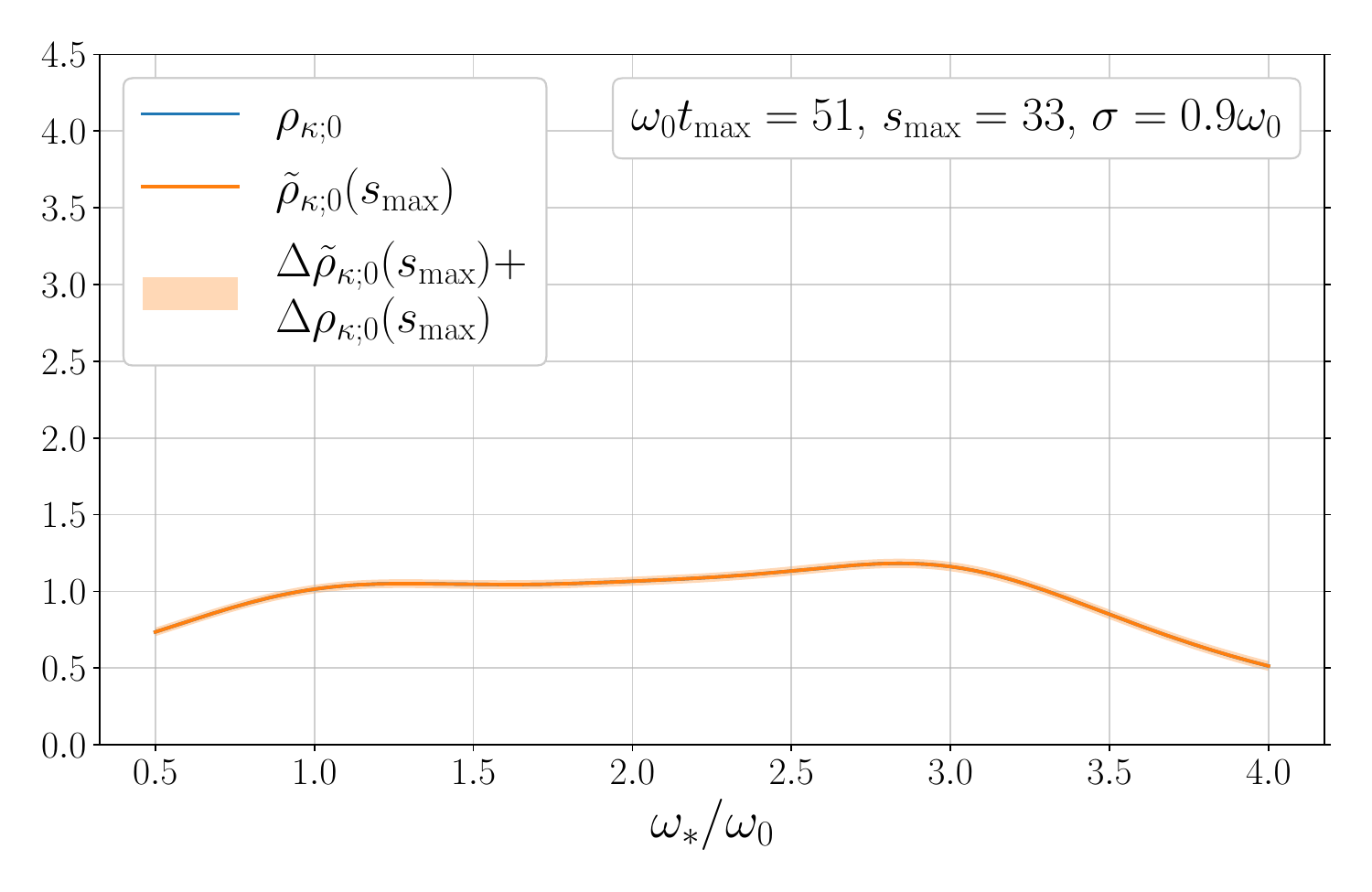}%
    \includegraphics[width=0.34\linewidth]{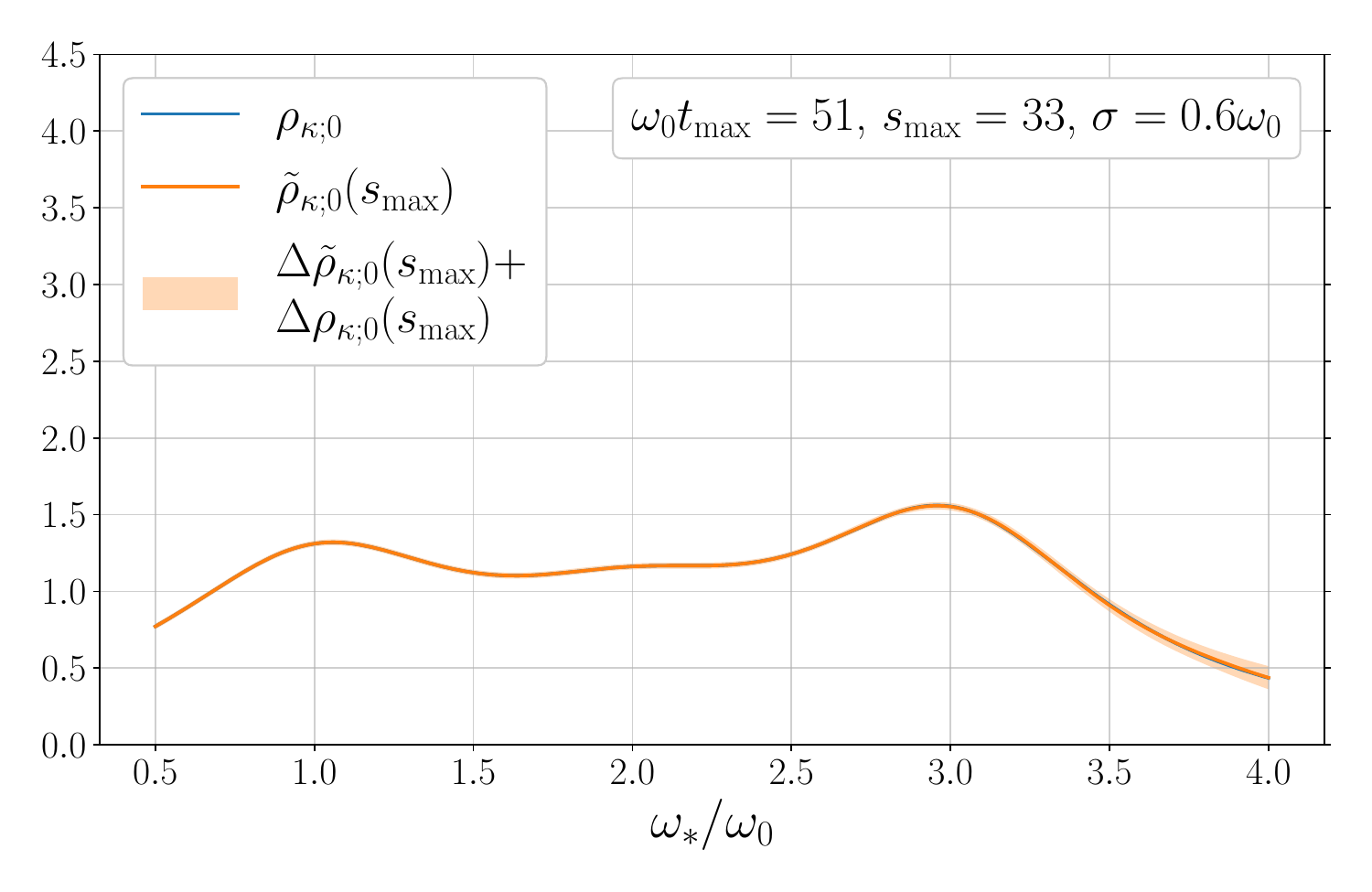}%
    \includegraphics[width=0.34\linewidth]{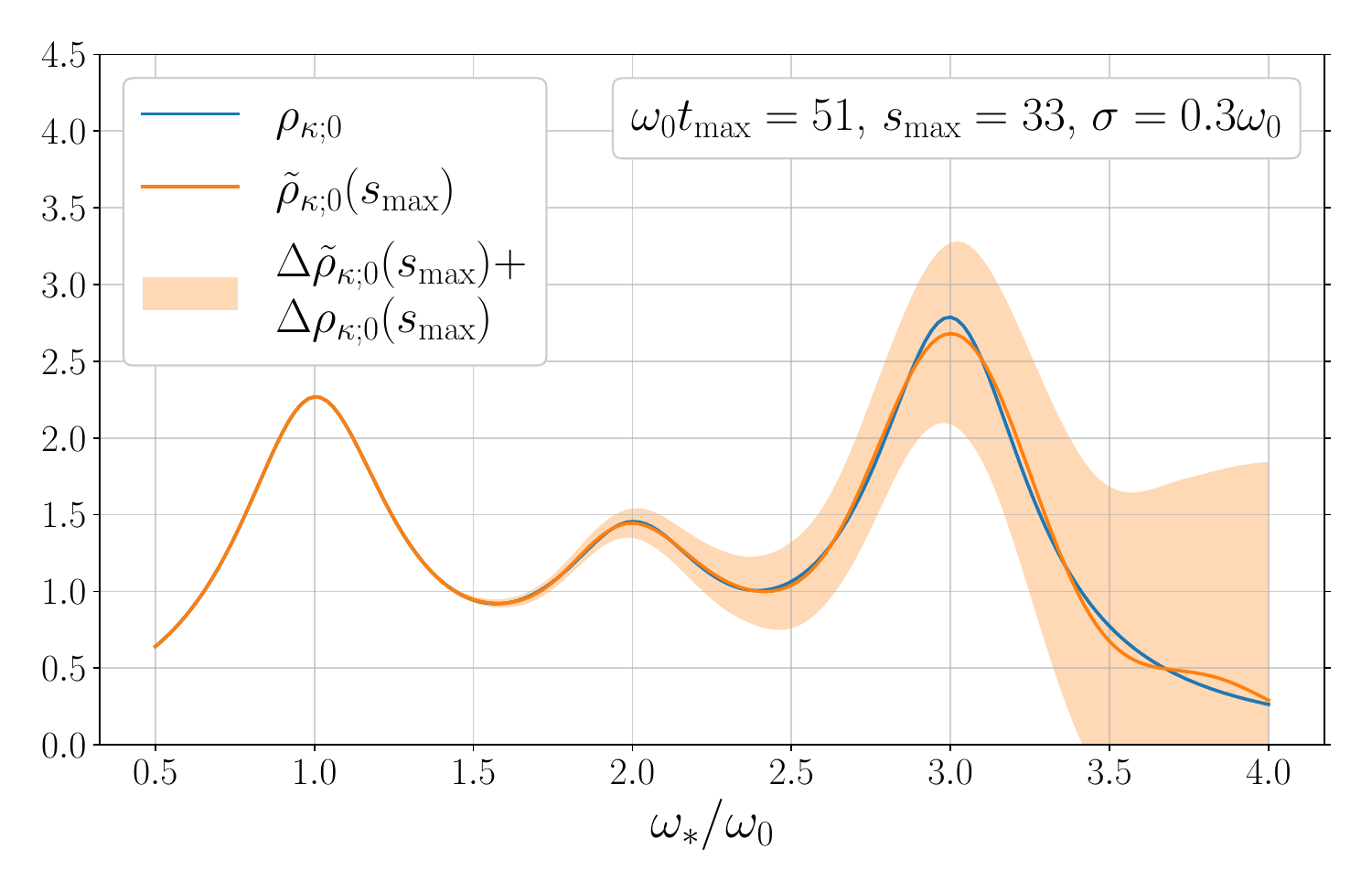}
    \caption{The approximated incomplete smeared spectral density
    $\tilde \rho_{\kappa;0}(s_{\rm max})$ with the total error estimated as
    in Eq.~(\ref{eq:syst}), against
the exact one $\rho_{\kappa;0}$ for the correlator in
Eq.~(\ref{eq:3exp}) and for a Breit-Wigner smearing centered in
$\omega_*$ and with width $\sigma/\omega_0=0.9$ (left), 
$0.6$ (center), and $0.3$ (right) for $\omega_0 t_{\rm max}=51$.}
\label{fig:rho_theo_discuss2}
\end{figure}
\begin{figure}[h!]
    \centering
    \includegraphics[width=0.34\linewidth]{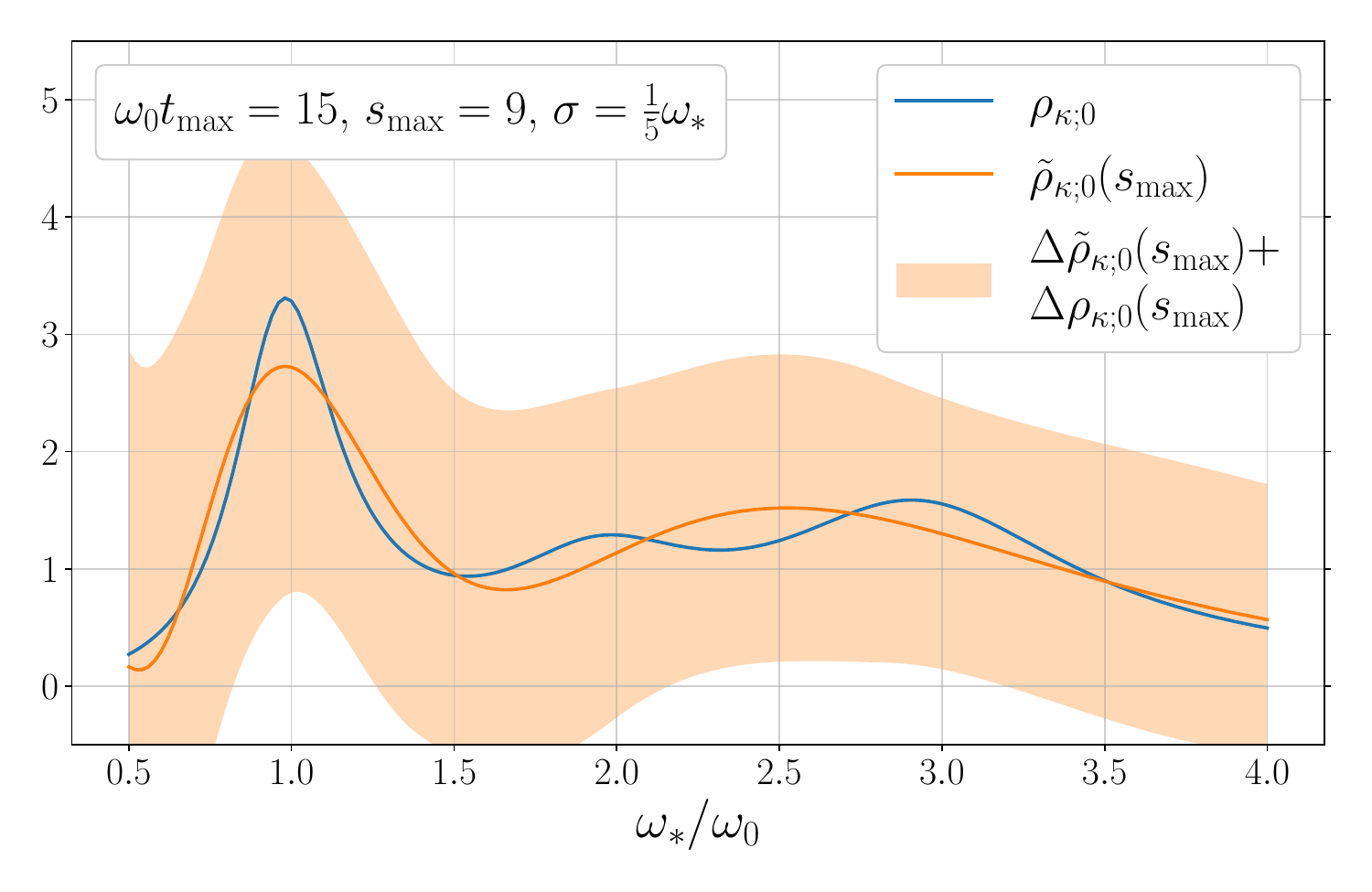}%
    \includegraphics[width=0.34\linewidth]{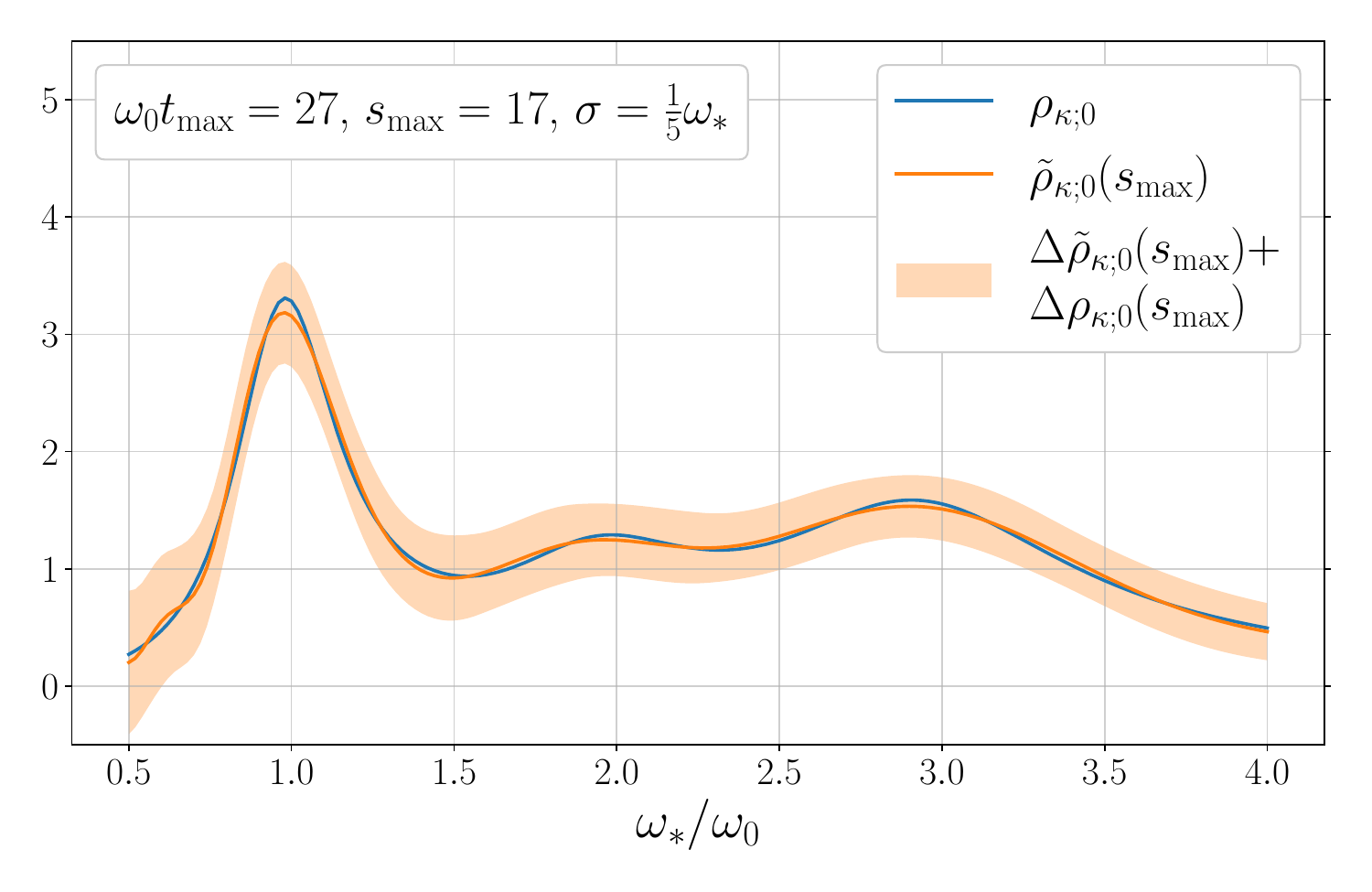}%
    \includegraphics[width=0.34\linewidth]{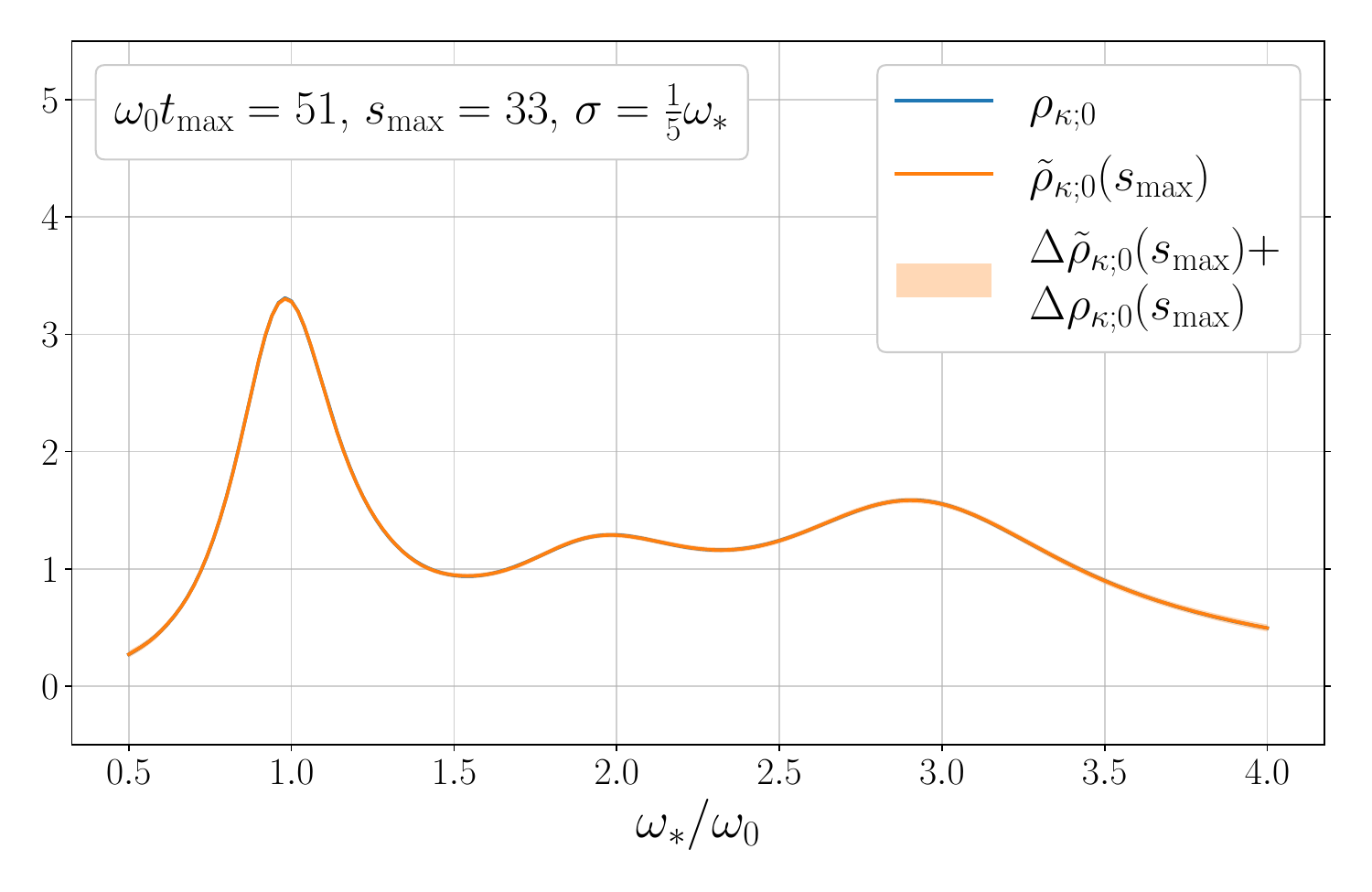}
    \caption{The approximated incomplete smeared spectral density
    $\tilde \rho_{\kappa;0}(s_{\rm max})$ with the total error estimated as
    in Eq.~(\ref{eq:syst}), against the exact one $\rho_{\kappa;0}$
    for the correlator in Eq.~(\ref{eq:3exp}) and for a Breit-Wigner
    smearing centered in $\omega_*$ and with width $\sigma/\omega_*=1/5$.
    The three panels correspond to  $\omega_0 t_{\rm max}=15$ (left),
    $27$ (center) and $51$ (right).}
    \label{fig:rho_theo_discuss3}
\end{figure}
\begin{figure}[h!]
    \centering
    \includegraphics[width=0.34\linewidth]{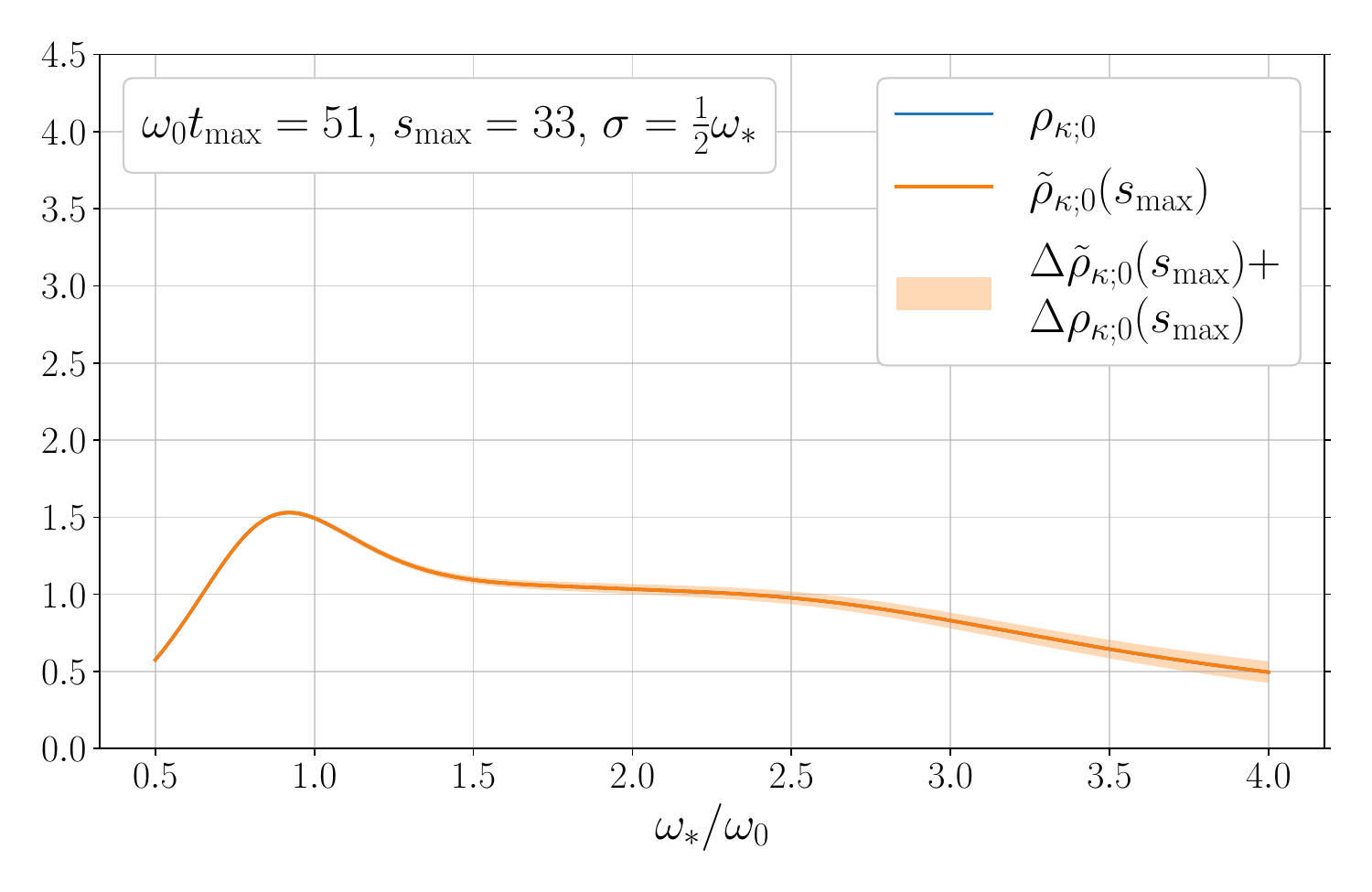}%
    \includegraphics[width=0.34\linewidth]{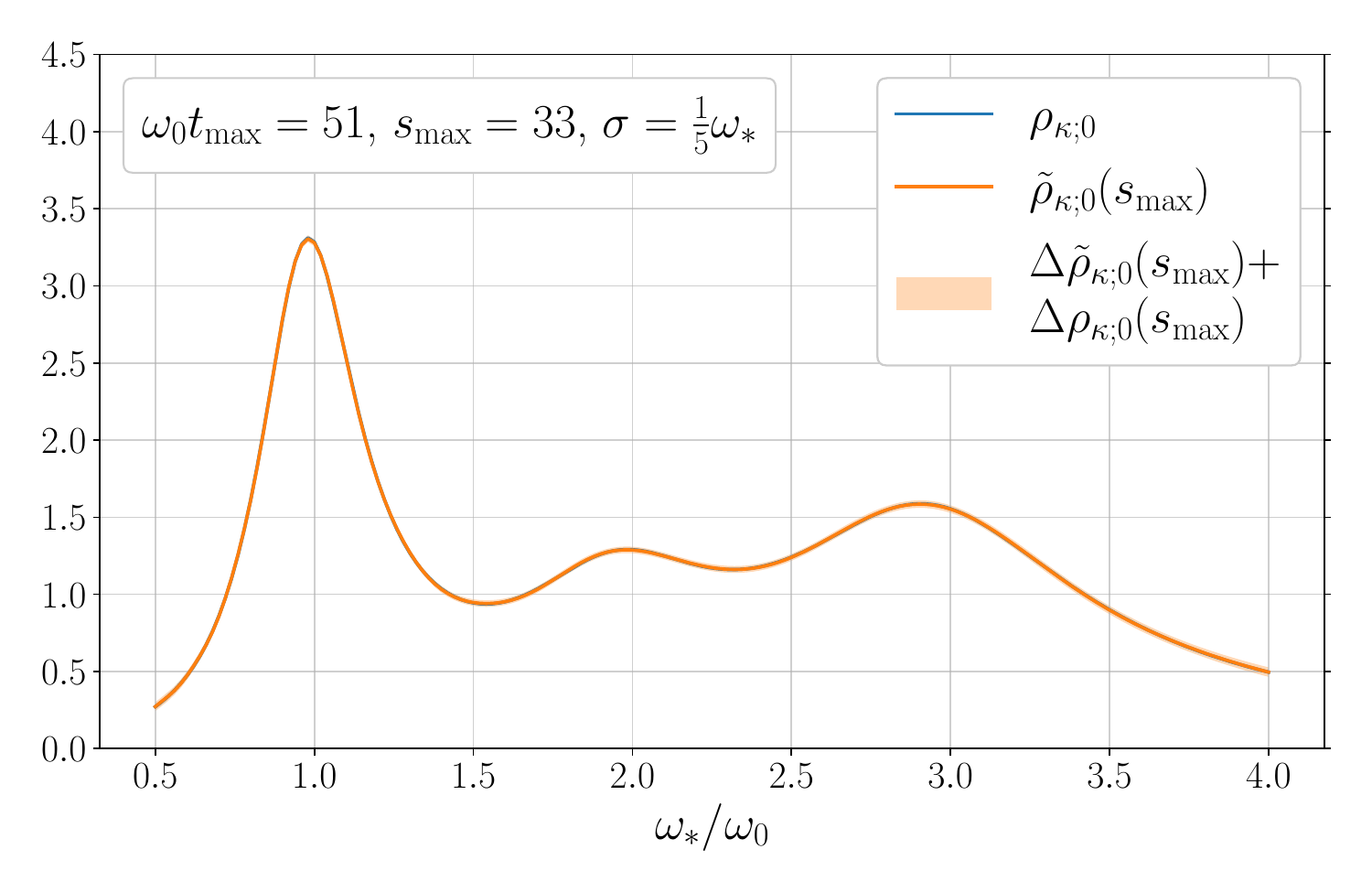}%
    \includegraphics[width=0.34\linewidth]{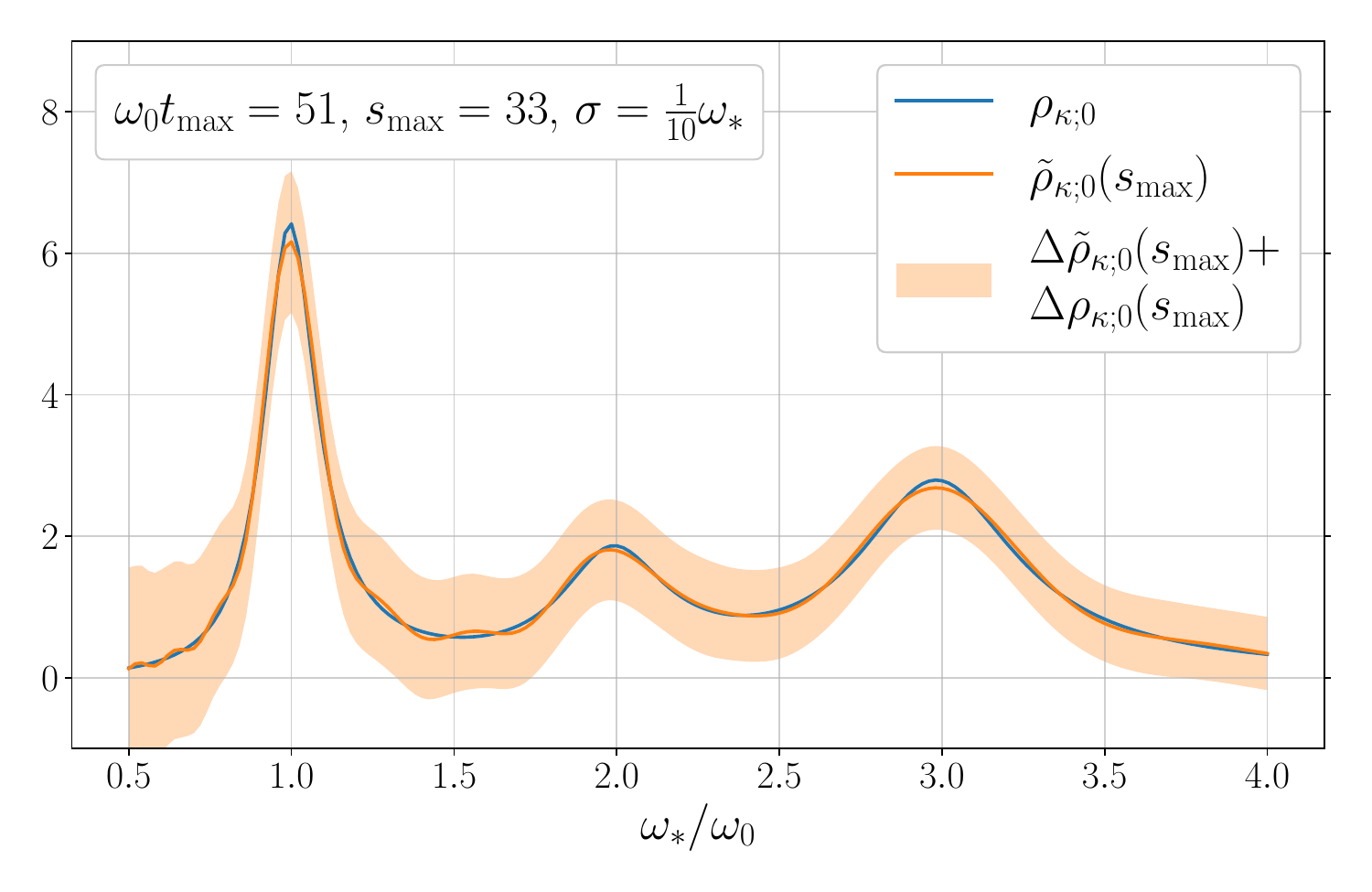}
    \caption{The approximated incomplete smeared spectral density
    $\tilde \rho_{\kappa;0}(s_{\rm max})$ with the total error estimated as
    in Eq.~(\ref{eq:syst}), against the exact one $\rho_{\kappa;0}$ for the
    correlator in Eq.~(\ref{eq:3exp}) and for a Breit-Wigner smearing centered in
    $\omega_*$ and with width $\sigma/\omega_*=1/2$ (left), 
    $1/5$ (center), and $1/10$ (right, different scale on vertical axis) for $\omega_0 t_{\rm max}=51$.}
    \label{fig:rho_theo_discuss4}
\end{figure}

In the three panels in Figure~\ref{fig:rho_theo_discuss1} we show 
the reconstructed smeared spectral density with constant width
$\sigma=0.6\, \omega_0$
for three values of $t_{\rm max}$ corresponding to
$\omega_0\, t_{\rm max}=15, 27$ and $51$. By comparing the exact
smeared spectral density (blue line) with the
incomplete one (orange line), it is clear that the density
can be reconstructed reliably with a rigorous bound of the
systematic error attached to it. For the Breit-Wigner, the error
decreases approximatively as\footnote{The choice of the
value of $s_{\rm max}$ gives a handle to minimize the error
in Eq.~(\ref{eq:syst}) by balancing the two competing contributions.
In the simple examples shown in this section we have not extensively
exploited this possibility.} $s^{-1}_{\rm max} \sim (\omega_0 t_{\rm max})^{-1}$,
and it turns out to be quite loose. In the three panels in
Figure~\ref{fig:rho_theo_discuss2} we show the very same reconstruction
at fixed $\omega_0\, t_{\rm max}=51$ but for three values of the smearing
width $\sigma/\omega_0=0.9$, $0.6$ and $0.3$. As expected, the three-peak
structure emerges when the smearing is smaller than the distance between
the peaks, with the bound on the error that increases significantly. 

In the three panels in Figure~\ref{fig:rho_theo_discuss3} we show,
for $\omega_0 t_{\rm max}=15, 27$ and $51$, 
the reconstruction of the smeared spectral density with varying width
but with the ratio $\sigma/\omega_*=1.5$ kept constant as a function
of $\omega_*$. The error bound is more uniform as a function of
$\omega_*$ than in the panels of Figure~\ref{fig:rho_theo_discuss1}.
It is again clear that the spectral density
can be reconstructed reliably with a rigorous estimate of the systematic error
attached to it. Finally, in Figure~\ref{fig:rho_theo_discuss4} we show
the very same reconstruction at fixed $\omega_0\, t_{\rm max}=51$ but for three
values of the ratio $\sigma/\omega_*=1/2$, $1/5$ and $1/10$. As expected, the
three-peak structure emerges when the smearing gets smaller and, at
variance of the analogous plots in Figure~\ref{fig:rho_theo_discuss2},
the error bound is more uniform as a function of $\omega_*$.
\begin{figure}[t]
    \centering
    \includegraphics[width=0.34\linewidth]{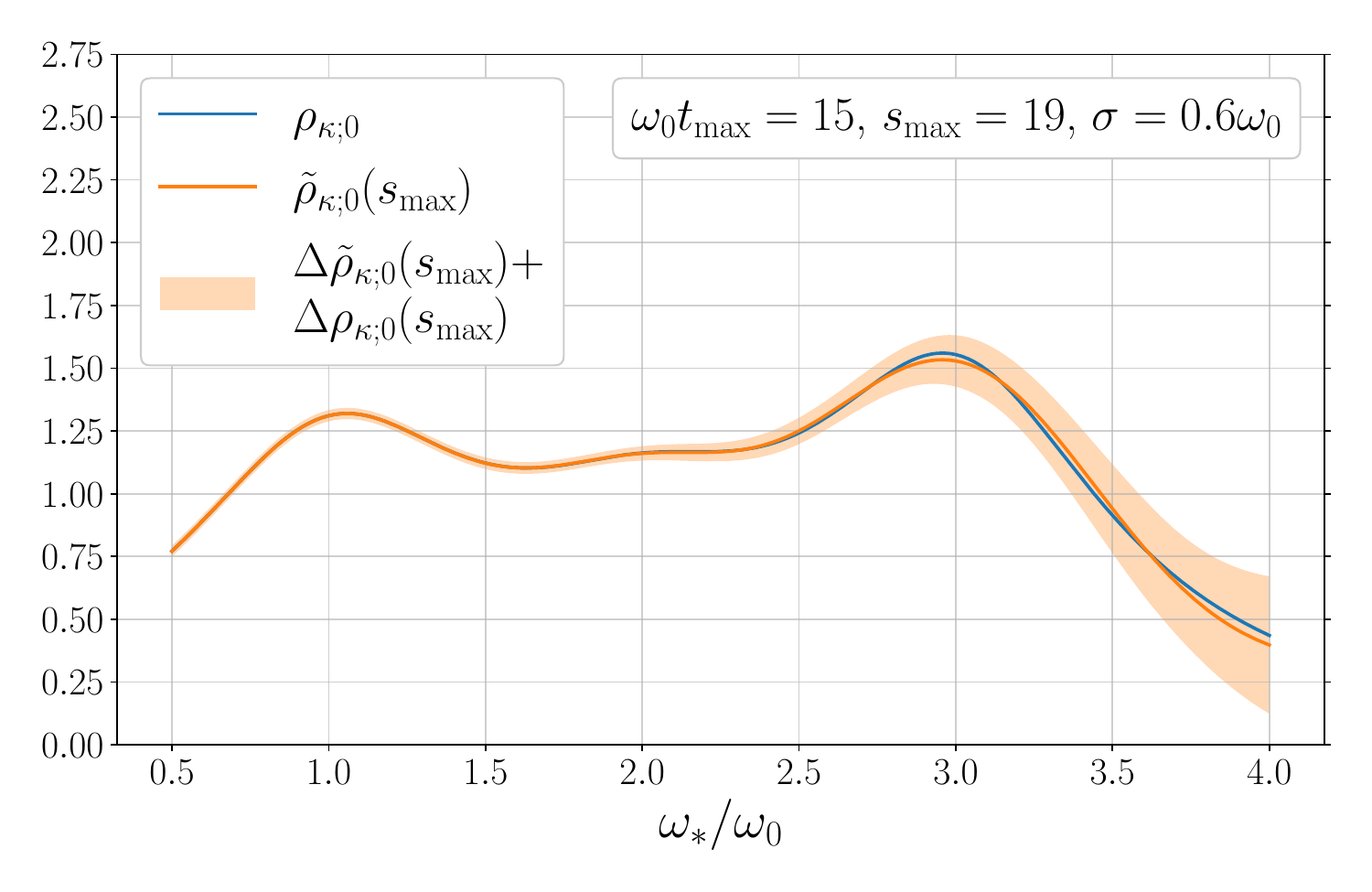}%
    \includegraphics[width=0.34\linewidth]{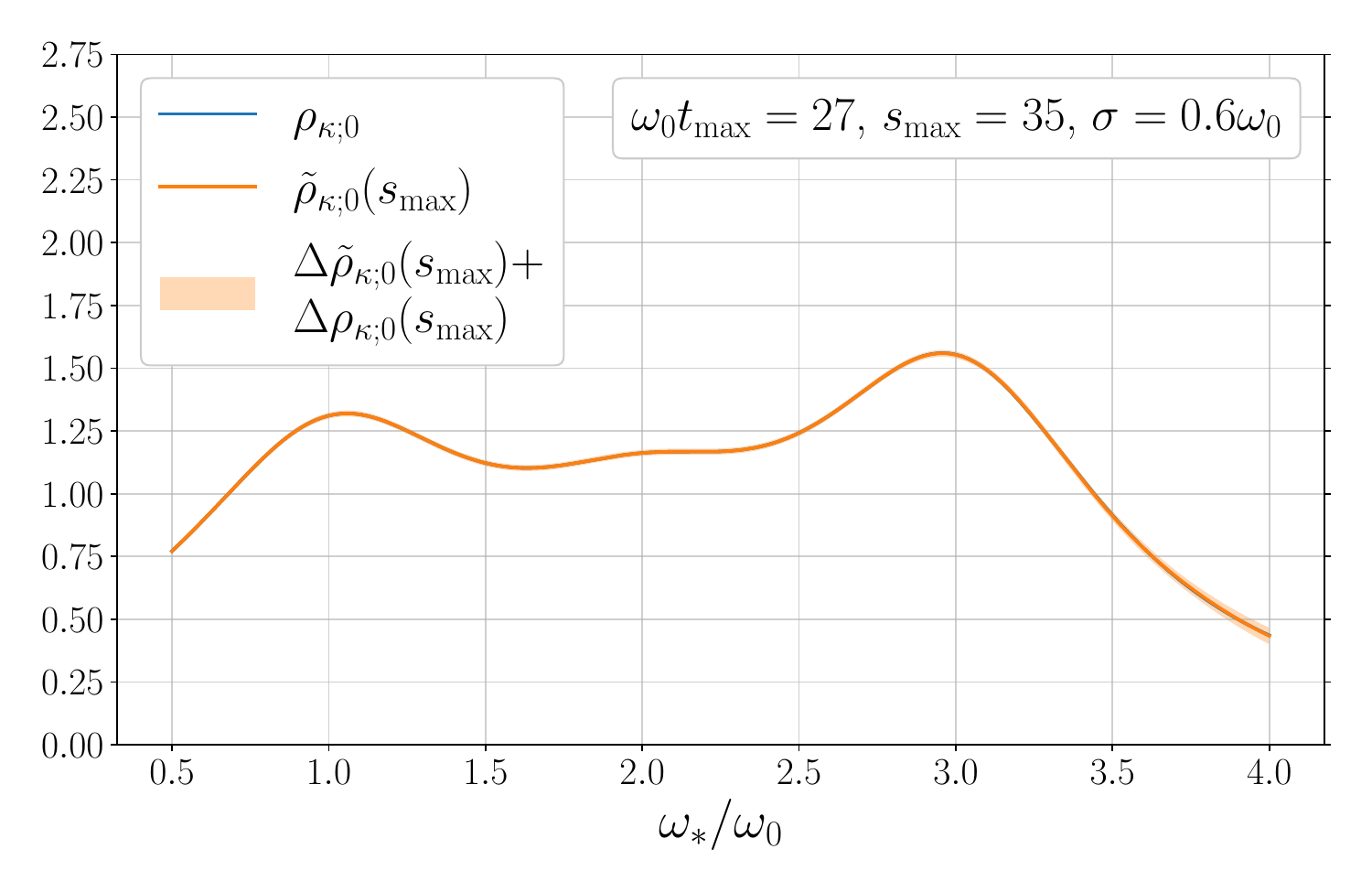}%
    \includegraphics[width=0.34\linewidth]{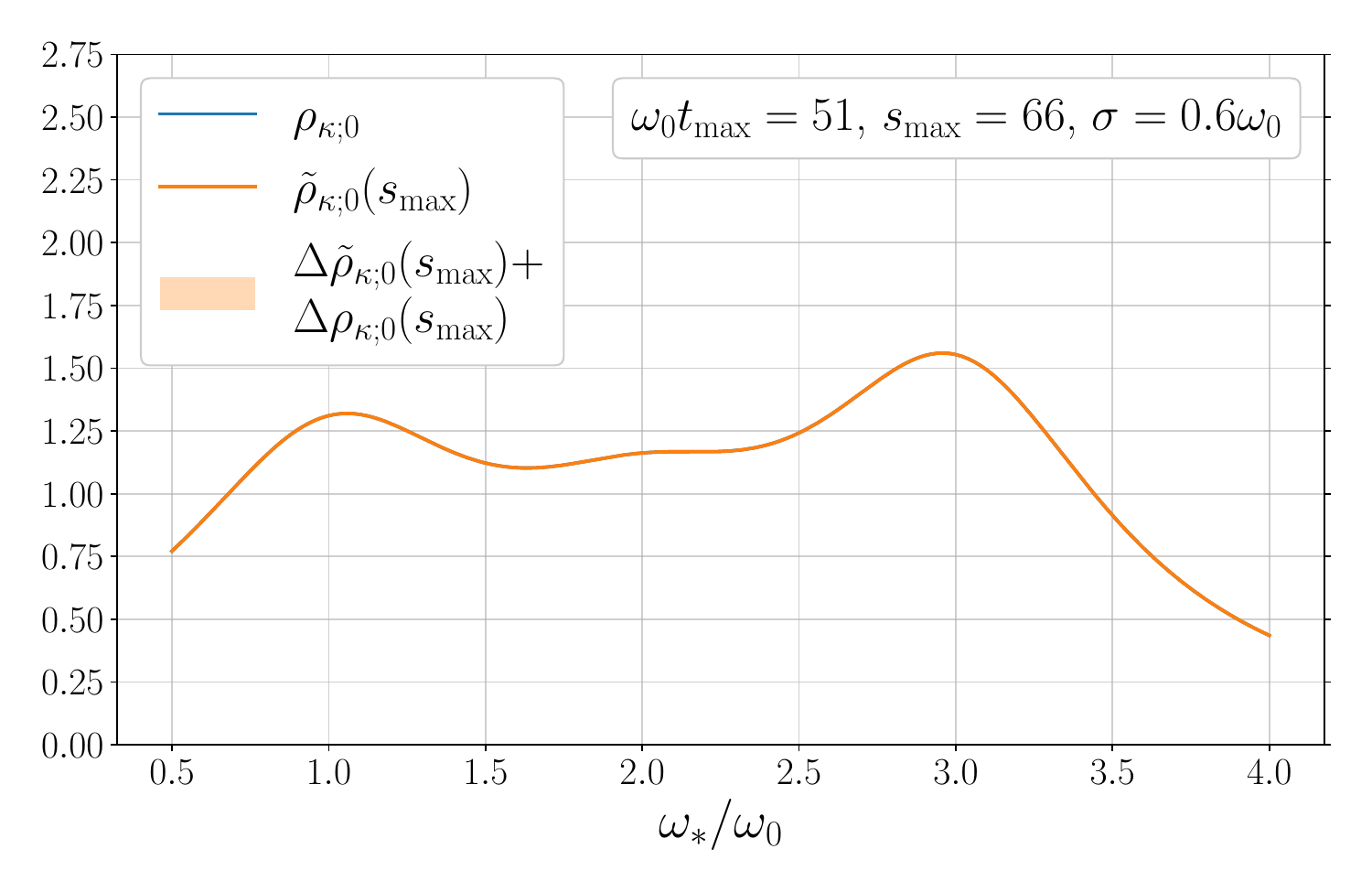}
    \caption{The approximated incomplete smeared spectral density
    $\tilde \rho_{\kappa;0}(s_{\rm max})$
with the total error estimated as in Eq.~(\ref{eq:syst}) computed
by preconditioning the correlator as explained in the main text,
against the exact one $\rho_{\kappa;0}$ for the correlator in
Eq.~(\ref{eq:3exp}) and for a Breit-Wigner smearing centered in
$\omega_*$ and with width $\sigma/\omega_0=0.6$. The three panels
correspond to  $\omega_0 t_{\rm max}=15$ (left), $27$ (center)
and $51$ (right).}
\label{fig:rho_theo_discuss5}
\end{figure}

If one or a few leading exponential states are known,
then they can be treated exactly and all considerations above apply to
the ``preconditioned'' correlation function, i.e. the correlation function
subtracted by the known exponentials~\cite{Kellermann:2025pzt,Morandi:2026nll}. In the three panels in
Figure~\ref{fig:rho_theo_discuss5} we show the reconstructed smeared spectral
density with constant width $\sigma=0.6\, \omega_0$
for three values of $t_{\rm max}$ corresponding to
$\omega_0\, t_{\rm max}=15, 27$ and $51$ obtained by subtracting the leading
exponential, by computing the spectral density, and then by adding back the contribution of the
leading exponential to the smeared spectral density. When compared
with the three panels in Figure~\ref{fig:rho_theo_discuss1}, the gain
is substantial and approximatively as expected. This time the relevant
parameter is $s_{\rm max}\sim \omega_1\, t_{\rm max}$ which is twice the
one without preconditioning. When the correlator is computed numerically, however,
whether
this is an advantage depends on how much $t_{\rm max}$ needs to be reduced once
the leading exponential(s) has (have) been subtracted from the original correlator.

\section{Conclusions}
Integral transforms offer a powerful theoretical framework
to compute the inverse Laplace transform of a correlator
and to study the associated systematics. Leading discretization effects,
systematics due to a finite temporal extent or to a finite
spatial volume can be disentangled and studied independently on the lattice and
in the continuum, respectively.

The solution proposed here, based on the Mehler-Fock and
Kontorovich–Lebedev transforms, guarantees that a spectral density
extracted from an $O(a)$-improved lattice correlator is $O(a)$-improved
as well. In practical applications, the correlator is known
on a finite time-interval only, e.g. up to $t\leq t_{\rm max}$. This
implies that we can only have access to an incomplete transform of the correlator
and therefore to an incomplete smeared spectral density. However, 
if the transform of the smearing function decays fast enough with
the conjugate variable $s$, the unknown contributions
can be bound and kept under control.

Here we have presented a generic derivation of the associated
error bound. Depending 
on the particular physics application one is interested in,
better bounds could be found by accelerating the convergence
to zero of the transform of the smearing function, by deriving a
more sophisticated formula for the bound, and/or by
preconditioning the correlation function.
Once all these optimizations are done, the improvement of
the reconstruction requires measuring the correlator at
larger and larger time-distances. This can be achieved by enhancing
the signal-to-noise ratio of the estimators of the correlators by variance
reduction techniques such as the multi-level Monte Carlo integration.

\begin{acknowledgments}
We thank M. Bruno for many discussions on the techniques and the
phenomenological applications of spectral density reconstruction.
This work is (partially) supported by ICSC - Centro Nazionale di
Ricerca in High Performance Computing, Big Data and Quantum Computing,
funded by European Union – NextGenerationEU. M.S. has been partially
supported by the Italian {\it Fondazione Angelo Della Riccia}.
\end{acknowledgments}

\appendix

\section{K\"all\'en-Lehmann representation and spectral densities}\label{app:KL}
In this appendix we collect well-known formulas for subtracted
dispersive relations in order to fix our notation and conventions for
regulated spectral densities. We focus on two-point connected correlation functions in Euclidean space-time
of the form
\be\label{eq:corr}
    C(x) = \langle {\mathcal O}_1(x) {\mathcal O}_2(0) \rangle_{\rm
      c}\, ,
\ee
where ${\mathcal O}_1$ and ${\mathcal O}_2$ indicate two renormalized scalar
fields\footnote{Generalizations to non-zero spin fields can be found in
many textbooks, see for instance Ref.~\cite{Weinberg:1995mt}.}. Thanks
to the K\"all\'en-Lehmann spectral decomposition, we can define
\be\label{eq:KL}
    \Pi(p^2) = \int C(x) \, e^{-ipx} d^4x  = \int_0^\infty 
    \frac{\rho(\sqrt s)}{p^2+s}\, ds\,, 
    \qquad p=(p_0,{\bf p})\,,
\ee
where $\rho(\sqrt s)$ is the spectral density associated to the correlator
${\rm C}(x)$
which has support over $[\sqrt{s}\geq \omega_0\! > \! 0,\infty)$ if the sector
has a mass gap $\omega_0$. Being an integrated correlation function, $\Pi(p^2)$ can suffer from
ultraviolet divergences. They can be removed by
subtracting local counter terms to the correlation function, i.e. by
defining the $m$-time subtracted function
$\Pi_{m}(p^2)$ ($m=1,2,\dots$) as
\be\label{eq:Pi_ns}
    \Pi(p^2) = \sum_{k=0}^{m-1} c_k (-p^2)^k + (-p^2)^m\, \Pi_{m}(p^2) \,,
\ee
where 
\be\label{eq:ck}
\Pi_{m}(p^2) = \int_0^\infty \frac{\rho_{2m}(\sqrt s)}{p^2+s}
\, ds\,, \quad
    \rho_m(\sqrt s) = \frac{\rho(\sqrt s)}{(\sqrt s)^{m}} \,, \quad 
    c_k = \int_0^\infty \frac{\rho(\sqrt s)}{s^{k+1}} ds \, , 
    \ee
with $\rho_m(\sqrt s)$ being the regulated spectral density. Each coefficient $c_k$ is, up to a sign, the derivative of order $k$ of $\Pi(p^2)$ with respect to $p^2$ computed at $p^2=0$.

\subsection{Spectral density as inverse Laplace transform}
By introducing the time-momentum representation of
the correlator
\be
C(x_0,{\bf p}) = \int {\rm C}(x)\, e^{-i{\bf p} {\bf x}}\, d^3{\bf x}
\, , \qquad x=({x_0,{\bf x}})\, , 
\ee
by using Eq.~(\ref{eq:KL}) and the known integral~\cite{Gradshteyn:1702455}
\be
    \int_{-\infty}^\infty \frac{\;\;\; e^{ip_0 x_0} }{p_0^2+\omega^2} \, \frac{dp_0}{2\pi} \, = 
    \frac1{2\omega} e^{-\omega |x_0|} \,,
\ee
it is easy to show that
\be
    C(x_0,{\bf p}) = \int_{_{\!\!\!\sqrt{{\bf p}^2}}}^\infty e^{-\omega x_0}  
    \rho(\sqrt{\omega^2-{\bf p}^2})\, d\omega \,  , \qquad x_0\geq 0\, , 
\ee
In Euclidean space-time, the spectral density can thus be
extracted as the ILT of the time-momentum representation of the
correlator. Following the same line of argumentation, one can relate
the subtracted correlators in the time-momentum representation with 
the regulated spectral densities as
\be
    C_{2m}(x_0,{\bf p}) = \int_{-\infty}^\infty \frac{dp_0}{2\pi}\,   e^{i
      p_0 x_0} \int_0^\infty 
    \frac{\rho_{2m}(\sqrt s)}{p_0^2+{\bf p^2}+s}\, ds  =
    \int_{_{\!\!\!\sqrt{{\bf p}^2}}}^\infty e^{-\omega x_0} 
    \rho_{2m}(\sqrt{\omega^2-{\bf p}^2})\, d\omega \, .
\ee

\section{Quasi-Carleman operators\label{app:carleman}}
We are interested in a class of Hankel operators, the so-called 
quasi-Carleman operators, defined as
\be
(A f) (x) = \int_0^{\infty} A(x+x') f(x') dx'
\ee
from the space of functions $f \in L^2(\mathbb R^+)$ onto itself with
kernels\footnote{We use the same symbol for operators and their
kernels since the ambiguity is easily resolved from the context.}  
\be\label{eq:quasi-carl}
A(x+x') = \frac{1}{x+x'+r}\; e^{-\beta (x+x')}\,,\qquad x,x'\geq 0 \, . 
\ee
The spectrum and the eigenfunctions of these operators for $r \geq 0$ and
$\beta=0$, and for $r=0$ and $\beta>0$ were derived many years
ago~\cite{Mehler:1881fgm,carleman,Magnus:1950wma,rosenblum,rosenblum2}, see also
Ref.~\cite{Titchmarsh:1946,Lebedev:1965,Yafaev:2014dry} for a review. Here we summarize
the results relevant for the present paper only.

\subsection{The Carleman operator}
The kernel of the Carleman operator is defined as in Eq.~(\ref{eq:quasi-carl}) with
$r = 0$ and $\beta=0$. Since
\be\label{eq:quasi-carl1}
A(t+t') =  \frac{1}{t+t'} = \int_0^{\infty} e^{-\omega (t+t')}\, d\omega \,,
\ee
by defining
\be\label{eq:MellinBasis}
u_s(t) = \frac{t^{is}}{\sqrt{2\pi t}}\,, \quad 
\lambda_s = \Gamma\left(\frac12+is\right)  \,, \quad s \in \mathbb R\,,
\ee
where $\Gamma$ is the Euler gamma function~\cite{Gradshteyn:1702455},
and by noticing
that 
\be\label{eq:utuw}
\int_0^{\infty} u_s(\omega) e^{-\omega t}\, d\omega = \lambda_s u_s^\ast(t)\, ,
\ee
it immediately follows that the $u_s(t)$ diagonalize the Carleman operator
\be\label{eq:carleig}
    \int_{0}^{\infty} \frac{1}{t+t'}\, u_s(t')\, dt' = \vert\lambda_s\vert^2 u_s(t) \,, \quad
    \vert\lambda_s\vert^2 = \frac\pi{\cosh(\pi s)} \,.
\ee
The orthonormality of the basis functions can be proven by considering the change in the integration variable
$\eta=\log(t)$ so that
\be\label{eq:us_ortho}
    \int_0^{\infty} u_s^\ast(t) u_{s'}(t)\, dt = 
    \int_{-\infty}^{\infty} e^{i \eta (s-s')}\, \frac{d\eta}{2\pi}  = \delta(s-s') \,.
\ee
The completeness is instead guaranteed by noticing that 
\be\label{eq:us_comp}
    \int_{-\infty}^{\infty} u_s^\ast(t) u_s(t')\, ds = \frac{\delta(\log(t/t'))}{\sqrt{tt'}} = \delta(t-t')\, ,
\ee
where $\delta(t-t')$ on the r.h.s. indicates a representation of the
$\delta$-function acting on the space of functions for which the
Mellin transform and its inverse exist, see Appendix~\ref{app:Trsf}.

In this basis, the kernel of the Carleman operator can be written in the diagonal form
\be\label{eq:A-us}
    A(t+t') =
    \int_{-\infty}^{\infty} \vert \lambda_s \vert^2\, u_s(t)\,
    u_s^\ast(t')\, ds \,.
\ee
Being the operator real, it is possible to choose a basis of real eigenfunctions
\begin{align}
    u_s^+(t) = \frac{u_s(t) + u_s^\ast(t)}{\sqrt 2} = \frac{\cos(s \log(t))}{\sqrt{\pi t}} \,, \\
    u_s^-(t) = \frac{u_s(t) - u_s^\ast(t)}{\sqrt 2 i} = \frac{\sin(s \log(t))}{\sqrt{\pi t}} \,,
\end{align}
with $s \in \mathbb R^+$. These functions satisfy the orthogonality relations
\be
\int_0^{\infty} u_s^\pm(t)\, u_{s'}^\pm(t)\, dt = \delta( s - s') \,,
\quad \int_0^{\infty} u_s^\pm(t)\, u_{s'}^\mp(t)\, dt = 0\, ,
\ee
while their completeness stems from
\be
    \int_{0}^{\infty} [ u_s^+(t) u_s^+(t') + u_s^-(t) u_s^-(t')
    ]\, d s = \delta(t-t') \,.
\ee
In this basis, the kernel can be written as 
\be
    A(t+t') =
    \int_{0}^{\infty} \vert \lambda_s \vert^2 \Big[ u_s^+(t)
    u_s^+(t') + u_s^-(t) u_s^-(t')\Big]\, ds \,,
\ee
making explicit the double degeneracy of the spectrum which originates from
the fact that the kernel of $A$ is singular when its argument approaches either zero or
infinity, see for instance~\cite{howlandII,Yafaev:2010lg}. Notice that, analogously to the case of
the Fourier transform, the two ensembles of
$u_s^+(t)$ and $u_s^-(t)$ are bases for the even and odd functions
under the transformation $t\rightarrow 1/t$, respectively. Indeed, to break the
degeneracy of the spectrum, one can introduce the additional operator
$\mathcal G(t,t')$ by starting from Eq.~(\ref{eq:carleig}) and by performing
the change of variable $t' \to 1/t'$ on the l.h.s so to obtain~\cite{howlandII}
\be
\int_{0}^{\infty} \mathcal G(t,t')\, u_s^\pm(t')\, dt' = \pm
\vert \lambda_s \vert^2\, u_s^\pm(t)\, 
\ee
with
\be\label{eq:G-app}
\mathcal G(t,t') =  \frac1{1+tt'} =
\int_{0}^{\infty} \vert\lambda_s\vert^2 \big[ u_s^+(t) u_s^+(t')
- u_s^-(t) u_s^-(t')\big]\, d s \,.
\ee

\subsection{The quasi-Carleman operator for $r=0$ and $\beta>0$}
For $r=0$ and $\beta=t_0>0$, the kernel in Eq.~(\ref{eq:quasi-carl}) reads
\be\label{eq:quasi-carl3}
A(\omega+\omega') = \frac{1}{\omega+\omega'}\; e^{-t_0 (\omega+\omega')} =
\int_{t_0}^{\infty} e^{-t(\omega+\omega')} dt\, . 
\ee
By defining
\be\label{eq:uhat}
\hat u_s(\omega,t_0) = \frac{\sqrt{2 s \sinh(\pi s)} }{\pi} \
\frac{K_{is}(\omega t_0)}{\sqrt{\omega}} 
\,,\qquad t_0>0 \,, \qquad s\in {\mathbb R}^+\, ,
\ee
where $K_{\nu}$ is the modified Bessel function of the second
kind~\cite{Gradshteyn:1702455}, from Eq.~(6.627) of
Ref.~\cite{Gradshteyn:1702455} it follows that
\be\label{eq:eiguhat}
    \int_{0}^{\infty} \frac{1}{\omega+\omega'}\; e^{-t_0 (\omega+\omega')}
    \, \hat u_s(\omega',t_0)\, d\omega' = \vert\lambda_s\vert^2 \hat u_s(\omega,t_0) \, ,
    \qquad \omega \geq 0\,,\;\; t_0>0\, .
\ee
The basis vectors $\hat u_s(\omega,t_0)$ satisfy the orthonormality
condition
\be\label{eq:orhouhat}
\int_{0}^{\infty} \hat u_s(\omega,t_0)\,\hat u_{s'}(\omega,t_0) \, d\omega = \delta(s-s') \,,
\ee
while their completeness stems from 
\be\label{eq:uhatcompl}
\int_0^{\infty} \hat u_s(\omega,t_0)\,\hat u_{s}(\omega',t_0) \, ds = \delta(\omega-\omega')\,
\qquad \omega,\omega'\geq 0\,,
\ee
where $\delta(\omega-\omega')$ on the r.h.s. indicates a representation of the
$\delta$-function acting on the space of functions for which the
Kontorovich–Lebedev transform and its inverse
exist, see Appendix~\ref{app:Trsf}. The kernel of the
quasi-Carleman operator in Eq.~(\ref{eq:quasi-carl3}) can then be written
in the diagonal form as 
\be
    A(\omega+\omega') =
    \int_{0}^{\infty} \vert \lambda_s \vert^2\,
    \hat u_s(\omega,t_0)\, \hat u_s(\omega',t_0)\, ds 
\ee
making explicit the non-degeneracy of the spectrum.

The Eq.~(\ref{eq:orhouhat}) can be derived by starting from 
Eq.~(6.576.4) of Ref.~\cite{Gradshteyn:1702455} which reads
\be\label{eq:biniz}
\int_{0}^{\infty} \frac{K_{is}(\omega) K_{is'}(\omega)}{\omega^{1-\epsilon}}  \, d\omega =
\frac{2^{\epsilon-3}}{\Gamma(\epsilon)}\,
\Big\vert\Gamma\Big(\frac{\epsilon+i(s+s')}{2}\Big)\Big\vert^2\;
\Big\vert\Gamma\Big(\frac{\epsilon+i(s-s')}{2}\Big)\Big\vert^2\, .
\ee
By using $\Gamma(z)=\Gamma(1+z)/z$, $|\Gamma(is)|^2=\pi/(s \sinh(\pi s))$,
and by noticing that the ratio $\epsilon/[\pi((s-s')^2+\epsilon^2)]$ is a representation of
the $\delta$-function when $\epsilon\rightarrow 0$, one arrives at~\cite{Passian:2009lg}
\be
\lim_{\epsilon\rightarrow 0}\int_{0}^{\infty}
\frac{K_{is}(\omega) K_{is'}(\omega)}{\omega^{1-\epsilon}}  \, d\omega = \frac{\pi^2}{2 s \sinh(\pi s)}
\delta(s-s')\, ,
\ee
from which it follows Eq.~(\ref{eq:orhouhat}). To understand how Eq.~(\ref{eq:uhatcompl}) can be
derived, we start by defining
\bea
I_\epsilon(\omega,\omega') & = &
\int_0^{\infty} e^{-\epsilon s}\, \hat u_s(\omega,t_0)\, \hat u_s(\omega',t_0)\, ds\nonumber\\[0.25cm]
& = & \frac{2}{\pi^2 \sqrt{\omega \omega'}} \int_0^{\infty} e^{-\epsilon s}\, s\,
\sinh(\pi s) K_{is}(\omega t_0)\, K_{is}(\omega' t_0)\, ds 
\eea
and, by using both representations in Eq.~(9.6.22) of Ref.~\cite{Abramowitz:lg}
\bea
K_{\nu}(x) & = & \frac{1}{\cos(\pi\nu/2)} \int_0^{\infty}\cos(x\sinh z)
\cosh(\nu z)\, dz \nonumber\\[0.25cm]
& = & \frac{1}{\sin(\pi\nu/2)}\int_0^{\infty}\sin(x\sinh z) \sinh(\nu z)\, dz \, , 
\eea
it follows that 
\bea
I_\epsilon(\omega,\omega') & = & \frac{2}{\pi^2\sqrt{\omega\omega'}}\int_0^\infty dz
\int_0^\infty dz' \sin(\omega t_0 \sinh z) \cos(\omega' t_0 \sinh z')\times\nonumber\\[0.325cm]
& &\hspace{1.5cm} \times \int_0^\infty e^{-\epsilon s} s \Big[ \sin(s(z+z')) + \sin(s(z-z'))\Big] d s\, .
\eea
By noticing that
\be
\int_0^{\infty} e^{-\epsilon s}\, s \sin(\alpha s)\, d s = -\pi
\frac{\partial}{\partial\alpha}\Big(\frac{\epsilon/\pi}{\alpha^2 + \epsilon^2}\Big)\, , 
\ee
and by choosing $\alpha=z\pm z'$, by replacing the derivative with respect to $\alpha$ with the one
with respect to $z$, and by integrating by parts we then obtain
\bea
I_\epsilon(\omega,\omega') & = & \frac{2 t_0}{\pi}\sqrt{\frac{\omega}{\omega'}} \int_0^\infty dz
\int_0^\infty dz' \cosh(z)\, \cos(\omega t_0 \sinh z)\, \cos(\omega' t_0 \sinh z')\,\times\nonumber\\[0.25cm]
& & \hspace{3.125cm} \times \Big[\,\frac{\epsilon/\pi}{(z+z')^2 + \epsilon^2} + \frac{\epsilon/\pi}{(z-z')^2 + \epsilon^2}\,\Big]\, .
\eea
We can now take the limit $\epsilon\rightarrow 0$ to obtain
\be\label{eq:Iepsfin}
\lim_{\epsilon\rightarrow 0} I_\epsilon(\omega,\omega') =
\frac{2 t_0}{\pi}\sqrt{\frac{\omega}{\omega'}}\int_0^\infty \cos(\omega t_0 x)
\cos(\omega' t_0 x)\, dx = \delta(\omega-\omega')\, , 
\ee
from which it follows Eq.~(\ref{eq:uhatcompl}).

\subsection{The quasi-Carleman operator for $r>0$ and $\beta=0$}
For $r=2 t_0 > 0$ and $\beta=0$ , the kernel in Eq.~(\ref{eq:quasi-carl}) reads
\be\label{eq:quasi-carl2}
A(t+t') = \frac{1}{t+t'+2 t_0} =
\int_0^{\infty} e^{-\omega (t+t'+ 2 t_0)}\, d\omega\,,\qquad t,t'\geq 0
\, .
\ee
By defining
\bea\label{eq:ubar}
\bar u_s(t,t_0) & = & \sqrt{\frac{s \tanh(\pi s)}{t_0}}\, P_{-\frac{1}{2}+is}\Big(\frac{t}{t_0}\Big)\nonumber\\[0.25cm]
& = & \sqrt{\frac{s \tanh(\pi s)}{t_0}}\, {}_2F_1\Big(\frac{1}{2}+is,\frac{1}{2}-is;1;\frac{t_0-t}{2 t_0}\Big)\,,
\;\; t \geq t_0>0\, \;\; s\in {\mathbb R}^+\, , 
\eea
where $P_{\nu}$ is the Legendre function and 
${}_2F_1$ is the Gauss's hypergeometric function~\cite{Gradshteyn:1702455}, it follows that 
\be\label{eq:poidisc}
    \int_{0}^{\infty} \frac{1}{t+t'+2\, t_0}\;
    \, \bar u_s(t'+t_0,t_0)\, dt' = \vert\lambda_s\vert^2\, \bar u_s(t+t_0,t_0) \,, \qquad
    t\geq 0\, , 
\ee
or equivalently
\be\label{eq:eigubar}
    \int_{t_0}^{\infty} \frac{1}{t+t'}\;
    \, \bar u_s(t',t_0)\, dt' = \vert\lambda_s\vert^2\, \bar u_s(t,t_0) \,,
    \qquad t\geq t_0\, .
\ee
The basis vectors $\bar u_s(t,t_0)$ satisfy the orthonormality
condition
\be\label{eq:ortubar}
\int_{t_0}^{\infty} \bar u_s(t,t_0)\,\bar u_{s'}(t,t_0) \, dt = \delta( s - s') \,,
\ee
while their completeness stems from
\be\label{eq:complubar}
\int_0^{\infty} \bar u_s(t,t_0)\,\bar u_{s}(t',t_0) \, ds = \delta(t - t') \,,
    \qquad t,t'\geq t_0>0\, ,
\ee
where $\delta(t-t')$ on the r.h.s. indicates a representation of the
$\delta$-function acting on the space of functions for which the
Mehler-Fock transform and its inverse exist, see
Appendix~\ref{app:Trsf}.
In this basis, the kernel of the quasi-Carleman operator in Eq.~(\ref{eq:quasi-carl2})
can be written in the diagonal form as
\be
    A(t+t') =
    \int_{0}^{\infty} \vert \lambda_s \vert^2\,
    \bar u_s(t,t_0)\, \bar u_s(t',t_0)\, ds \,,\qquad t,t'\geq t_0> 0 
\ee
making explicit the non-degeneracy of the spectrum.

The Eqs.~(\ref{eq:eigubar})-(\ref{eq:complubar}) can be derived directly from the analogous
ones~(\ref{eq:eiguhat})-(\ref{eq:uhatcompl}) in the previous subsection by noticing
that, see Eqs.~(6.621.3), (8.82) and (8.82.7) in Ref.~\cite{Gradshteyn:1702455} and 
Eq.~(15.3.4) in Ref.~\cite{Abramowitz:lg}, the $\bar u_s(t,t_0)$ is the Laplace
transform of $\hat u_s(\omega,t_0)$, i.e. 
\be\label{eq:lapuhat}
\int_{0}^{\infty}\, e^{-t\omega} \hat u_s(\omega,t_0)\, d\omega =
\vert \lambda_s \vert\, \bar u_s(t,t_0)\, . 
\ee
By computing an incomplete Laplace transform of the two sides of Eq.~(\ref{eq:lapuhat}), and 
by using Eqs.~(\ref{eq:quasi-carl3}) and Eq.~(\ref{eq:eiguhat}), it follows that
\be\label{eq:ubartouhat}
\int_{t_0}^{\infty} e^{-\omega t}\, \bar u_s(t,t_0)\, dt = \vert \lambda_s \vert\, \hat u_s(\omega,t_0)\, . 
\ee
The Eq.~(\ref{eq:eigubar}) can be derived\footnote{A direct derivation can also be obtained by
using Eq.~(\ref{eq:ubar}) and Eq.~(7.512.10) in Ref.~\cite{Gradshteyn:1702455}.} by performing the
Laplace transform of both sides of
Eq.~(\ref{eq:eiguhat}), and then by using Eq.~(\ref{eq:quasi-carl3}). The Eq.~(\ref{eq:ortubar})
is obtained by writing each of the $\bar u_s(t,t_0)$ as in Eq.~(\ref{eq:lapuhat}), and then by using
Eqs.~(\ref{eq:quasi-carl3}) and (\ref{eq:eiguhat}). A similar procedure applies to derive Eq.~(\ref{eq:complubar})
from Eq.~(\ref{eq:uhatcompl}). For completeness we notice that Eqs.~(\ref{eq:ortubar}) and (\ref{eq:complubar})
can also be obtained with a derivation analogous to the one in
Eqs.~(\ref{eq:biniz})--(\ref{eq:Iepsfin}) \cite{Sneddon_1972}.

\section{Integral transforms\label{app:Trsf}}
In this appendix we collect a few textbook formulas on the integral transforms
that are used in the paper.

\subsection{Mellin transform}
Given a function $f(x)$
which is piecewise continuous for $x \in {\mathbb R}^+$ and is of bounded variation in every
finite subinterval, its Mellin transform\footnote{With respect to the most common definition, we
introduce a normalization factor $1/\sqrt{2\pi}$.} reads~\cite{Wong:1989ll,Flajolet:1995ll}
\be
\tilde f_{\bf s} =  \frac{1}{\sqrt{2\pi}} \int_{0}^{\infty} f(x)\, x^{{\bf s}-1} dx =
\int_{0}^{\infty} f(x)\, x^{q-1/2} u_s(x) dx\, ,
\ee
where ${\bf s}\in {\mathbb C}$ is often defined as
${\bf s}=q+is$ with $q,s \in {\mathbb R}$, while $u_s(x)$ is given
in Eq.(\ref{eq:MellinBasis}). If
\be\label{eq:asympt}
\lim_{x\rightarrow 0^+} f(x) = O(x^{-u})\,, \qquad \lim_{x\rightarrow
  + \infty} f(x) = O(x^{-v}) \, ,
\ee
then $\tilde f_{\bf s}$ exists in the ``fundamental'' open strip of the complex plane
$q\in (u,v)$, $\tilde f_{\bf s}$ depends
analytically on ${\bf s}$ in that strip, and then the inversion
formula reads\footnote{For functions which are $L^2(\mathbb R^+)$, 
the $u_s(x)$ are the basis functions of choice and $q=1/2$.}
\be\label{eq:invmellin}
f(x) =  \frac{1}{\sqrt{2\pi} i} \int_{q-i\infty}^{q+i\infty} \tilde f_{\bf s}\,
x^{-{\bf s}} d {\bf s} =
x^{1/2-q}  \int_{-\infty}^{+\infty} \tilde f_{\bf s}\,
u^*_s(x) d s\,, \qquad q\in (u,v)\, .  \\[0.25cm]
\ee
One important property of the Mellin transform is the
rescaling rule which states that the transform of  $g(x) = f(\mu x)$
is   
\be\label{eq:rescal}
\tilde g_{\bf s} = \frac{1}{\mu^{{\bf s}}} \tilde f_{\bf s}\, ,
\qquad \mu>0\, .   
\ee
As an example relevant for this paper, the Mellin transform of
the simple exponential
\be\label{eq:exptilde}
f(x) = \exp{-x} \quad {\rm is} \quad \tilde f_{\bf s} = \frac{\Gamma({\bf s})}{\sqrt{2\pi}}\,,  
\ee
and is defined in the strip $q\in (0,+\infty)$. 

\subsection{Kontorovich–Lebedev transform}
The Kontorovich–Lebedev transform of a function
$f(\omega)$ is defined as\footnote{With respect to the standard definition
which assumes $t_0=1$, we let the value of $t_0>0$ be generic and leave the
dependence on it implicit for the clarity of the
presentation.}~\cite{Kontorovich1938lg,Kontorovich1939lg,Lebedev:1965,Glaeske:2006lg}
\be
\hat f_s = \int_{0}^{\infty} f(\omega)\, \hat u_s(\omega,t_0) d\omega\, ,\qquad s\in {\mathbb R}^+\,  
\ee
where $\hat u_s(\omega,t_0)$ is defined in Eq.~(\ref{eq:uhat}).
If $f(\omega)$, defined for  $\omega \in {\mathbb R}^+$, is piecewise continuous, of bounded
variation in every finite subinterval, and it satisfies 
\be\label{eq:KLasympt}
\lim_{\omega\rightarrow 0^+} f(\omega) = O(\omega^{-u})\,, \qquad \lim_{\omega\rightarrow
  + \infty} f(\omega) = O(\omega^{-v}) \, ,
\ee
then $\hat f_s$ and the pointwise inversion formula exist if $u<1/2$ and $v>1$~\cite{Lebedev:1965}. 
By using the completeness relation of the basis functions~(\ref{eq:uhatcompl}), the inversion 
formula then reads
\be
f(\omega) = \int_{0}^{\infty}\, \hat f_s\, \hat u_s(\omega,t_0)\, ds\, ,\qquad t_0>0\, .
\ee
As a consequence, the functions which satisfy the conditions above are $L^2(\mathbb R^+)$ but the
viceversa in not true, i.e. not all the functions in $L^2(\mathbb R^+)$ admit the inverse of
the Kontorovich–Lebedev transform which requires a stronger convergence to zero at infinity.

\subsection{Mehler-Fock transform}
The Mehler-Fock transform of a function
$f(t)$~\cite{Mehler:1881fgm,Fock:1943,Lebedev:1965,Glaeske:2006lg} is
defined as\footnote{We let again the value of $t_0>0$ be generic and
leave the dependence on it implicit.}
\be
\bar f_s = \int_{t_0}^{\infty} f(t)\, \bar u_s(t,t_0)\, dt\, ,\qquad s\in {\mathbb R}^+\, ,
\ee
where $\bar u_s(t,t_0)$ is defined in Eq.~(\ref{eq:ubar}).
If $f(t)$, defined for  $t\geq t_0$, is piecewise continuous, of bounded
variation in every finite subinterval, and it satisfies 
\be\label{eq:MFasympt}
\lim_{t\rightarrow t_0^+} f(t) = O((t-t_0)^{-u})\,, \qquad \lim_{t\rightarrow
  + \infty} f(t) = O(t^{-v}) \, ,
\ee
then $\bar f_s$ and the pointwise inversion formula exist if $u<1/4$ and
$v>1/2$~\cite{Lebedev:1965}. By using the completeness relation of the basis
functions~(\ref{eq:complubar}), the inversion  formula then reads
\be
f(t) = \int_{0}^{\infty} \bar f_s\, \bar u_s(t,t_0)\, ds\, . 
\ee
As a consequence, the functions which satisfy the conditions above are
$L^2(t_0,\infty) $ but the viceversa in not true, i.e. not all the functions in
$L^2(t_0,\infty)$ admit the inverse of the Mehler-Fock transform which
requires a weaker divergence in $t_0$.

\section{Quasi-Carleman operators on discrete\label{app:hilbert}}
With the intent of studying the discretized counterpart of the quasi-Carleman operators
introduced in Appendix~\ref{app:carleman}, here we summarize known facts
of the infinite Hilbert matrix~\cite{Magnus:1950wma,rosenblum,rosenblum2}
from which we derive properties of the discretized
quasi-Carleman operators of interest for this paper.

\subsection{Hilbert matrix}
The infinite Hilbert matrix  
\be
\A_{n n' } = \frac{1}{n+n'+\lambda}\, , \qquad n,n'=0,1,2,\dots\, , \quad \lambda\in {\mathbb R}\,,\quad \lambda\neq 0,-1,-2,\dots
\ee
can be regarded as an operator acting on $\ell^2(\mathbb Z^+)$. We begin from its diagonalization~\cite{Hill} 
\be\label{eq:hilbertC}
    \sum_{n'=0}^\infty A_{n n' }\, x_{n'}(\lambda,\mu) = 
    \frac{\pi}{\sin(\pi \mu)}\, x_{n}(\lambda,\mu) \,,
\ee
where $0< {\rm Re}\, \mu \leq 1/2$, with the components of the eigenvectors
given by 
\be\label{eq:xn}
    x_n(\lambda,\mu) = \sum_{k=0}^n \binom{n}{k} (-1)^k\,
    \frac{\Gamma(k+\mu) \Gamma(k+1-\mu)}{\Gamma(k+\lambda)
    \Gamma(k+1)} \,, \qquad n=0,1,2,\dots\, .
\ee
By noticing that
\be
\binom{n}{k} (-1)^k = \frac{(-n)_k}{k!}=\frac{1}{k!}(-n)(-n+1)\dots(-n+k-1)\, ,
\ee
with the rising factorial (Pochhammer symbol) defined as 
\be
    (x)_{k} = x\, (x+1)\dots(x+k-1) \,,
\ee
and by using the definition of the generalized hypergeometric series in Eq.~(9.14.1) of
Ref.~\cite{Gradshteyn:1702455}, it is easy to show that 
\be\label{eq:xnlammu}
    x_n(\lambda,\mu) = \frac{\Gamma(1-\mu)\Gamma(\mu)}{\Gamma(\lambda)} \ {}_3F_2 (-n, \mu, 1-\mu;1,\lambda;1)\, .
\ee
For the case of interest, $\mu=\frac12+is$, the eigenvalues in Eq.~(\ref{eq:hilbertC}) coincide
with those in the continuum, i.e. $\pi/\sin(\pi \mu) = \vert \lambda_s \vert^2$,
see Eq.~(\ref{eq:carleig}). To verify the completeness of the basis of eigenvectors, we notice that
the $x_n(\lambda,\frac12+is)$ are related to the so-called continuous
dual Hahn polynomials, defined as~\cite{hahn1,hahn2,hahn3}
\be\label{eq:hahn_poly}
    S_n(s^2;a,b,c) = (a+b)_n (a+c)_n\; {}_3F_2(-n,a+is,a-is;a+b,a+c,1)\,.
\ee
They satisfy the orthogonality relation, see for instance Ref.~\cite{hahn1},
\be
\begin{aligned}
    \frac1{2\pi} \int_0^\infty ds \, \vert W_s(a,b,c) \vert^2 & \, S_m(s^2;a,b,c) \, S_n(s^2;a,b,c)   \\
    & = n!\, \Gamma(n+a+b) \Gamma(n+a+c) \Gamma(n+b+c)\, \delta_{nm} \,,
    \label{eq:ortho1}
\end{aligned}
\ee
with
\be
    W_s(a,b,c) = \frac{\Gamma(a+is) \Gamma(b+is) \Gamma(c+is)}{\Gamma(2is)} \,.
\ee
By choosing $a=b=1/2$ and $c=\lambda-1/2$, we obtain the completeness relation for the normalized
eigenvectors of the Hilbert matrix
\be\label{eq:xmu_comp}
  \frac{1}{2\pi}  \int_0^\infty \Big\vert\frac{\Gamma(\lambda -1/2 + is)}{\Gamma(2 is)}\Big\vert^2
    x_n\big(\lambda,\frac12+is\big) \, x_m\big(\lambda,\frac12+is\big)\, ds  
    = \delta_{nm}\, , 
\ee
from which one can verify explicitly the orthogonality condition
\be\label{eq:normxn}
\sum_{n=0}^\infty x_n\big(\lambda,\frac12+is\big)\, x_n\big(\lambda,\frac12+is'\big) = 2\pi
\Big\vert\frac{\Gamma(2 is)}{\Gamma(\lambda -1/2 + is)}\Big\vert^2
\delta(s-s')\,.
\ee

\subsection{Discretized quasi-Carleman operator for $r>0$ and $\beta=0$}
We consider the discretization of the kernel of the quasi-Carleman operator in
Eq.~(\ref{eq:quasi-carl2}) defined as 
\be\label{eq:AaC}
    \A(n + n') = \frac{1}{a(n+n')+2 t_0} =
    \int_0^{\infty} e^{-\omega (a n + a n'+ 2 t_0)}\, d\omega\,, \quad (n,n'=0,1,2,\dots)\, ,    
\ee
where $a$ is the lattice spacing. By using Eq.~(\ref{eq:hilbertC}),
we obtain
\be\label{eq:discr-sonc-t}
a\!\! \sum_{n'=0}^\infty \A(n + n') \, \overline v_s(n',t_0,a) =\!\! 
\sum_{n'=0}^\infty \frac{1}{n+n'+2 t_0/a} \, \overline v_s(n',t_0,a)
= \vert\lambda_s\vert^2\, \overline v_s(n,t_0,a) \, ,
\ee
with $s \in\mathbb R^+$ and, thanks to Eqs.~(\ref{eq:xnlammu}), (\ref{eq:normxn}), and (\ref{eq:AaC}), the eigenvectors
are given by 
\be\label{eq:vs_vsbar}
\overline v_s(n,t_0,a) =  \sqrt{\frac{s \tanh(\pi s)}{t_0}}\,
\frac{\big\vert\Gamma(2t_0/a\!-\!1/2\!+\!is)\big\vert}{\sqrt{a/(2 t_0)}\, \Gamma(2t_0/a)}\,
{}_3F_2\Big(-n,\frac{1}{2}\!+\!is,\frac{1}{2}\!-\!is;1,\frac{2t_0}{a};1\Big)\, .
\ee
The Eq.~(\ref{eq:discr-sonc-t}) is the discretized version of Eq.~(\ref{eq:poidisc}) that we adopt.
As expected, the eigenvalues of $\A$ coincide with
those in the continuum, i.e. $\vert\lambda_s\vert^2$, since 
for $2 t_0=a$ the matrix $\A(n + n')$ corresponds to a discrete
representation of the quasi-Carleman operator in
Eq.~(\ref{eq:quasi-carl3}), see Ref.~\cite{Yafaev:2017dry} for a
recent review.

Thanks to Eqs.~(\ref{eq:xmu_comp}) and (\ref{eq:normxn}), the following
completeness and orthogonality conditions hold
\be\label{eq:complortodisc}
    \int_{0}^{\infty} \overline v_s(n,t_0,a) \overline v_s(n',t_0,a)\, d s =
    a^{-1} \delta_{nn'}\,, \quad 
    a \sum_{n=0}^\infty \overline v_s(n,t_0,a) \overline v_{s'}(n,t_0,a) =
    \delta(s-s')\,.
\ee

\noindent {\it Continuum limit at fixed $t_0$}\\[0.125cm]
We study the approach to the continuum limit of $\overline v_s(n,t_0,a)$
when $t_0$ and $t=a n+ t_0$ are kept constant. To this
aim, we begin with the representation of 
the generalized hypergeometric series in Eq.~(2.2.2) of Ref.~\cite{andrews_1999}
\be\label{eq:3F2to2F1}
\hspace{-0.25cm} {}_3F_2\Big(\lambda z,\frac{1}{2}\!+\! is,\frac{1}{2}\!-\! is;1,\lambda;1\Big)\! =\!
\frac{1}{\vert \lambda_s \vert^2}
\int_0^1\!\!\!  x^{-\frac{1}{2} - is} (1\!-\!x)^{-\frac{1}{2}+ is}
{}_2F_1\Big(\lambda z,\frac{1}{2}\!+\! is;\lambda;x\Big)\, dx\, ,
\ee
where $\lambda=2t_0/a$ and $z=(t-t_0)/(2t_0)$. We then represent the Gauss hypergeometric function
as in Eq.~(3.197.1) of Ref.~\cite{Gradshteyn:1702455}, but with the change
of integration variable $y\rightarrow y/(1-x)$, i.e.
\be
\hspace{-0.25cm} {}_2F_1\Big(\lambda z,\frac{1}{2}\!+\!is;\lambda;x\Big) =
\frac{\Gamma(\lambda)}{\lambda_s \Gamma(\lambda\! -\! \frac{1}{2}\! -\! i s)}
\int_0^\infty\!\!\!\! y^{-\frac{1}{2} + i s}\, e^{-\lambda [z \log(1+(1-x) y) + (1-z) \log(1+y)]}\, dy\, .
\ee
By noticing that the expression in the squared brackets at the exponent,
for $z<0$ and $x\in(0,1)$, has its minimum for $y=0$,
we use the Laplace method at the next-to-leading order in $1/\lambda$
to obtain
\bea
& &\hspace{-0.75cm} {}_2F_1\Big(\lambda z,\frac{1}{2} + is;\lambda;x\Big) = 
\frac{\Gamma(\lambda) \lambda^{-\frac{1}{2}- is}}{\Gamma(\lambda\!-\!\frac{1}{2}\!-\! is)}\times\nonumber\\[0.25cm]
& & \times \Big\{\Big[1+ \frac{1}{2\lambda}\,\Big[\frac{3}{4}+ 2 i s - s^2 +z(1-z)
\frac{\partial^2}{\partial z^2}\Big] \Big]\,
(1-xz)^{-\frac{1}{2} - is} + O\Big(\frac{1}{\lambda^2} \Big)\Big\}\, .
\eea
By inserting this equation in Eq.~(\ref{eq:3F2to2F1}), using Eq.~(9.111)  in Ref.~\cite{Gradshteyn:1702455},
and then plugging in the result in Eq.~(\ref{eq:vs_vsbar}), we obtain
\be
\overline v_s(n,t_0,a) = \frac{\vert \Gamma(\lambda - \frac{1}{2} + i s) \vert}{\Gamma(\lambda - \frac{1}{2} - i s)}
\lambda^{-i s} \Big\{\Big[1 + \frac{1}{2\lambda}\Big[1+ 2is+ 2 t \frac{\partial}{\partial t}\Big] \Big] \, \bar u_s(t,t_0)
+ O\Big(\frac{1}{\lambda^2}\Big)\Big\}\, , 
\ee
where $t=an+t_0$ and we have used the differential equation for the Legendre
functions in Eq.~(8.1.1) of Ref.~\cite{Abramowitz:lg}
\be
\Big[(t_0^2 - t^2)\frac{\partial^2}{\partial t^2}- 2 t \frac{\partial}{\partial t} - \frac{1}{4} - s^2\Big]\, \bar u_s(t,t_0)=0\, .
\ee
By using Eqs.~(6.1.27) and (6.3.18) of Ref.~\cite{Abramowitz:lg} we obtain
\be
\frac{\vert \Gamma(\lambda - \frac{1}{2} + i s) \vert}{\Gamma(\lambda - \frac{1}{2} - i s)}
\lambda^{-i s} = 1 - \frac{is}{\lambda} + O\Big(\frac{1}{\lambda^2}\Big)\, , 
\ee
which finally leads to
\be\label{eq:fiuuu1}
\overline v_s(n,t_0,a) = \Big\{1 + \frac{a}{4 t_0}\, \Big[1+2
t \frac{\partial}{\partial t}\Big]\Big\}\, \bar u_s(t,t_0) + O(a^2)\, ,
\ee
from which the derivative of $\overline v_s(n,t_0,a)$ with respect to $a$
in the limit $a \rightarrow 0$ is readily obtained. As expected the basis vectors
$\overline v_s(n,t_0,a)$ tend to the continuum ones $\bar u_s(t,t_0)$, whereas
discretization effects start at $O(a)$.

\subsection{Discretized quasi-Carleman operator for $r=0$ and $\beta>0$}
By adopting a naive discretization of the integral on the r.h.s of Eq.~(\ref{eq:quasi-carl3}),
the kernel of that quasi-Carleman operator is modified as
\be\label{eq:quasi-carl3d}
\A(\omega+\omega') = \frac{a\, e^{-t_0(\omega+\omega')}}{1-e^{-a (\omega+\omega')}} = 
a \sum_{n=0}^{\infty} e^{-(a n+t_0)(\omega+\omega')}\, . 
\ee
The Eq.~(\ref{eq:eiguhat}) is then replaced by 
\be\label{eq:eigvhat}
    \int_{0}^{\infty} \frac{a\, e^{-t_0(\omega+\omega')}}{1-e^{-a (\omega+\omega')}}
    \, \hat v_s(\omega',t_0,a)\, d\omega' = \vert\lambda_s\vert^2 \hat v_s(\omega,t_0,a) \, ,
    \qquad \omega \geq 0\,,\;\; t_0>0\, ,
\ee
with the eigenfunctions given by 
\bea\label{eq:vhat}
\hat v_s(\omega,t_0,a) &= & \frac{\sqrt{2 s \sinh(\pi s)} }{\pi}
\frac{\big\vert\Gamma(2t_0/a\!-\!1/2\!+\!is)\big\vert}{\sqrt{a/(2 t_0)}\, \Gamma(2t_0/a)}\,
\times\nonumber\\[0.25cm]
& & \times \sqrt{\frac{\pi}{2 t_0}}\, e^{- \omega t_0}\, \frac{a}{1-e^{-a\omega}}\; 
{}_2F_1\Big(\frac{1}{2}+is,\frac{1}{2}-is;\frac{2t_0}{a};-\frac{e^{-a \omega}}{1-e^{-a \omega}}\Big)\, .
\eea
The $\hat v_s(\omega,t_0,a)$ and $\overline v_s(t,t_0,a)$ are related by
the equation 
\be\label{eq:lapdisc1}
 a \sum_{n=0}^{\infty} e^{- \omega (an+t_0)}\, \bar v_s(n,t_0,a) =
 \vert\lambda_s\vert \hat v_s(\omega,t_0,a)\, 
\ee
which is a discretized version of the Laplace transform in Eq.~(\ref{eq:ubartouhat}).
It can be derived by combining Eqs.~(\ref{eq:xn}) and (\ref{eq:xnlammu}) so to have
\bea
\sum_{n=0}^\infty \zeta^{n}\, {}_3F_2 (-n, \mu, 1\!-\!\mu;1,\lambda;1)\! & = &\! \frac{\Gamma(\lambda)}{\Gamma(1\!-\!\mu)\Gamma(\mu)}
\sum_{k=0}^\infty (-1)^k\,
    \frac{\Gamma(k\!+\!\mu) \Gamma(k\!+\!1\!-\!\mu)}{\Gamma(k\!+\!\lambda)
    \Gamma(k\!+\!1)}\sum_{n=k}^\infty  \binom{n}{k} \zeta^{n}\nonumber\\[0.25cm]
    & = & \frac{1}{1-\zeta}\; {}_2F_1 \Big(\mu, 1-\mu;\lambda;-\frac{\zeta}{1-\zeta}\Big)\, , 
\eea
where we have used the identity
\be
\sum_{n=k}^\infty  \binom{n}{k} \zeta^{n} = \frac{\zeta^k}{(1-\zeta)^{k+1}}
\ee
obtained by deriving $k$ times the summation of the geometric series
with respect to $\zeta$. Then Eq.~(\ref{eq:vhat}) is obtained from
Eqs.~(15.1.1) in Ref.~\cite{Abramowitz:lg} by choosing $\zeta = e^{-a\omega}$.

The Eq.~(\ref{eq:eigvhat}) can easily be derived by starting from Eq.~(\ref{eq:discr-sonc-t}),
by representing the kernel as on the r.h.s of Eq.~(\ref{eq:AaC}), then by
performing a discrete Laplace transform on both sides of the equation, and finally by
using Eq.~(\ref{eq:quasi-carl3d}). By performing the Laplace transform on both sides of
Eq.~(\ref{eq:lapdisc1}), and by using Eqs.~(\ref{eq:AaC}) and Eq.~(\ref{eq:discr-sonc-t}) it
follows that
\be\label{eq:lapdisc2}
\int_{0}^{\infty} e^{- (an+t_0) \omega}\,
\hat v_s(\omega,t_0,a)\, d\omega\,  =
\vert\lambda_s\vert\, \bar v_s(n,t_0,a)\, , 
\ee
the analogous of Eq.~(\ref{eq:lapuhat}).
By using Eqs.~(\ref{eq:AaC}) and (\ref{eq:discr-sonc-t}), and
the completeness and orthonormality conditions in Eq.~(\ref{eq:complortodisc})
it follows that 
\be
\hspace{-0.25cm} \int_{0}^{\infty}\!\!\! \hat v_s(\omega,t_0,a)\, \hat v_{s'}(\omega,t_0,a)\, d\omega =
\delta(s-s')\, ,\;\;
\int_{0}^{\infty}\!\!\! \hat v_s(\omega,t_0,a)\, \hat v_{s}(\omega',t_0,a)\, d s =
\delta(\omega-\omega')\, ,
\ee
where $\delta(\omega-\omega')$  on the r.h.s. indicates a representation of the $\delta$-function
acting on the space of functions for which Eq.~(\ref{eq:ctd}) can be uniquely inverted.\\

\noindent {\it Continuum limit at fixed $t_0$}\\[0.125cm]
To study the approach to the continuum limit
of the functions $\hat v_s(\omega,t_0,a)$ defined in Eq.~(\ref{eq:vhat}), we start from Eq.~(15.4.16) of
Ref.~\cite{Abramowitz:lg} which, by defining $\lambda=2 t_0/a$, reads
\be\label{eq:2F1toP}
{}_2F_1\Big(\frac{1}{2}+is,\frac{1}{2}-is;\lambda;-\frac{e^{-2 \omega t_0/\lambda}}{1-e^{-2 \omega t_0/\lambda}}\Big) =
\Gamma(\lambda)\, e^{\omega t_0}\, e^{-\omega t_0/\lambda}\,
P^{1-\lambda}_{-\frac{1}{2} - is}\Big(\coth\Big(\frac{\omega t_0}{\lambda}\Big)\Big)\, .
\ee
Thanks to Eq.~(8.713.3) of Ref.~\cite{Gradshteyn:1702455}, the associated Legendre function of first
kind can be represented as
\be\label{eq:PtoJ}
P^{1-\lambda}_{-\frac{1}{2} - is}(z) = \sqrt{\frac{2}{\pi}}\,
\frac{\Gamma(\bar\lambda)}{\vert\Gamma(\bar\lambda - is)\vert^2}\, J(z,\bar\lambda)\, , 
\quad \bar\lambda = \lambda - \frac{1}{2}\, ,\quad z= \coth\Big(\frac{\omega t_0}{\lambda}\Big)\, ,
\ee
where
\be
 \quad J(z,\bar\lambda) = (z^2-1)^{\frac{\bar\lambda}{2}-\frac{1}{4}}\,
\int_0^\infty \cosh(isx)\, e^{-\bar\lambda \log(z+\cosh x)}\, dx\, .
\ee
By defining $y^2=(\cosh x -1)$, then 
\be
J(z,\bar\lambda) = 2\, (z^2-1)^{\frac{\bar\lambda}{2}-\frac{1}{4}} \int_0^\infty 
\cosh(is x(y))\, e^{-\bar\lambda \log(z+1+y^2)}\, \frac{dy}{\sqrt{2+y^2}}\, , 
\ee
and by using the Laplace method at the next-to-leading order in $1/\bar\lambda$
we obtain
\bea
\hspace{-0.25cm} J(z,\bar\lambda) & = &  (z^2-1)^{\frac{\bar\lambda}{2}-\frac{1}{4}}\,
(z+1)^{-\bar\lambda}\, e^{\alpha} \times \nonumber\\[0.25cm]
& &  \times \Big\{\Big[1 + \frac{\bar\lambda}{2(z + 1)^2} \Big[1 + 
2\frac{\partial}{\partial \alpha} + \frac{\partial^2}{\partial \alpha^2}\Big] \Big]
K_{is}(\alpha)\Big|_{\alpha=\frac{\bar\lambda}{z+1}}+ O\Big(\frac{1}{\lambda^2}\Big)\Big\}\, ,
\eea
where we have used the integral representation of the modified Bessel function
in  Eq.~(9.6.24) of Ref.~\cite{Abramowitz:lg}
\be
K_{is}(\alpha) = \int_0^\infty \cosh(isx)\, e^{-\alpha \cosh x}\, dx \, .
\ee
By expanding the various terms in $1/\lambda$,  we obtain 
\be
J(z,\bar\lambda) = \sqrt{\frac{t_0}{\lambda}}\, \Big\{\omega\, \Big[1 - \frac{1}{2 \lambda} \Big[1+s^2 +2
\omega \frac{\partial}{\partial\omega}\Big]\Big] \frac{K_{is}(\omega t_0)}{\sqrt{\omega}} +
O\Big(\frac{1}{\lambda^2}\Big)\Big\}\, ,
\ee
where we have used the differential equation for the modified Bessel function in Eq.~(9.6.1) of Ref.~\cite{Abramowitz:lg}
\be
\Big[\omega^2 \frac{\partial^2}{\partial \omega^2} + \omega \frac{\partial}{\partial\omega}
- (\omega t_0)^2 +s^2\Big]\, K_{is}(\omega t_0) = 0\, .
\ee
By expanding at the next-to-leading order in $1/\lambda$ the prefactors
in Eqs.~(\ref{eq:vhat}) and (\ref{eq:PtoJ}), and by setting $\lambda=2 t_0/a$ we finally obtain
\be\label{eq:fiuuu2}
\hat v_s(\omega,t_0,a) = \Big\{1 - \frac{a}{4 t_0}\, \Big[1+2
\omega \frac{\partial}{\partial\omega}\Big]\Big\}\, \hat u_s(\omega,t_0) + O(a^2)\, ,
\ee
from which the derivative of $\hat v_s(\omega,t_0,a)$ with respect to $a$
in the limit $a \rightarrow 0$ is readily obtained. As expected the basis vectors
$\hat v_s(\omega,t_0,a)$ tend to the continuum ones $\hat u_s(\omega,t_0)$,
whereas discretization effects start at $O(a)$. As further checks, we have verified that, order
by order, the r.h.s. of Eq.~(\ref{eq:fiuuu1}) is obtained from the r.h.s. of 
Eq.~(\ref{eq:fiuuu2}) by using Eq.~(\ref{eq:lapdisc2}) and viceversa by using
Eq.~(\ref{eq:lapdisc1}).

\section{Alternative strategy for (smeared) regulated spectral densities\label{app:appsub}}
In this appendix, we present an alternative method to compute the regulated spectral
density $\rho_m(\omega)$. The idea is to modify the ultraviolet behavior of $C(t)$
so that, in the presence of a mass gap, the modified correlation function
\begin{equation}\label{eq:rho_bar}
    C_{(m)}(t) = 
    C(t)\, t^m 
    = \int_0^\infty \rho(\omega)\, \Big( - \frac{\partial}{\partial\omega}\Big)^m e^{-\omega t}\, d\omega
    = \int_0^\infty \rho_{(m)}(\omega)\, e^{-\omega t}\, d\omega 
\end{equation}
is associated to the spectral density
\begin{equation}\label{eq:rho_m_def}
    {\rho}_{(m)}(\omega) = \frac{\partial^m}{\partial\omega^m} \big[ \rho(\omega) \big] = \frac{\partial^m}{\partial\omega^m} \big[ \omega^m \rho_m(\omega) \big] \,.
\end{equation}
This relation can be inverted similarly to Eq.~(\ref{eq:Cmdiff}). By noticing that, in
the presence of a mass gap, $\rho_m(\omega)$ and its derivatives vanish at $\omega=0$, then
the regulated spectral density can be obtained as
\begin{equation}\label{eq:rho_m}
    \rho_m(\omega) = \frac{1}{\Gamma(m)\, \omega^m} \int_0^\omega {\rho}_{(m)}(\omega')\,
    (\omega - \omega')^{m-1} d \omega' \, \,.
\end{equation}
The smeared spectral density
$\rho_{\kappa;m}$, see Eq.~(\ref{eq:rhokappa}), directly follows 
from integrating $\rho_m(\omega)$, written as in Eq.~(\ref{eq:rho_m}), with $\kappa(\omega)$
\begin{equation}\label{eq:rho_m_smea}
    \rho_{\kappa;m} = \int_0^\infty \rho_{(m)}(\omega) \kappa_{(m)}(\omega)\, d\omega\, ,
\end{equation}
where
\begin{equation}
    \kappa_{(m)}(\omega) = \frac1{\Gamma(m)} \int_{\omega}^\infty 
    \kappa(\omega') \frac{(\omega'-\omega)^{m-1}}{\omega^{' m}}\, d\omega' \, .
\end{equation}
For the Breit-Wigner kernel considered in this work, see Eq.~(\ref{eq:BWsm}),
this integral amounts to
\begin{equation}
\kappa_{(m)}(\omega) = \frac{1}{\pi\, \Gamma(1+m)\, \omega} \, \text{Im} \left\{ {}_2 F_1\left(
        1,1;1+m;\frac{\omega_\ast+i\sigma}{\omega}
    \right)\right\} \,,
\end{equation}
where in the derivation Eqs.~(3.197.2) and (15.3.3) of Refs.~\cite{Gradshteyn:1702455} and \cite{Abramowitz:lg}
have been used, respectively.

\bibliographystyle{JHEP.bst}
\bibliography{biblio.bib}

\end{document}